\documentclass[twocolumn]{aastex62}

\usepackage{my_astro}
\usepackage{courier}
\usepackage[super]{nth}
\usepackage{hyperref}
\usepackage{hyperref}

\newcommand{\md}{M~dwarf}
\newcommand{\mds}{M~dwarfs}

\newcommand{\smoothn}{100}
\newcommand{\bandwidth}{200 km \pers}

\newcommand{\interlineDivide}{1430 \AA}
\newcommand{\interlineLoLim}{1170 \AA}
\newcommand{\fuvlo}{912 \AA}
\newcommand{\fuvhi}{1700 \AA}
\newcommand{\statsSNcut}{2}
\newcommand{\fuvwaverough}{1170~--~1270~$+$~1330~--~1430~\AA}
\newcommand{\gfuv}{FUV$_{130}$}

\newcommand{\epseri}{$\mathrm{\epsilon}$ Eri}
\newcommand{\dcrit}{\delta_\mathrm{crit}}
\newcommand{\tform}{T_\mathrm{form}}

\newcommand{\bbratio}{160}
\newcommand{\bbT}{9000~K}
\newcommand{\fidSiivqF}{1~\fluxcgs}
\newcommand{\fidSiivqM}{0.1~\fluxcgs}
\newcommand{\EUVinactive}{\nu = 0.75\ \mathrm{d}^{-1}\ (E/10^{30}\ \mathrm{erg})^{-0.5}}
\newcommand{\EUVpew}{\nu = 0.6\ \mathrm{d}^{-1}\ (\delta/1000\ \mathrm{s})^{-0.5}}

\newcommand{\fidurl}{https://github.com/parkus/fiducial_flare}
\newcommand{\ffdurl}{http://www.github.com/parkus/ffd}
\newcommand{\Fsfc}{$F_\mathrm{sfc}$}
\newcommand{\EfEq}{$E_f/E_q$}

\received{2018 March 1}
\revised{2018 August 27}
\accepted{2018 September 17}
\submitjournal{ApJ}

\begin{document}

\title{The MUSCLES Treasury Survey. V. FUV Flares on Active and Inactive M Dwarfs
\footnote{Based on observations made with the NASA/ESA Hubble Space Telescope, obtained from the data archive at the Space Telescope Science Institute. STScI is operated by the Association of Universities for Research in Astronomy, Inc. under NASA contract NAS 5-26555.}
\footnote{The scientific results reported in this article are based in part on observations made by the Chandra X-ray Observatory.}
\footnote{Based in part on observations obtained with XMM-Newton, an ESA science mission with instruments and contributions directly funded by ESA Member States and NASA}}

\correspondingauthor{R. O. Parke Loyd}
\email{robert.loyd@colorado.edu}

\author{R. O. Parke Loyd}
\affiliation{School of Earth and Space Exploration, Arizona State University, Tempe, AZ 85287}
\affiliation{Laboratory for Atmospheric and Space Physics, Boulder, CO 80309}

\author{Kevin France}
\affiliation{Laboratory for Atmospheric and Space Physics, Boulder, CO 80309}

\author{Allison Youngblood}
\affiliation{Goddard Space Flight Center, Greenbelt, MD 20771}
\affiliation{Laboratory for Atmospheric and Space Physics, Boulder, CO 80309}

\author{Christian Schneider}
\affiliation{Hamburger Sternwarte, Gojenbergsweg 112, 21029, Hamburg, Germany}
\affiliation{Scientific Support Office, Directorate of Science, European Space Research and Technology Center (ESA/ESTEC), Keplerlaan 1, 2201, AZNoordwijk, The Netherlands}

\author{Alexander Brown}
\affiliation{Center for Astrophysics and Space Astronomy, University of Colorado, 389 UCB, Boulder, CO 80309, USA}

\author{Renyu Hu}
\affiliation{Jet Propulsion Laboratory, California Institute of Technology, Pasadena, CA 91109}
\affiliation{Division of Geological and Planetary Sciences, California Institute of Technology, Pasadena, CA 91125}

\author{Ant\'igona Segura}
\affiliation{NASA Astrobiology Institute Virtual Planetary Laboratory, Box 351580, U.W. Seattle, WA 98195, USA}
\affiliation{Instituto de Ciencias Nucleares, Universidad Nacional Aut\'onoma de M\'exico, M\'exico}

\author{Jeffrey Linsky}
\affiliation{JILA, University of Colorado and NIST, 440 UCB, Boulder, CO 80309}

\author{Seth Redfield}
\affiliation{Astronomy Department and Van Vleck Observatory, Wesleyan University, Middletown, CT 06459, USA}

\author{Feng Tian}
\affiliation{Department of Earth System Science, Tsinghua University, Beijing 100084, China}

\author{Sarah Rugheimer}
\affiliation{Centre for Exoplanet Science, University of St. Andrews, School of Earth and Environmental Sciences, Irvine Building, North Street, St. Andrews, KY16 9AL, UK}
\affiliation{Harvard Smithsonian Center for Astrophysics, 60 Garden Street, Cambridge, MA 02138, USA}
\affiliation{Carl Sagan Institute, Department of Astronomy, Cornell University, Ithaca, NY 14853, USA}

\author{Yamila Miguel}
\affiliation{Leiden Observatory, NL-2333 CA Leiden, The Netherlands}

\author{Cynthia S. Froning}
\affiliation{Department of Astronomy/McDonald Observatory, C1400, University of Texas at Austin, Austin, TX 78712, USA}

\begin{abstract}
M dwarf stars are known for their vigorous flaring.
This flaring could impact the climate of orbiting planets, making it important to characterize M dwarf flares at the short wavelengths that drive atmospheric chemistry and escape.
We conducted a far-ultraviolet flare survey of 6 M dwarfs from the recent MUSCLES (Measurements of the Ultraviolet Spectral Characteristics 
of Low-mass Exoplanetary Systems) observations, as well as 4 highly-active M dwarfs with archival data.
When comparing absolute flare energies, we found the active-M-star flares to be about 10$\times$ more energetic than inactive-M-star flares.
However, when flare energies were normalized by the star's quiescent flux, the active and inactive samples exhibited identical flare distributions, with a power-law index of -$0.76^{+0.1}_{-0.09}$ (cumulative distribution).
The rate and distribution of flares are such that they could dominate the FUV energy budget of M dwarfs, assuming the same distribution holds to flares as energetic as those cataloged by \textit{Kepler} and ground-based surveys. 
We used the observed events to create an idealized model flare with realistic spectral and temporal energy budgets to be used in photochemical simulations of exoplanet atmospheres.
Applied to our own simulation of direct photolysis by photons alone (no particles), we find the most energetic observed flares have little effect on an Earth-like atmosphere, photolyzing $\sim$0.01\% of the total \OIII\ column. 
The observations were too limited temporally (73 h cumulative exposure) to catch rare, highly energetic flares.
Those that the power-law fit predicts occur monthly would photolyze $\sim$1\% of the \OIII\ column and those it predicts occur yearly would photolyze the full \OIII\ column.
Whether such energetic flares occur at the rate predicted is an open question.
\end{abstract}

\section{Introduction}
\label{sec:intro}
Exoplanet science is swiftly advancing toward an answer to the question ``How typical is Earth?''
Results from the \textit{Kepler} mission have shown 10-60\% of F~--~M stars harbor a planet of super-Earth size or smaller orbiting in the liquid-water habitable zone (e.g., \citealt{traub12,gaidos13,dressing15}), establishing that planets the size, mass, and equilibrium temperature of Earth are common. What remains to be learned is whether the Earth's atmosphere and corresponding climate are common as well.

The atmospheric evolution of a planet is influenced by both its intrinsic properties and its space environment.
If most terrestrial planets in the habitable zone orbited Sun-like stars, one might assume their space-environment would pose no major challenges to evolving an atmosphere like Earth's. 
However, most habitable-zone planets orbit \mds\ -- a consequence of the plurality of M dwarfs \citep{henry06,bochanski10} and the weak, possibly inverse, relationship between planet occurrence rates and stellar mass \citep{howard12,fressin13}.

The prevalence of \mds, in concert with several detection biases favoring their planets, places them in the limelight of exoplanet science now and through the next decade. (See, e.g.,\citealt{tarter07, scalo07, shields16b} for discussions of M dwarf exoplanet science and their potential to host planets with life.)
Understanding the space environment these stars provide, therefore, is paramount. 

Of particular importance is the radiative output of M dwarfs at short wavelengths. 
While this radiation contributes only negligibly to a star's bolometric luminosity, it has a vastly disproportionate impact on a planetary atmosphere.
X-ray and extreme UV photons (X-ray,~$<$~100~\AA; EUV,~100~--~912~\AA; together XUV) ionize and heat atmospheric gas above roughly the nanobar level, powering thermal atmospheric escape (e.g., \citealt{murray09,koskinen13}).
For close-in planets, the rate of energy deposition can be sufficient to power outflowing ``planetary winds'' that eject enough gas as to be easily observed (e.g., the hot-Neptune orbiting the \md\ GJ~436; \citealt{kulow14,ehrenreich15}).

At longer wavelengths, namely the far UV (FUV,~912~\AA~--~1700~\AA) and near UV (NUV,~1700~\AA~--~3200~\AA), stellar radiation dissociates and heats planetary atmospheres down to roughly the millibar level, resulting in nonthermal chemistry (i.e., photochemistry).
It is this process which produces Earth's stratospheric ozone, among other effects.
In this way, the UV emission from M dwarfs perturbs the thermochemical equilibrium of their planets' atmospheres (e.g., \citealt{miguel15}), with potentially detectable changes in spectral features \citep{rugheimer15}.
This photochemical forcing could lead to the loss of oceans \citep{luger15,tian15a} and the buildup of tens to hundreds of bars of abiotic \OII\ and \OIII\ \citep{luger15,tian15b,schaefer16} for rocky M dwarf planets.

Lately, the role of flares in shaping the atmospheres of planets has received increasing attention.
Analyses have found that flares and (possibly) associated energetic particle showers could drastically alter the composition and retention of Earth-like atmospheres \citep{lammer07,segura10,venot16,airapetian17,lingam17,tilley18}.
However, these analyses have been forced to rely on observations from a single well-characterized M dwarf flare observed at FUV wavelengths together with scalings from the Sun and scalings from M dwarf observations at optical wavelengths.
There is a paucity of direct FUV data on M dwarf flares. 

Thus far, efforts to better characterize the high-energy radiation of \mds\ have focused on its long-term evolution and present state.
This includes the earlier work of the MUSCLES Treasury Program (described in detail below), of which this paper is a part.
MUSCLES addresses the present high energy radiation environment of cool stars.
Another program, HAZMAT (HAbitable Zones and M dwarf Activity across Time), has used \textit{GALEX} (\textit{Galactic Evolution Explorer}) survey data to explore the evolution of \md\ ultraviolet activity with age \citep{shkolnik14,schneider18}, finding saturated activity to 0.1~--~1~Gyr followed by a $t^{-1}$ decline akin to the trends previously observed in coronal X-ray and chromospheric optical emission (e.g., \citealt{vaughan80,walter82,vilhu84}).

There are several challenges to observations, both time-integrated and time-resolved, at UV and shorter wavelengths.
Below the hydrogen ionization edge at 912~\AA, stellar emission is strongly attenuated by the interstellar medium (ISM). This attenuation abates below $\sim$400~\AA\ for some nearby objects with hydrogen columns $\lesssim$10$^{18}$~cm$^{-2}$, but the greatest coverage of any currently-operating  astronomical observatory in this range is limited to $<$120~\AA\ (Chandra LETGS, e.g., \citealt{ness04}).
Light at both X-ray and UV wavelengths longward of 912~\AA\ is accessible only above Earth's atmosphere, namely with the heavily-subscribed \textit{Chandra} and \textit{XMM-Newton} observatories for X-ray wavelengths and \textit{HST} for UV wavelengths.

Given the scarcity of observing resources, most X-ray and UV flare observations have been limited to single targets known for exhibiting spectacular flares, such as the panchromatic flare data for the \mds\ AD~Leo and EV~Lac \citep{hawley03,osten05}.
However, \cite{miles17} leveraged the voluminous \textit{GALEX} dataset to examine overall variability for a sample of M stars in short-exposure, broadband NUV and FUV measurements, finding greater variability in the NUV toward later types and evidence for a much stronger flare response in the \textit{GALEX} FUV versus NUV band.
\cite{welsh07} have also leveraged \textit{GALEX} data for a time-domain study of M dwarfs, finding that the UV flares of earlier-type (M0 to M5) dwarfs are roughly 5 times more energetic than those of later (M6 to M8) type stars.  
Prior to \textit{GALEX} and \textit{HST}, the \textit{Far-Ultraviolet Spectrographic Explorer (FUSE)} and \textit{Extreme-Ultraviolet Explorer (EUVE)} observatories enabled studies of flares at UV wavelengths.
These were limited to the bright \mds\ AD~Leo (e.g., \citep{hawley95,christian06}), AU~Mic (e.g., \citep{cully93,bloomfield02,redfield02}), AB Dor \citep{dupree05}, and EV~Lac. 

Other wavelength regimes, namely the visible, have recently benefited from time-domain survey missions, such as \textit{MOST} and \textit{Kepler}.
The massive statistical sample provided by \textit{Kepler} has permitted surveys of white-light flares on \mds, revealing greater rates of flaring on active \mds\ \citep{hawley14} and confirming  a greater fraction of \mds\ versus Sun-like stars exhibit white-light flares \citep{davenport16a}.
These flares are ubiquitous even to L0 spectral types \citep{paudel18}.

The present work is one in a series from the MUSCLES Treasury Survey (Measurements of the Ultraviolet Spectral Characteristics 
of Low-mass Exoplanetary Systems; \citealt{france16}), a program that aims to characterize the high energy radiation environment that cool stars provide to their planets.
Paper~I \citep{france16} provided a general overview of the program and some of the most impactful results, including  FUV and XUV fluxes in the habitable zone of the surveyed stars; stellar FUV/NUV ratios that drive the balance of \OII\ and \OIII\ populations in planetary atmospheres; and correlations of FUV and XUV emission with \Mgii\ and \Siiv\ emission line fluxes.
Paper~II \citep{youngblood16} described the reconstruction of the \lya\ line profile for these stars and the estimation of EUV fluxes, presented empirical relations between \lya\ and \Mgii\ flux and \lya\ flux and rotation period, and constrained H column densities along the line of sight to the targets.
Paper~III \citep{loyd16} presented a library of X-ray to IR SEDs for the sample stars, intended for use in steady-irradiance photochemical modeling, computed wavelength-dependent photodissociation ($J$) values, and showed evidence of a Si$^{+}$ to Si ionization edge in the FUV continuum of the K star \epseri. 
Paper~IV \citep{youngblood17} related \lya\ fluxes with an optical indicator of activity, \Caii~K emission, and developed a solar scaling that permits the estimation of energetic particle fluxes based on the \Heii~1640~\AA\ and \Siiv~1400~\AA\ energy of a stellar flare.

The work presented here expands the MUSCLES legacy by providing the first statistical constraints on the FUV flaring behavior of a sample of \md\ exoplanet host stars.
This has revealed an intriguing consistency in the flares of M dwarfs of differing \Caii~K activity levels as well as new constraints on the energetics M dwarf upper atmospheres.
These flares have been observed in unprecedented detail in time and wavelength, enabling a detailed breakdown of the flare energy budget and an examination of relationships between differing sources of emission.
Accompanying some observations are rare simultaneous X-ray data.
From the flare sample, tools are established for the benefit of future forays into modeling the effects of M dwarf flares on planetary atmospheres, and some initial modeling is presented that explores the potential impact of the observed and predicted flares. 

Because of the volume of this work, we have attempted to partition the paper with ample headings and subheadings so that the reader can quickly scan the paper and identify the section(s) most relevant to their interests or needs. 
We begin with a description of the dataset and methods for detecting and characterizing flares in Section \ref{sec:data}.
We then examine the population of observed flares from several angles: 
In Section \ref{sec:ffds}, we focus on the frequency distribution of flares in the broadband FUV and the implications for stellar physics.
In Section \ref{sec:lineflares}, we isolate flares to specific emission lines.
In Section \ref{sec:startrends}, we explore relationships with stellar properties. 
In Section \ref{sec:shapes} we examine flare lightcurves and spectral energy budgets.
The paper then turns its focus to the application of these data to planets.  
Section \ref{sec:fiducialflare} describes a framework for generating simplified, synthetic EUV~--~NUV flares based on the sample of FUV flares in hand, intended for community use in modeling planetary atmospheres. 
Section \ref{sec:planets} describes the results of applying this framework to gauge the potential for flares like those observed to photolyze molecules in an Earth-like atmosphere.
The work is summarized in Section \ref{sec:conclusions}. 

\section{Data and Reduction}
\label{sec:data}

\subsection{Observations}
The sample stars and those of their properties that are expected to correlate with flare activity are given in Table \ref{tbl:starprops}.
We conducted the flare analysis primarily on two stellar populations, the MUSCLES M dwarfs (the ``inactive'' sample) and the well-known M dwarf flare stars AD~Leo, Prox~Cen, EV~Lac, and AU~Mic (the ``active'' sample).
There is roughly an order-of-magnitude separation in the optical chromospheric emission of the inactive and active samples, with \Caii~K equivalent widths $<$2~\AA\ for the inactive stars and $>$10~\AA\  for the active stars \citep{youngblood17}.
These values are corrected for differences in the surrounding continuum due to differing stellar effective temperatures, and positive values indicate emission. 
Only the K line of the \Caii~H~\&~K pair is used because the H line can be contaminated by H$\epsilon$ emission in low resolution spectra. 

\begin{deluxetable*}{llrlrlrlrlrrrl}
\tablewidth{0pt}
\tabletypesize{\scriptsize}
\rotate
\tablecaption{Selected properties of the stars in the sample. \label{tbl:starprops}}

\tablehead{ \colhead{Star} & \colhead{Type\tablenotemark{a}} & \colhead{$T_\mathrm{eff}$} & \colhead{ref} & \colhead{$P_\mathrm{rot}$} & \colhead{ref} & \colhead{$W_\lambda$ \Caii\ K \tablenotemark{b}} & \colhead{$F$(X-ray)\tablenotemark{c}} & \colhead{Known\tablenotemark{d}} & \colhead{Observation} & \colhead{Exposure} & \colhead{Instrument} \\ \colhead{} & \colhead{} & \colhead{[K]} & \colhead{} & \colhead{[day]} & \colhead{} & \colhead{[\AA]} & \colhead{[erg s$^{-1}$ cm$^{-2}$]} & \colhead{Planets} & \colhead{Epochs} & \colhead{Time [ks]} & \colhead{\& Grating} }
\startdata
\cutinhead{MUSCLES Stars -- the ``Inactive" Sample}
GJ 667C & M1.5 & $ 3445 \pm 110 $ & 1 & $ 103.9 \pm 0.7 $ & 2 & $ 0.44 \pm 0.11 $ & $3.9 \pm 0.3 \times10^{-14}$ & 5 & 2015-08-07 & 12.7 & COS G130M\\
GJ 176 & M2.5 & $ 3679 \pm 77 $ & 3 & $ 39.3 \pm 0.1 $ & 2 & $ 1.76 \pm 0.27 $ & $4.8 \pm 0.3 \times10^{-14}$ & 1 & 2015-03-02 & 12.6 & COS G130M\\
GJ 832 & M2/3 & $ 3416 \pm 50 $ & 4 & $ 45.7 \pm 9.3 $ & 2 & $ 0.88 \pm 0.09 $ & $6.2_{-0.7}^{+0.8}\times10^{-14}$ & 2 & 2012-07-28, 2014-10-11 & 15.1 & COS G130M\\
GJ 436 & M3 & $3416_{-61}^{+54}$ & 5 & $ 39.9 \pm 0.8 $ & 2 & $ 0.58 \pm 0.07 $ & $1.2 \pm 0.1 \times10^{-14}$ & 1 & 2012-06-23, 2015-06-26 & 15.5 & COS G130M\\
GJ 581 & M3 & $ 3442 \pm 54 $ & 6 & $ 132.5 \pm 6.3 $ & 2 & $ 0.36 \pm 0.08 $ & $1.8 \pm 0.2 \times10^{-14}$ & 3 & 2011-07-20, 2015-08-11 & 13.8 & COS G130M\\
GJ 876 & M3.5 & $ 3129 \pm 19 $ & 3 & $ 87.3 \pm 5.7 $ & 2 & $ 0.82 \pm 0.15 $ & $9.1 \pm 0.8 \times10^{-14}$ & 4 & 2012-01-05, 2015-07-07 & 14.8 & COS G130M\\
\cutinhead{Flare Stars -- the ``Active" Sample\tablenotemark{e}}
AU Mic\tablenotemark{f} & M1 & 3650 & 8 & $ 4.85 \pm 0.02 $ & 9 & $ 12.1 \pm 2.2 $ & \nodata & 0 & 1998-09-06 & 17.6 & STIS E140M\\
EV Lac & M4.0 & $ 3325 \pm 100 $ & 10 & 4.4 & 11 & $ 14.9 \pm 2.5 $ & \nodata & 0 & 2001-09-20 & 10.9 & STIS E140M\\
AD Leo & M4.0 & $ 3414 \pm 100 $ & 10 & 2.6 & 11 & $ 11.6 \pm 1.6 $ & \nodata & 0 & 2000-03-12, 2002-06-01 & 67.0 & STIS E140M\\
Prox Cen & M5.5 & $ 3098 \pm 56 $ & 12 & 82.5 & 13 & $ 13.7 \pm 5.9 $ & \nodata & 1 & 2000-05-08, 2017-05-31 & 48.0 & STIS E140M\\
\enddata

\tablenotetext{a}{Spectral types taken from SIMBAD, \url{http://simbad.u-strasbg.fr/simbad/}.}
\tablenotetext{b}{All Ca II K equivalent widths from \cite{youngblood17}.}
\tablenotetext{c}{Mean soft X-ray flux from \textit{XMM-Newton} or \textit{Chandra} observations presented in \cite{loyd16} and searched for flares in this work.}
\tablenotetext{d}{Planet count retrieved from NASA Exoplanet Archive, \url{https://exoplanetarchive.ipac.caltech.edu}.}
\tablenotetext{e}{As categorized in SIMBAD. Data from these stars is archival; they were not included in the MUSCLES survey.}
\tablenotetext{f}{Pre main-sequence star.}

\tablerefs{(1) \citealt{neves14}; (2) \citealt{suarez15}; (3) \citealt{braun14}; (4) \citealt{houdebine10}; (5) \citealt{braun12}; (6) \citealt{boyajian12}; (7) \citealt{newton17}; (8) \citealt{mccarthy12}; (9) \citealt{messina11}; (10) \citealt{houdebine16}; (11) \citealt{hempelmann95}; (12) \citealt{demory09}; (13) \citealt{kiraga07}}

\end{deluxetable*}

The MUSCLES Treasury Survey, \textit{HST} observing program 13650, obtained photon-counting (TIME-TAG mode) FUV data using the COS G130M spectrograph for 5 \textit{HST} orbits per target ($\sim$3.5~h of exposure within a span of $\sim$8~h), with the specific intent of monitoring stellar variability.
We augmented these data with all available COS G130M data on the MUSCLES targets in the \textit{HST} archive  as of 2017 Sep (observing programs 12034, 12035, 12464, and 13020).
We discarded all GJ~1214 data from the analysis, including that of the MUSCLES program, due to low S/N. 

The MUSCLES survey also obtained contemporaneous and occasionally simultaneous X-ray data for the targets.
For GJ~176, GJ~436, GJ~581, GJ~667C, and GJ~876, these observations were made with the \textit{Chandra X-ray Observatory} (\textit{CXO}; proposals 15200539 and 16200943) using the ACIS-S instrument.
For GJ~832 and \epseri\ (a K star discussed further only in Section \ref{sec:xrays}), the survey employed \textit{XMM-Newton} (observation 0748010201) with the EPIC instrument.
These observations varied from 2.8 to 5.6 h.

For the flare stars, all FUV data are archival aside from some recent observations of Prox~Cen (program 14860, PI Schneider). 
We did not retrieve any archival X-ray data. 
A previous survey of flares in the archival \textit{HST} FUV data exists \citep{loyd14}.
That work focused on constraining variability in FUV emission to assess its impact on transit observations.
In comparison, the present work is devoted to the flares themselves and their contribution to the space environment to which planets are exposed. 
We reanalyzed the archival data (observing programs 7556, 8040, 8613, 8880, and 9271) using the methods presented here to ensure homogeneity.

\subsection{UV Lightcurve Creation}

For the COS and STIS UV data, we created lightcurves over a given bandpass using the process described in \cite{loyd14}.
In brief, this involves binning detector events within a ribbon covering the signal trace over the desired wavelengths.
Regions offset from the signal trace at the same spectral location are used to make an estimate of the background count rate that is then scaled according to area and subtracted from the signal count rate.
The flux calibration from the full exposure is then applied to the sub-exposure count tallies to create a lightcurve in flux units.
We did not attempt a subtraction of the continuum because it is negligible for these cool stars in at FUV wavelengths.
The lightcurves all contain $\sim$45~min gaps between sequences of exposures due to regular occultations of the target by Earth during \textit{HST's} orbit.
These are noteworthy because they frequently truncate the beginning or end of a flare. 

The photon-counting data allow lightcurve bandpasses to be defined arbitrarily within the limits of the spectrograph wavelength range and resolution.
Wavelength uncertainties are well below the bandpass widths for the medium-resolution gratings used for the bulk of this work.
For each exposure, we adjusted the photon wavelengths by using strong emission lines to define a wavelength offset  that was a linear function of wavelength (or a constant offset when only a single reference line could be used), thus removing the stellar radial velocity and mitigating some systematic errors in the instrumental wavelength solution. 

For emission lines, we used bandpasses of \bandwidth\ (full width) intended to capture the bulk of the line flux with limited contamination from any surrounding continuum and adjacent lines.
Although Doppler shifts resulting from mass motions are a factor, we did not observe any significant emission beyond this band in our observations (see Section \ref{sec:profiles}).
For multiplets, we integrated flux over the union of the \bandwidth\ bands of each individual line.
The \lya\ line has significant emission beyond the default band, so we employ a wider band spanning 1214.45~--~1216.89~\AA\ for it.
Note that we analyzed \lya\ and \Oi\ only for the STIS observations due to contamination by geocoronal airglow in the COS observations. 
Wavelengths of the lines we examined in this analysis are given in Table \ref{tbl:COSlines}.

\begin{deluxetable}{llr}
\tablecaption{Selected stellar emission lines in the \textit{HST} COS G130M bandpass (FUV).\label{tbl:COSlines}}
\tabletypesize{\footnotesize}
\tablewidth{0pt}
\tablehead{\colhead{Ion} & \colhead{$\lambda_\mathrm{rest}$} & \colhead{$\log_{10} (T_\mathrm{peak}/\mathrm{K})$}\tablenotemark{a}\\
 & \colhead{\AA} & }
\startdata
\Ciii & 1174.93, 1175.26, 1175.59,  & 4.8 \\
    & 1175.71, 1175.99, 1176.37 &  \\
\Siiii & 1206.51 & 4.7 \\
\Hi\tablenotemark{b} & 1215.67 & 4.5 \\
\Nv & 1238.82, 1242.80 & 5.2 \\
\Oi\tablenotemark{b} & 1302.17, 1304.86, 1306.03 & 3.8 \\
\Cii & 1334.53, 1335.71 & 4.5 \\
\Siiv & 1393.76, 1402.77 & 4.9 \\
\Civ & 1548.20, 1550.774 & 4.8 \\
\Heii & 1640.4 & 4.9 \\
\Ci & 1656.27, 1656.93, 1657.01,  & 3.8 \\
 & 1657.38, 1657.91, 1658.12 & \\
\enddata
\tablenotetext{a}{Peak formation temperatures of the C, O, and H lines are from \cite{avrett08}, using the values at line center. Other lines are from a CHIANTI spectral synthesis using a differential emission measure curve estimated from data during an M2 class solar flare (retrieved from \url{http://www.chiantidatabase.org/chianti_linelist.html} on 2017 July 31; \citealt{dere09}).}
\tablenotetext{b}{Also emitted by Earth's upper atmosphere (``geocorona''), contaminating COS observations. These lines are only observable with STIS, the instrument used by the archival flare star observations.}

\end{deluxetable}

We also defined broad bandpasses encompassing all flux captured by various instrument configurations, omitting regions contaminated by airglow and detector edges that are inconsistently covered due to instrument dithering. 
Of these, the band covered by the greatest quantity of exposure time is the COS G130M bandpass, which is a subset of the STIS E140M bandpass.
This extends from roughly \fuvwaverough, and we label it \gfuv.
Specifically, \gfuv\ refers to flux integrated in the ranges 1173.65~--~1198.49,  1201.71~--~1212.16,  1219.18~--~1274.04,  1329.25~--~1354.49,  1356.71~--~1357.59,  and 1359.51~--~1428.90~\AA. 

\subsubsection{``Count-binned'' Lightcurves}
\label{sec:countbinning}
Because the STIS and COS detectors are photon counters, there is great flexibility in the spectral and temporal binning of the data.
We utilized this flexibility to create lightcurves where the time-binning changes in accordance with the flux to maintain a roughly constant S/N in each time bin.
We do this by measuring the time taken for a set number of events to occur rather than counting the number of events during a set interval, leading us to call these ``count-binned'' lightcurves.
These lightcurves are useful for visually examining flares and measuring their peak flux and FWHM (full width at half maximum; used here to denote width in time, not wavelength).
However, the statistical distribution this method produces has a greater skew than the corresponding Poisson distribution, so we do not use these lightcurves for identifying or integrating flares.

\subsection{X-ray Lightcurve Creation}
Similar to the UV lightcurve creation, X-ray lightcurves were created by integrating all detector events within a signal region and subtracting area-corrected event counts from a nearby background region, chosen to be devoid of other sources.
Events of all recorded energies within the detector bandpass were integrated.
The \textit{CXO} \replaced{ASIS-S}{ACIS-S} bandpass is roughly 1~--~40~\AA\ and the \textit{XMM-Newton} EPIC bandpass is roughly 1~--~60~\AA.
Unlike the FUV spectra, we did not estimate absolute fluxes from the X-ray count rates. 
The count rate conversion factors (counts s$^{-1}$ to erg \pers cm$^{-2}$) sensitively (factors of a few) depend on the assumed plasma temperature, a parameter that is expected to change considerably during the flares. 
Since the X-ray data are insufficient to accurately determine the plasma temperature on short time scales, we utilize only photon count rates.
X-ray data were never count binned; time-binned lightcurves were used for all X-ray flare characterization. 

\begin{figure*}
\includegraphics{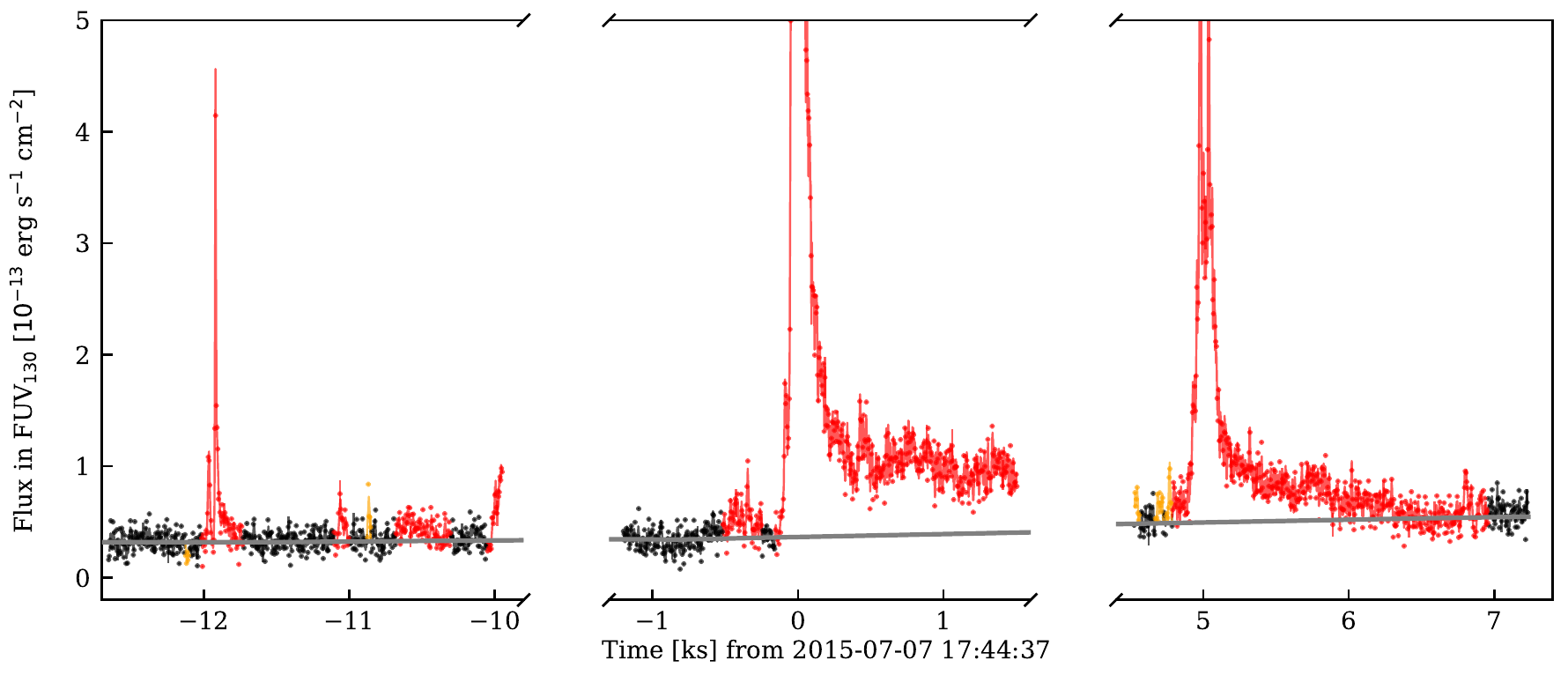}
\caption{Example identification of flares in three exposures of the GJ~876 data using the \gfuv\ bandpass.
Points show the lightcurve binning used in the identification process (Section \ref{sec:flareident}) and the jagged line underlying the points is a ``count-binned'' lightcurve (see Section \ref{sec:countbinning}).
The smooth thick gray line shows the Gaussian process fit to quiescence.
Red data has been identified as belonging to a flare and orange data has been flagged as anomalous.
Both were excluded in fitting the quiescence.}
\label{fig:ex_ident}
\end{figure*}

\subsection{Flare Identification with FLAIIL}
\label{sec:flareident}

We developed a custom algorithm for identifying flares in both the FUV and X-ray  data that we have named Flare Identification in Intermittent Lightcurves (FLAIIL)\footnote{\url{https://github.com/parkus/flaiil}}.
Using an automated pipeline provided consistency in the treatment of all datasets and the ability to rapidly reanalyze the data following upstream changes to the pipeline. 
A variety of shcemes for identifying flares have been developed and employed by  previous analyses, such as the cross-correlation method used by \cite{davenport16a} on \textit{Kepler} data.
However, the gappinness of the data and the highly variable time profiles of flares in  FUV emission led us to develop a custom pipeline for this dataset.
We briefly describe the identification algorithm here, with additional details provided in Appendix \ref{app:flare_ident}.

Because of the diversity in time profiles of flares, we specifically designed our pipeline to be agnostic to the flare shape.
The pipeline identifies flares based on the area of ``runs,'' consecutive points above and below quiescence.
Quiescent variations are modeled using a Gaussian Process with a covariance kernel that describes correlations as exponentially decaying with time, employing the code \texttt{celerite} for this purpose \citep{foreman17}\footnote{\url{http://celerite.readthedocs.io}}.
The variance and decay time constant are free parameters.
Table \ref{tbl:qvars} gives the best-fit values of these parameters for each star. 
If the likelihood of a white-noise model with constant mean comes within a factor of two of the best-fit Gaussian Process model, it is used instead.
Following the quiescence fit, anomalous runs are masked out, the quiescence is refit,  and the process is iterated to convergence.

It is possible for flares to overlap, with physically distinct events superposing in a lightcurve of the star's disk-integrated emission. 
The algorithm makes no attempt to separate overlapping events, as the diversity of FUV flare light curves would make a consistent disentanglement nearly impossible. 
It is also the case that many flares are truncated by exposure gaps.
Again, because of the inconsistency in flare light curves, no attempt is made to reconstruct the unobserved portions.

Figure \ref{fig:ex_ident} shows the end result of applying this algorithm for three exposures of the GJ~876 data.
Several clear, large flares are identified, as well as a number of smaller deviations from quiescence. 
Following identification, each event is characterized using a number of metrics, discussed in the next section.

\subsection{Flare Metrics}
\label{sec:metrics}

\begin{figure*}
\includegraphics{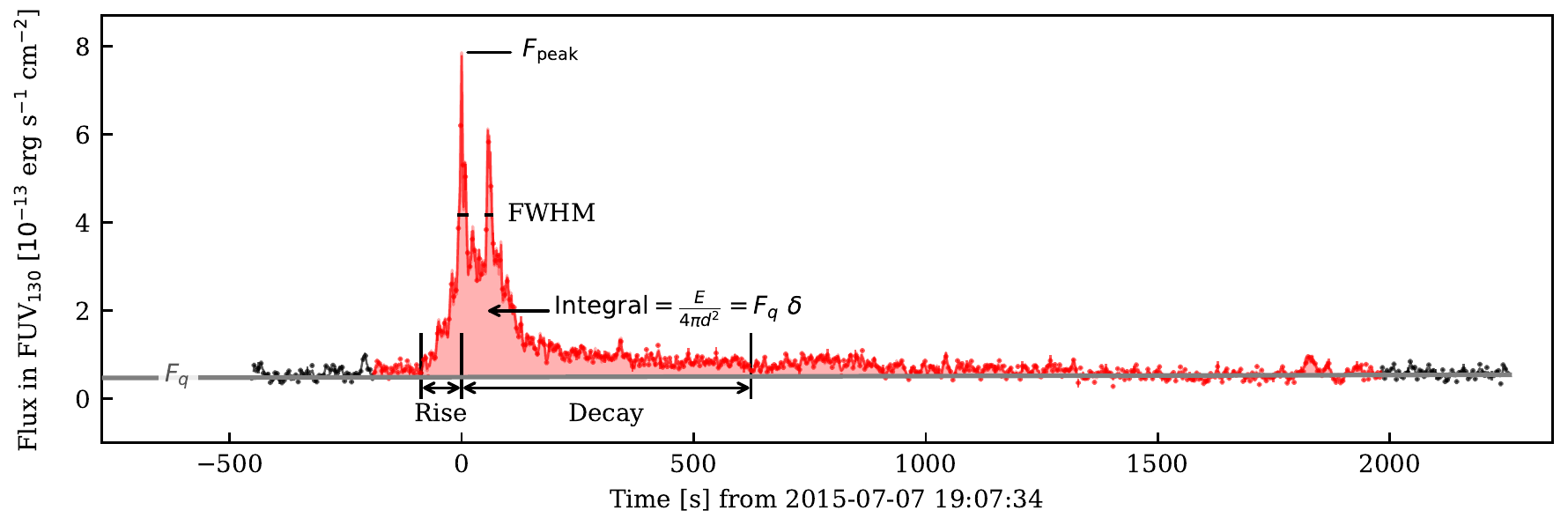}
\caption{Visual explanation of the various metrics recorded for each flare, using a well-resolved flare that occurred on GJ~876.}
\label{fig:annotated_flare}
\end{figure*}

We cataloged a variety of metrics for each flare, including peak flux, FWHM, presence of multiple peaks, absolute energy, and equivalent duration.
Though mostly straightforward, there are some nuances to their computation.
We define each metric below and provide an annotated plot of a flare in Figure \ref{fig:annotated_flare} to aid the reader in visualizing the various flare metrics.
The parameters of the 20 flares with the largest equivalent duration in \gfuv\ emission are provided in Table \ref{tbl:flares}.

\subsubsection{Peak Flux}
We use lightcurves count-binned to \smoothn\ counts to measure the flare peak.
Count-binning mitigates the chances  the peak flux will be underestimated because it was not temporally resolved. 
In cases where the count rate is too low for the count-binned lightcurve to provide superior sampling, we revert to the time-binned lightcurve.
The STIS data for all flare stars show a high-frequency signal with peaks at periods of 0.35 and 0.5~s in the autocorrelation function that we suspect is an instrumental effect.
Therefore, we do not allow bins less than 1~s in duration for these data. 
We note these differences in binning will result in different estimates of the peak, as larger bins will tend to dilute the peak.

\subsubsection{Full Width at Half-Maximum (FWHM) and Multipeaked Classification}
As with the peak flux, we again use a count-binned lightcurve to compute the FWHM of the FUV flares.
Measuring the FWHM is complicated by noise and secondary peaks that cause the lightcurve to cross the half-max flux value many times.
To mitigate this, we take the FWHM to be the sum of all time spans in which flux was above the half-maximum value during the flare, including secondary peaks.
We flagged flares as complex where multiple distinct peaks could be identified by eye. 

\subsubsection{Rise, Decay, and Duration}
Using the count-binned lightcurve, we recorded the rise and decay times.
We define the rise time as the time between the point at which the flux peaked and the closest preceding time at which it first rose above the quiescent flux.
Similarly, we define the decay time as the time required for the flux to have first dipped below the quiescent level following the flare peak.
The duration is simply the sum of these figures.
These values will be biased by the noise level of the lightcurve (more noise results in more quiescence-crossings), but we retain these definitions for ease of interpretation.
They are also agnostic of the flare shape, a useful feature given the complexity of some of the observed flares.
However, future work might implement a decay metric that finds the time-constant of an exponential fit to the tail of the flare after the last major peak.

\subsubsection{Absolute Energy and Equivalent Duration}
\label{sec:pew}
We computed the absolute energy of the flare, $E$, as
\begin{equation}
E = 4\pi d^2 \int_\mathrm{flare} (F - F_q)dt,
\end{equation}
where $d$ is the distance to the star, $F$ is the measured flux, and $F_q$ is the estimated quiescent flux.
The integral is nominally taken over the full region flagged as flaring, i.e. all of the red area in Figure \ref{fig:annotated_flare} (see Section \ref{sec:flareident}).
In cases where the tail of the flare only increases noise without significantly increasing the integral, the extent of the integral is shortened accordingly.
We do not estimate bolometric flare energies in this work, therefore discussions of energy are tied to specific bandpasses.

We also computed the equivalent duration, $\delta$, of each flare, essentially a measure of the flare's energy normalized by the quiescent luminosity of the star in the same bandpass \citep{gershberg72}.
It is analogous to the equivalent width of a spectral line, sometimes occasioning the use of the term ``photometric equivalent width.''
In this analogy, the flare substitutes for an emission line and the quiescent lightcurve substitutes for the the spectral continuum.
Mathematically,
\begin{equation}
\delta = \int_\mathrm{flare} \frac{F - F_q}{F_q}dt.
\end{equation}
\cite{hawley14} include a useful schematic of this value as their Figure 6.

\begin{deluxetable*}{rrrrrrrrrll}
\tablewidth{0pt}
\tabletypesize{\scriptsize}
\tablecaption{Selected measurements from the 20 flares with greatest $\delta$ in the \gfuv\ band. \label{tbl:flares}}
\tablehead{ \colhead{Star} & \colhead{$\delta$} & \colhead{$E$} & \colhead{$t_\mathrm{peak}$} & \colhead{$F_{\mathrm{peak}}$} & \colhead{$\frac{F_\mathrm{peak}}{F_q}$\tablenotemark{a}} & \colhead{Rise Time} & \colhead{FWHM} & \colhead{Decay Time} & \colhead{Complex?\tablenotemark{b}} \\ \colhead{} & \colhead{s} & \colhead{$10^{27}$ erg} & \colhead{MJD} & \colhead{$10^{-13}\ \frac{\mathrm{erg}}{\mathrm{cm}^2\mathrm{s\ \AA}}$} & \colhead{} & \colhead{s} & \colhead{s} & \colhead{s} & \colhead{} }
\startdata
Prox Cen & $ 14973 \pm 289 $ & $ 316.0 \pm 5.8 $ & 51673.1049 & $ 128 \pm 10 $ & $ 124 \pm 12 $ & 48 & 40 & 600 & N\\
Prox Cen & $ 11556 \pm 209 $ & $ 453.2 \pm 7.7 $ & 57904.9613 & $ 138 \pm 12 $ & $ 74.5 \pm 7.7 $ & 48 & 78 & 450 & Y\\
GJ 876 & $ 6801 \pm 55 $ & $ 665.0 \pm 4.6 $ & 57210.7393 & $ 20.1 \pm 2.1 $ & $ 56.1 \pm 6.2 $ & 120 & 74\tablenotemark{c} & \nodata & Y\\
GJ 832 & $ 4060 \pm 59 $ & $ 275.6 \pm 2.9 $ & 56941.5122 & $ 5.45 \pm 0.63 $ & $ 24.5 \pm 3.6 $ & 150 & 140\tablenotemark{c} & \nodata & Y\\
AD Leo & $ 3443 \pm 53 $ & $ 6887 \pm 101 $ & 51616.1046 & $ 463 \pm 18 $ & $ 62.0 \pm 2.9 $ & 57 & 22 & 430 & N\\
GJ 876 & $ 1725 \pm 28 $ & $ 227.7 \pm 3.7 $ & 57210.7969 & $ 7.87 \pm 0.89 $ & $ 17 \pm 2 $ & 87 & 28 & 620 & Y\\
AD Leo & $ 1721 \pm 44 $ & $ 3397 \pm 86 $ & 51615.2245 & $ 82 \pm 14 $ & $ 11.9 \pm 2.1 $ & 25 & 31 & 230 & Y\\
Prox Cen & $ 1682 \pm 84 $ & $ 92.2 \pm 4.6 $ & 57905.0773 & $ 69 \pm 14 $ & $ 26.6 \pm 5.6 $ & 12 & 4.4 & 22 & \nodata\\
Prox Cen & $ 1427 \pm 113 $ & $ 27.5 \pm 2.2 $ & 51673.0718 & $ 24.5 \pm 4.8 $ & $ 27.0 \pm 5.5 $ & 21 & 17 & 49 & N\\
AD Leo & $ 1398 \pm 34 $ & $ 3438 \pm 84 $ & 51614.2162 & $ 100 \pm 9 $ & $ 11.7 \pm 1.1 $ & 230 & 73 & 110 & Y\\
AD Leo & $ 1388 \pm 33 $ & $ 2943 \pm 70 $ & 51614.4263 & $ 207 \pm 12 $ & $ 26.7 \pm 1.7 $ & 98 & 24 & 150 & N\\
Prox Cen & $ 1328 \pm 131 $ & $ 18.9 \pm 1.8 $ & 51672.0746 & $ 18.5 \pm 3.6 $ & $ 27.3 \pm 5.6 $ & 13 & 14 & 43 & \nodata\\
GJ 176 & $ 1005 \pm 42 $ & $ 240.6 \pm 9.8 $ & 57083.2087 & $ 0.91 \pm 0.12 $ & $ 4.90 \pm 0.85 $ & 140 & 110 & 110 & \nodata\\
Prox Cen & $ 967 \pm 162 $ & $ 11.4 \pm 1.9 $ & 51672.2840 & $ 13 \pm 3 $ & $ 23.7 \pm 5.9 $ & 7.6 & 21 & 26 & \nodata\\
GJ 876 & $ 919 \pm 22 $ & $ 130.3 \pm 2.5 $ & 55931.1241 & $ 9.8 \pm 1.1 $ & $ 19.5 \pm 2.6 $ & \nodata & 28\tablenotemark{c} & 330 & \nodata\\
Prox Cen & $ 919 \pm 119 $ & $ 13.1 \pm 1.7 $ & 51672.0860 & $ 9.0 \pm 2.5 $ & $ 14 \pm 4 $ & 15 & 46 & 32 & \nodata\\
AD Leo & $ 842 \pm 28 $ & $ 1635 \pm 54 $ & 51615.1698 & $ 133 \pm 20 $ & $ 19.1 \pm 2.9 $ & 22 & 12 & 150 & Y\\
Prox Cen & $ 813 \pm 83 $ & $ 16.6 \pm 1.7 $ & 51673.0910 & $ 12.1 \pm 3.1 $ & $ 13.0 \pm 3.4 $ & 5.3 & 33 & 10 & \nodata\\
Prox Cen & $ 802 \pm 51 $ & $ 42.9 \pm 2.6 $ & 57905.0905 & $ 165 \pm 13 $ & $ 63.5 \pm 5.9 $ & 14 & 8.8 & 19 & N\\
GJ 581 & $ 795 \pm 184 $ & $ 9.5 \pm 1.1 $ & 57245.8531 & $ 0.454 \pm 0.071 $ & $ 19 \pm 16 $ & 33 & 26 & 29 & N\\
\enddata

\tablenotetext{a}{Ratio of peak flux to quiescent flux.}
\tablenotetext{b}{Subjective determination of the complexity of the flare shape based on its deviation from an impulse-decay, generally due to multiple peaks. No data indicates the flare was not well-enough resolved or the classification was particularly ambiguous.}
\tablenotetext{c}{Flare cut off by the start or end of an exposure.}

\tablecomments{Uncertainties are statistical and do not reflect systematic effects due to choices made in the flare identification and measurement algorithm. See Appendix \ref{app:inject} for an assessment of systematic errors in energy. }
\end{deluxetable*}

\begin{deluxetable*}{lrrrrr}
\tablewidth{0pt}
\tabletypesize{\scriptsize}
\tablecaption{Fits to quiescent FUV emission and literature variability metrics. \label{tbl:qvars}}
\tablehead{\colhead{Star} & \colhead{Epoch} & \colhead{$\sigma_{x, \mathrm{GP}}$\tablenotemark{a}} & \colhead{$\tau_\mathrm{GP}$\tablenotemark{a}} & \colhead{$\sigma_{x, \mathrm{LF14}}$\tablenotemark{b}} & \colhead{MAD$_\mathrm{rel}$\tablenotemark{c}} \\ \colhead{} & \colhead{} & \colhead{} & \colhead{s} & \colhead{} & \colhead{}}
\startdata
GJ 667C & 2015-08-07 & $0.272_{-0.069}^{+0.043}$ & \nodata & $0.250_{-0.040}^{+0.048}$ & $ 0.209 \pm 0.018 $\\
GJ 176 & 2015-03-02 & $0.111_{-0.026}^{+0.016}$ & \nodata & $0.146_{-0.017}^{+0.019}$ & $ 0.171 \pm 0.011 $\\
GJ 832 & 2012-07-28 & $0.124_{-0.052}^{+0.028}$ & \nodata & $0.170_{-0.046}^{+0.064}$ & $ 0.119 \pm 0.023 $\\
 & 2014-10-11 & $0.087_{-0.031}^{+0.017}$ & \nodata & $0.113_{-0.017}^{+0.016}$ & $ 0.362 \pm 0.007 $\\
GJ 436 & 2012-06-23 & $1.18_{-0.26}^{+0.22}$ & \nodata & $0.97_{-0.22}^{+0.31}$ & $ 0.526 \pm 0.097 $\\
 & 2015-06-25 & $0.214_{-0.122}^{+0.053}$ & \nodata & $0.274_{-0.041}^{+0.045}$ & $ 0.200 \pm 0.013 $\\
GJ 581 & 2011-07-20 & $0.25_{-0.13}^{+0.45}$ & \nodata & $0.80_{-0.25}^{+0.43}$ & $ 0.304 \pm 0.092 $\\
 & 2015-08-11 & $0.84_{-0.10}^{+0.09}$ & \nodata & $0.622_{-0.079}^{+0.091}$ & $ 0.349 \pm 0.034 $\\
GJ 876 & 2012-01-05 & $0.45_{-0.09}^{+1.53}$ & $31043_{-4590}^{+404953}$ & $0.53_{-0.12}^{+0.20}$ & $ 1.58 \pm 0.19 $\\
 & 2015-07-07 & $0.194_{-0.026}^{+0.337}$ & $75553_{-17195}^{+418943}$ & $0.213_{-0.023}^{+0.028}$ & $ 0.767 \pm 0.042 $\\
AU Mic & 1998-09-06 & $0.0580_{-0.0069}^{+0.1685}$ & $11690_{-1095}^{+237424}$ & $0.189_{-0.026}^{+0.031}$ & $ 0.177 \pm 0.017 $\\
EV Lac & 2001-09-20 & $0.259_{-0.026}^{+1.439}$ & $13203_{-1536}^{+299657}$ & $0.434_{-0.047}^{+0.052}$ & $ 0.275 \pm 0.027 $\\
AD Leo & 2000-03-10 & $0.0760_{-0.0036}^{+0.1350}$ & $39428_{-2395}^{+294628}$ & $ 0.189 \pm 0.011 $ & $ 0.2745 \pm 0.0075 $\\
 & 2002-06-01 & $0.107_{-0.022}^{+0.014}$ & \nodata & $0.138_{-0.016}^{+0.019}$ & $ 0.1158 \pm 0.0089 $\\
Prox Cen & 2000-05-08 & $0.269_{-0.052}^{+0.311}$ & $151736_{-42159}^{+441448}$ & $0.723_{-0.060}^{+0.065}$ & $ 0.874 \pm 0.021 $\\
 & 2017-05-31 & $0.208_{-0.016}^{+0.672}$ & $33368_{-3713}^{+411548}$ & $0.511_{-0.059}^{+0.071}$ & $ 1.02 \pm 0.05 $\\
\enddata

\tablenotetext{a}{Pertains to covariance kernel function, $\sigma_x^2 e^{-\Delta t / \tau}$, of the Guassian Process used to model quiescent variations, normalized by the mean flux of the model. Values and uncertainties are based on the \nth{16}, \nth{50}, and \nth{84} percentiles of the MCMC samples. When no value is given for $\tau$, this indicates that a quiescent model including correlated noise had a likelihood ratio less than 2$\times$ that of white noise. In these cases, the quiescence was modeled as constant with white noise equal to the quadrature sum of the measurement noise and $\sigma_x$.}
\tablenotetext{b}{``Excess noise'' at 60 s cadence per \cite{loyd14}. Values and uncertainties are based on the \nth{16}, \nth{50}, and \nth{84} percentiles of the analytical solution  of the posterior distribution.}
\tablenotetext{c}{Median Absolute Deviation per \cite{miles17}. Uncertainties are based on the \nth{16}, \nth{50}, and \nth{84} percentiles from bootstrapped samples. Uses a 100~s cadence and includes flares.}

\end{deluxetable*}

\section{The Frequency Distribution of FUV Flares and Its Implications}
\label{sec:ffds}

\subsection{FUV Flare Frequency Distributions and Power-Law Fits}
\label{sec:ffd_fits}

We fit the cumulative energy-frequency distribution of the flares (flare frequency distributions, FFDs) with power-law models, specifically
\begin{equation}
\label{eqn:pow}
\nu = \mu \left( \frac{\delta}{\delta_\mathrm{ref}} \right)^{-\alpha}
\end{equation}
and
\begin{equation}
\label{eqn:powE}
\nu = \mu \left( \frac{E}{E_\mathrm{ref}} \right)^{-\alpha}.
\end{equation}
where $\nu$ is the occurrence rate of flares with equivalent durations above $\delta$ or energies above $E$, $\mu$ is a rate constant, and $\alpha$ is the power-law index.
We introduce the reference values $\delta_\mathrm{ref}$ and $E_\mathrm{ref}$ to remove any ambiguity concerning units and mitigate problematically high correlations between parameters when fitting FFDs.
For this work, we use  $E_\mathrm{ref} = 10^{30}$ erg and $\delta_\mathrm{ref} = 1000$ s.
Smaller $\alpha$ values correspond to higher rates of high energy flares and lower rates of low energy flares.
However, low energy flares are always more prevalent in number so long as $\alpha>0$.

The free parameters of the power law models are $\mu$ and $\alpha$.
They are tightly correlated, analogous to the slope and y-intercept of a linear fit to data.
Because of this, we employed an MCMC sampler (via the Python module \texttt{emcee}\footnote{\url{http://dfm.io/emcee}}; \citealt{foreman13}) to sample the parameter space. 
The fit procedure works directly from the discrete flare events (i.e., does not fit the binned FFD curves) and accounts for the varying detection limits when events from multiple datasets are aggregated.
We estimated the detection limits using injection/recovery tests that account for multiple events.
The fitting algorithm and injection/recovery process are described further in Appendices \ref{app:powfit} and \ref{app:inject} and the code we developed has been made available online.\footnote{\url{\ffdurl}}
To mitigate overprecision in the power law fits given systematic errors from flare overlap and flare truncation, we carried out 9 flare identification runs with FLAIIL using reasonable changes to the algorithm parameters, then combined the MCMC chains from separate fits to each of the resulting flare samples.

We divided the flare samples into seven groups with separate fits to each. 
These consisted of the flares on the individual stars AD~Leo, Prox~Cen, GJ~176, and GJ~876,  as well as all inactive stars, all active stars, and all stars.
Attempts at fitting FFDs to the flares of individual objects aside from GJ~176, GJ~876, AD~Leo, and  Prox~Cen provided inconsistent results given  the relatively small number of detected flares.
However, meaningful constraints on the rate of flares for these stars is still possible if an assumption is made regarding the power law index, $\alpha$.
Therefore, to constrain the rate of flares in equivalent duration on individual stars other than GJ~176, GJ~876, Prox~Cen, and AD~Leo, we set the following priors on $\alpha$:\begin{itemize}
\item all stars, equivalent duration: the posterior on $\alpha$ resulting from the power-law fit to events aggregated from all stars
\item inactive stars, absolute energy: the posterior on $\alpha$ resulting from the power-law fit to events aggregated from the inactive stars
\item active stars, absolute energy: the posterior on $\alpha$ resulting from the power-law fit to events from AD~Leo.
\end{itemize}
Applying a prior on $\alpha$ allowed the MCMC walkers to explore the posterior on the rate constant $\mu$ within the confines of the $\alpha$ prior. 

Tables \ref{tbl:pew_stats} and \ref{tbl:energy_stats} give the parameters of the power-law fits.
The tables also list a variety of derived quantities, the most direct of which is the rate of  flares with $E$ or $\delta$ greater than three characteristic thresholds:
\begin{itemize}
\item Equivalent durations of $>$10~s represent frequent but often undetectable flares, with about 100 events per day.
\item Flares with equivalent durations of  $>$1000~s are easily discernible  in FUV data, with peak fluxes 10s of times above quiescence, and occur a few times per day.
\item Dramatic (and as yet unobserved) events with equivalent  durations of $>10^{6}$~s might occur about once a month.
\end{itemize}
As a reference point, we estimate the Great AD~Leo Flare \citep{hawley91} had an equivalent duration of a few $\times10^{4}$ to $10^{5}$~ks in the FUV. 
The thresholds in energy for the flare rate predictions in Table \ref{tbl:energy_stats} follow the same pattern, however rates at the various thresholds vary between active and inactive stars (Section \ref{sec:absrel}).
The largest energy threshold, $10^{33}$~erg, represents an event where the energy  emitted in the  FUV alone  would  designate it a ``superflare'' (a flare with energy greater than  any solar flares yet observed).
It is important to note that the highest thresholds in $\delta$ and $E$ represent extrapolations.
Assuming such extrapolations hold, statistical uncertainties nonetheless balloon as the power-laws are extrapolated further from the range of observed events.
In consequence, the waiting time between FUV superflares can only be constrained to a range of decades to weeks. 
Flare surveys in the FUV have not reached sufficient durations to measure the true rate at which such energetic, infrequent events occur. 

Another of the quantities derived from the power law fits is the predicted ratio of FUV energy emitted by flares to that emitted by quiescence. 
Loosely worded, this amounts to an integral of the $\delta$ FFD within a chosen  range under the assumption that the FFD is well-described by a  single power law  within that range. 
Considering a range of $10 < \delta < 10^{6}$~s  yields a cumulative energy output anywhere from a tenth to a few times the quiescent emission of the star.
This suggests a star's flares could dominate FUV emission, a question  we explore further with another derived quantity, $\delta_\mathrm{crit}$, discussed in greater detail  in Section \ref{sec:fvq}.

As a means of comparing the absolute energy  output  of a star's flares while accounting  for differences  in the stellar surface area available for magnetic processes, we have also computed an FUV flare ``surface flux.''
This  averages the  integrated energy of  flares within a  given energy  range  over  both time and the stellar  surface area.
Hence, a large value of the  flare surface flux could be interpreted as indicating greater heating by magnetic reconnection per unit area on the star.
We  computed this  value for flares within the rough energy range identified in this analysis, $10^{27}$~--~$10^{31}$~erg. 
We  consider the FUV  flare surface flux to be an absolute  metric of a star's flare activity, while the aforementioned ratio of flare to quiescent emission is a corresponding relative metric. 

For each power-law fit, we assess the goodness-of-fit with a stabilized Kolmogorov-Smirnov (KS)  test \citep{maschberger09}.
The stabilized KS test was second most sensitive test in discriminating non power-law behavior in the comparison carried out by \cite{maschberger09} and was readily adaptable for application to events aggregated from multiple datasets with differing detection limits.
We compare to Monte-Carlo simulations of data drawn from actual power laws to determine a $p$-value for the statistic.
The $p$-value represents the likelihood that a power-law could explain the observed flare energies or equivalent durations.
One might  reasonably take  any value above 0.05 to indicate an acceptable  fit.
Lower values indicate increasingly poor fits.
Having presented the methodology and results of the FFD fits, we devote the remainder of this section to a discussion of their various implications. \\ \\

\begin{deluxetable*}{lrrrrrrrrrrr}
\rotate
\tablewidth{0pt}
\tabletypesize{\scriptsize}
\tablecaption{Parameters of fits to a cumulative flare distribution of the form $\nu = \mu (\delta / 1000\ \mathrm{s})^{-\alpha}$  in the \gfuv\ bandpass and derived quantities.\label{tbl:pew_stats}}
\tablehead{\colhead{Star} & \colhead{$N_\mathrm{fit}$\tablenotemark{a}} & \colhead{$N_\mathrm{all}$\tablenotemark{b}} & \colhead{$\alpha$} & \colhead{KS Test\tablenotemark{c}} & \colhead{$\log\left(\mu\right)$} & \colhead{$\log\left(\nu(>10 \mathrm{\ s})\right)$} & \colhead{$\log\left(\nu(>10^3 \mathrm{\ s})\right)$} & \colhead{$\log\left(\nu(>10^6 \mathrm{\ s})\right)$} & \colhead{$\log\left(E_f/E_q\right)$\tablenotemark{d}} & \colhead{$\log\left(\delta_\mathrm{crit}\right)$\tablenotemark{e}} & \colhead{$\delta_\mathrm{min}$\tablenotemark{f}} \\ \colhead{} & \colhead{} & \colhead{} & \colhead{} & \colhead{$p$-value} & \colhead{$\log(\mathrm{d}^{-1}$)} & \colhead{$\log(\mathrm{d}^{-1})$} & \colhead{$\log(\mathrm{d}^{-1})$} & \colhead{$\log(\mathrm{d}^{-1})$} & \colhead{} & \colhead{$\log(\mathrm{s})$} & \colhead{s}}
\startdata
GJ 667C & 1 & 1 & \nodata & \nodata & $0.22_{-0.58}^{+0.43}$ & $1.76_{-0.59}^{+0.45}$ & $0.22_{-0.58}^{+0.43}$ & $-2.10_{-0.68}^{+0.57}$ & $-0.52_{-0.59}^{+0.45}$ & $7.0_{-1.9}^{+8.9}$ & 260\\
GJ 176 & 6 & 6 & $0.98_{-0.34}^{+0.43}$ & 1.0 & $0.33_{-0.57}^{+0.47}$ & $2.32_{-0.31}^{+0.35}$ & $0.33_{-0.57}^{+0.47}$ & $-2.6_{-1.8}^{+1.4}$ & $-0.48_{-0.32}^{+0.64}$ & $5.3_{-2.7}^{+5.0}$ & 54\\
GJ 832 & 4 & 4 & \nodata & \nodata & $0.20_{-0.35}^{+0.28}$ & $1.73_{-0.33}^{+0.27}$ & $0.20_{-0.35}^{+0.28}$ & $-2.11_{-0.54}^{+0.49}$ & $-0.55_{-0.38}^{+0.34}$ & $7.2_{-1.3}^{+7.5}$ & 48, 57\\
GJ 436 & 0 & 1 & \nodata & \nodata & $<1.0$ & $<2.8$ & $<1.0$ & $<-1.2$ & $<0.29$ & $6_{-5}^{+11}$ & 1000, 510\\
GJ 581 & 0 & 1 & \nodata & \nodata & $<1.4$ & $<3.2$ & $<1.4$ & $<-0.95$ & $<0.64$ & $5_{-16}^{+8}$ & 1200, 1600\\
GJ 876 & 6 & 8 & $0.58_{-0.21}^{+0.23}$ & 0.4 & $0.75_{-0.40}^{+0.32}$ & $ 1.91 \pm 0.24 $ & $0.75_{-0.40}^{+0.32}$ & $-1.0_{-1.1}^{+0.9}$ & $0.21_{-0.63}^{+0.55}$ & $4.9_{-0.4}^{+4.6}$ & 63, 47\\
AU Mic & 0 & 2 & \nodata & \nodata & $<0.90$ & $<2.6$ & $<0.90$ & $<-1.3$ & $<0.16$ & $7_{-2}^{+12}$ & 230\\
EV Lac & 0 & 2 & \nodata & \nodata & $<1.2$ & $<2.9$ & $<1.2$ & $<-1.1$ & $<0.42$ & $5_{-7}^{+10}$ & 580\\
AD Leo & 20 & 34 & $0.93_{-0.19}^{+0.22}$ & 0.5 & $0.45_{-0.25}^{+0.22}$ & $2.30_{-0.22}^{+0.27}$ & $0.45_{-0.25}^{+0.22}$ & $-2.33_{-0.86}^{+0.78}$ & $-0.38_{-0.24}^{+0.32}$ & $6_{-15}^{+3}$ & 100, 64\\
Prox Cen & 6 & 20 & $0.87_{-0.28}^{+0.34}$ & 0.3 & $0.92_{-0.23}^{+0.21}$ & $2.67_{-0.58}^{+0.65}$ & $0.92_{-0.23}^{+0.21}$ & $-1.7_{-1.1}^{+0.9}$ & $0.22_{-0.27}^{+0.30}$ & $5_{-20}^{+2}$ & 1500, 370\\
Inactive\tablenotemark{g} & 17 & 22 & $0.77_{-0.15}^{+0.17}$ & 0.7 & $0.45_{-0.23}^{+0.20}$ & $2.00_{-0.19}^{+0.18}$ & $0.45_{-0.23}^{+0.20}$ & $-1.86_{-0.73}^{+0.63}$ & $-0.30_{-0.33}^{+0.34}$ & $6.0_{-0.8}^{+4.9}$ & \nodata\\
Active\tablenotemark{h} & 26 & 58 & $0.80_{-0.13}^{+0.14}$ & 0.9 & $0.60_{-0.17}^{+0.16}$ & $2.21_{-0.19}^{+0.22}$ & $0.60_{-0.17}^{+0.16}$ & $-1.80_{-0.56}^{+0.50}$ & $-0.17_{-0.24}^{+0.25}$ & $5.9_{-2.3}^{+2.5}$ & \nodata\\
All & 43 & 80 & $0.76_{-0.09}^{+0.10}$ & 0.5 & $ 0.57 \pm 0.14 $ & $2.08_{-0.12}^{+0.15}$ & $ 0.57 \pm 0.14 $ & $-1.71_{-0.43}^{+0.38}$ & $ -0.17 \pm 0.21 $ & $6.2_{-0.8}^{+2.7}$ & \nodata\\
\enddata

\tablenotetext{a}{Number of flares used in the FFD fit, i.e. only those with equivalent durations where the survey was deemed sufficiently complete (Appendix \ref{app:inject}).}
\tablenotetext{b}{Total number of flares detected. If $N_\mathrm{all} > N_\mathrm{fit}$, then the difference represents flares not used in the FFD fits because they had equivalent durations below the threshold where the survey was deemed sufficiently complete (Appendix \ref{app:inject}).}
\tablenotetext{c}{Stabilized KS test from \cite{maschberger09}. Lower $p$-values imply a lower probability of the events having been generated by a power law.}
\tablenotetext{d}{Ratio of flare to quiescent energy emitted averaged over very long timescales based on the power-law fit, integrating over an equivalent duration range of 10~--~$10^6$~s.}
\tablenotetext{e}{Critical equivalent duration beyond which, if the power-law model holds, energy emitted by flares over long timescales will exceed the integrated quiescent emission. Error bars are defined by the location of the 5th adn 95th percentiles.}
\tablenotetext{f}{Detection limit of each dataset.}
\tablenotetext{g}{Stars from the MUSCLES survey, $EW_\mathrm{Ca\ II\ K} < 2$~\AA.}
\tablenotetext{h}{Flare stars with archival data, $EW_\mathrm{Ca\ II\ K} > 10$~\AA.}

\tablecomments{Values are quoted as the median value of the MCMC samples with error bars defined by the central 68\% of the distribution.}
\end{deluxetable*}

\begin{deluxetable*}{lrrrrrrrrrr}
\rotate
\tablewidth{0pt}
\tabletypesize{\scriptsize}
\tablecaption{Parameters of fits to a cumulative flare distribution of the form $\nu = \mu (E / 10^{30}\ \mathrm{erg})^{-\alpha}$ in the \gfuv\ bandpass and derived quantities.\label{tbl:energy_stats}}
\tablehead{\colhead{Star} & \colhead{$N_\mathrm{fit}$\tablenotemark{a}} & \colhead{$N_\mathrm{all}$\tablenotemark{b}} & \colhead{$\alpha$} & \colhead{KS Test\tablenotemark{c}} & \colhead{$\log\left(\mu\right)$} & \colhead{$\log\left(\nu(>10^{27} \mathrm{erg})\right)$} & \colhead{$\log\left(\nu(>10^{30} \mathrm{erg})\right)$} & \colhead{$\log\left(\nu(>10^{33} \mathrm{erg})\right)$} & \colhead{$\log\left(F_\mathrm{sfc}\right)$\tablenotemark{d}} & \colhead{$E_\mathrm{min}$\tablenotemark{e}} \\ \colhead{} & \colhead{} & \colhead{} & \colhead{} & \colhead{$p$-value} & \colhead{$\log(\mathrm{d}^{-1})$} & \colhead{$\log(\mathrm{d}^{-1})$} & \colhead{$\log(\mathrm{d}^{-1})$} & \colhead{$\log(\mathrm{d}^{-1})$} & \colhead{$\log$(erg s$^{-1}$ cm$^{-2})$} & \colhead{10$^{27}$ erg}}
\startdata
GJ 667C & 1 & 1 & \nodata & \nodata & $-0.90_{-0.68}^{+0.57}$ & $1.37_{-0.62}^{+0.46}$ & $-0.90_{-0.68}^{+0.57}$ & $-3.1_{-1.1}^{+0.9}$ & $2.77_{-0.62}^{+0.47}$ & 9.0\\
GJ 176 & 6 & 6 & $0.95_{-0.32}^{+0.41}$ & 0.97 & $-0.25_{-0.81}^{+0.65}$ & $2.62_{-0.40}^{+0.48}$ & $-0.25_{-0.81}^{+0.65}$ & $-3.1_{-2.0}^{+1.6}$ & $3.73_{-0.24}^{+0.29}$ & 12\\
GJ 832 & 4 & 4 & \nodata & \nodata & $-0.59_{-0.50}^{+0.45}$ & $1.69_{-0.28}^{+0.25}$ & $-0.59_{-0.50}^{+0.45}$ & $-2.84_{-0.99}^{+0.88}$ & $3.06_{-0.35}^{+0.34}$ & 2.6, 3.8\\
GJ 436 & 0 & 1 & \nodata & \nodata & $<-1.0$ & $<2.4$ & $<-1.0$ & $<-4.4$ & $<3.2$ & 24, 34\\
GJ 581 & 0 & 1 & \nodata & \nodata & $<-1.2$ & $<2.2$ & $<-1.2$ & $<-4.6$ & $<3.3$ & 10, 19\\
GJ 876 & 6 & 8 & $0.60_{-0.20}^{+0.27}$ & 0.62 & $0.15_{-0.63}^{+0.50}$ & $1.98_{-0.25}^{+0.27}$ & $0.15_{-0.63}^{+0.50}$ & $-1.7_{-1.4}^{+1.1}$ & $3.86_{-0.39}^{+0.36}$ & 6.7, 5.6\\
AU Mic & 0 & 2 & \nodata & \nodata & $<1.2$ & $<4.7$ & $<1.2$ & $<-1.9$ & $<4.9$ & 1800\\
EV Lac & 0 & 2 & \nodata & \nodata & $<-0.0073$ & $<3.6$ & $<-0.0073$ & $<-3.6$ & $<4.4$ & 350\\
AD Leo & 20 & 34 & $0.93_{-0.20}^{+0.22}$ & 0.52 & $0.74_{-0.19}^{+0.17}$ & $3.54_{-0.49}^{+0.55}$ & $0.74_{-0.19}^{+0.17}$ & $-2.06_{-0.80}^{+0.72}$ & $4.63_{-0.15}^{+0.24}$ & 210, 130\\
Prox Cen & 6 & 20 & $0.87_{-0.28}^{+0.37}$ & 0.52 & $-0.39_{-0.64}^{+0.50}$ & $2.23_{-0.43}^{+0.51}$ & $-0.39_{-0.64}^{+0.50}$ & $-3.0_{-1.7}^{+1.3}$ & $4.48_{-0.22}^{+0.23}$ & 28, 16\\
Inactive\tablenotemark{f} & 17 & 22 & $0.74_{-0.15}^{+0.18}$ & 0.45 & $-0.31_{-0.39}^{+0.34}$ & $ 1.91 \pm 0.18 $ & $-0.31_{-0.39}^{+0.34}$ & $-2.52_{-0.91}^{+0.79}$ & \nodata & \nodata\\
Active\tablenotemark{g} & 26 & 58 & $0.525_{-0.082}^{+0.089}$ & 0.022 & $0.64_{-0.16}^{+0.15}$ & $2.22_{-0.18}^{+0.19}$ & $0.64_{-0.16}^{+0.15}$ & $-0.92_{-0.41}^{+0.36}$ & \nodata & \nodata\\
\enddata

\tablenotetext{a}{Number of flares used in the FFD fit, i.e. only those with equivalent durations where the survey was deemed sufficiently complete (Appendix \ref{app:inject}).}
\tablenotetext{b}{Total number of flares detected. If $N_\mathrm{all} > N_\mathrm{fit}$, then the difference represents flares not used in the FFD fits because they had equivalent durations below the threshold where the survey was deemed sufficiently complete (Appendix \ref{app:inject}).}
\tablenotetext{c}{Stabilized KS test from \cite{maschberger09}. Lower $p$-values imply a lower probability of the events having been generated by a power law.}
\tablenotetext{d}{Average $FUV_{130}$ surface flux from flares averaged over very long timescales based on power law fit integrated across the observed flare energy range.}
\tablenotetext{e}{Detection limit of each dataset.}
\tablenotetext{f}{Stars from the MUSCLES survey, $EW_\mathrm{Ca\ II\ K} < 2$~\AA.}
\tablenotetext{g}{Flare stars with archival data, $EW_\mathrm{Ca\ II\ K} > 10$~\AA.}

\tablecomments{Fits to flares aggregated from both inactive and active stars are not included due to the strong differences in flare rates and energies, see text. Values are quoted as the median value of the MCMC samples with error bars defined by the central 68\% of the distribution. }
\end{deluxetable*}

\begin{figure*}
\includegraphics{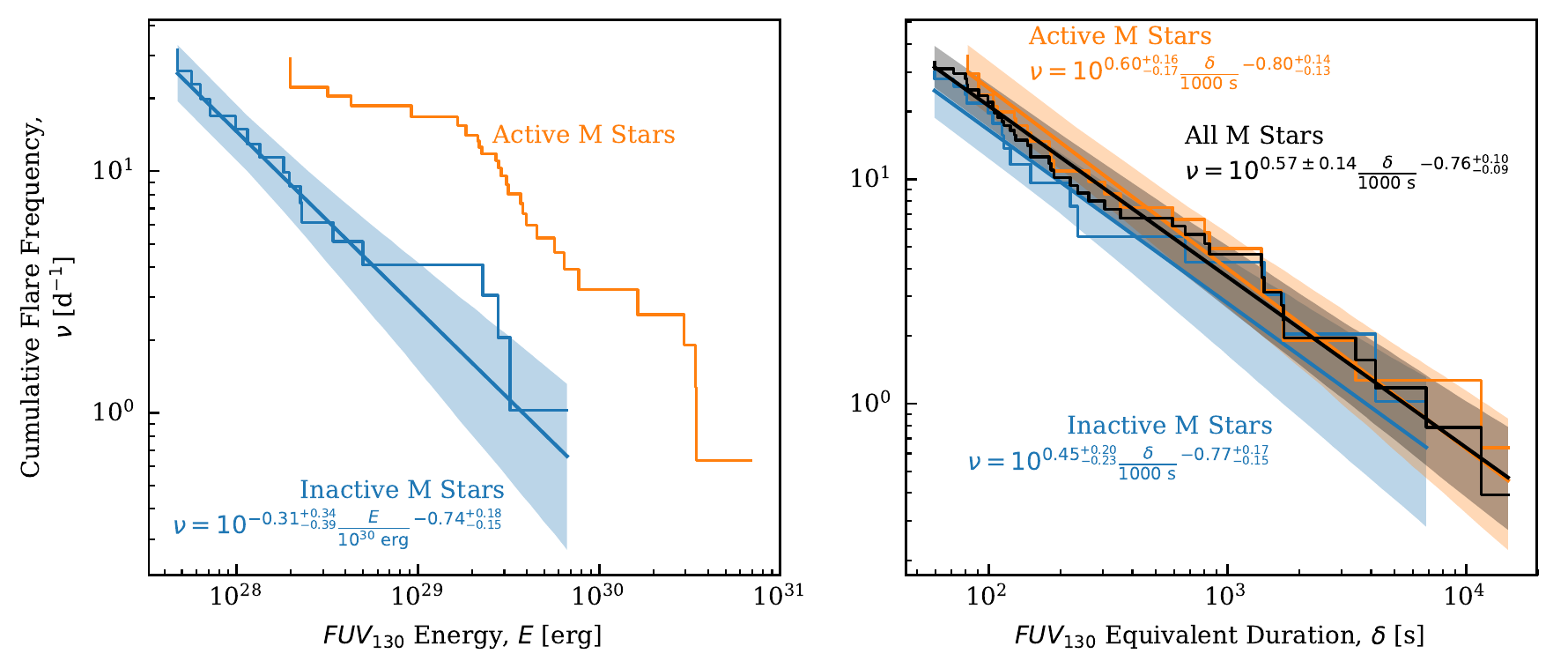}
\caption{Flare frequency distributions and power-law fits in absolute energy, $E$ (left), and equivalent duration (energy relative to quiescent emission), $\delta$ (right), in the \gfuv\ bandpass.
Flares are aggregated from two groups, ``active'' ($\mathrm{EW}_\mathrm{Ca\ II\ K} > 10$~\AA, Section \ref{sec:data}) and ``inactive'' ($\mathrm{EW}_\mathrm{Ca\ II\ K} < 2$~\AA) M stars.
For flares occurring at a given rate, the flares of active stars are about an order of magnitude more energetic, but in relative units  the two distributions are statistically indistinguishable.
Note that the fits are not made directly to the stairstep lines and a fit to the active-star flares in absolute  energy  is not shown due to substantial bias (see Section \ref{sec:absrel}). }
\label{fig:ffd_abs_v_rel}
\end{figure*}

\subsection{M-Dwarf Flares: Absolutely Different, Relatively the Same}
\label{sec:absrel}
The FFDs of the inactive ($\mathrm{EW}_\mathrm{Ca\ II\ K} < 2$~\AA) and active ($\mathrm{EW}_\mathrm{Ca\ II\ K} > 10$~\AA) star flares, plotted in Figure \ref{fig:ffd_abs_v_rel}, are well separated in energy.
For a given flare frequency, the energy of the active-star flares is about an order of magnitude larger than those of the inactive stars. 
This result is consistent with previous studies that show greater flare activity in active stars based on absolute flare energy comparisons \citep{hilton11,hawley14}.
However, Prox~Cen is an exception, having a rate of  10$^{30}$~erg flares about an order of magnitude below AD~Leo and AU~Mic. 
(For EV~Lac only an upper limit is possible.)
This could be due to the comparative youth of AD~Leo ($<$300~Myr; \citealt{shkolnik09}) and AU~Mic (12~Myr; \citealt{plavchan09}) versus Prox~Cen (5.8~Gyr; \citealt{yildiz07}).

The observations of Prox~Cen and AD~Leo dominate the active-star sample, but, due to Prox~Cen's nearness, flares of lower energy could be sampled than for AD~Leo.
Given Prox~Cen's order-of-magnitude lower rate of $10^{30}$~erg flares, aggregating flares from all active stars results in a paucity of flares at the low-energy end of the distribution and a highly biased power-law fit with an index of 0.5, below that of either Prox~Cen or AD~Leo.
This results in a poor fit to a power-law as indicated by its low KS test $p$-value (Table~\ref{tbl:energy_stats}), and we exclude this fit from Figure~\ref{fig:ffd_abs_v_rel}.

The power-law fit describing the inactive star flares has an index of 0.74, within the range of values estimated by M star flare studies in other bandpasses.
In comparison, \cite{hilton11} obtained a value of 0.5 for M3~--~M5 stars (SDSS U band); \cite{davenport16a} obtained values of 0.5~--~0.9 for the 49 targets with masses in the range of 0.2~--~0.5~$M_\sun$ (Kepler band); and \cite{hawley14} obtained indices of 0.5 and 0.8 for two inactive M1 and M2 dwarfs and 0.7 and 1.0 for two active M4 and M5 dwarfs.
For Prox~Cen, there are well-determined energy FFDs in the visible from Evryscope and \textit{MOST} observations, yielding indices of 0.7 and 1.0 in comparison to 0.9 in this work \citep{davenport16b,howard18}. 

FFDs in different bands for the same object provide an avenue for estimating the average energy budget of a flare in lieu of simultaneous observations.
The difference in the energy of flares occurring at the same rate gives the ratio of the energy emitted by flares in the observed bands, assuming the observations are cataloging the same root phenomenon (i.e. that white-light flares do not result from a different physical process than FUV flares).
An opportunity for this comparison is afforded by Prox~Cen's FFDs in the \gfuv, Evryscope, and \textit{MOST} bands.
From the spacing of these FFDs in energy, we infer that white-light M-star flares observed in the optical correspond to flares emitting about an order of magnitude less energy in the \gfuv\ band. 

Remarkably, when the flares are characterized in relative units, i.e., equivalent durations, the FFDs lie on top of one another.
The power-law fits to these FFDs are statistically indistinguishable in rate constant and index. 
This is in spite of the supposed differing levels of magnetic activity on these stars that results in disparate levels of emission from chromospheric lines like \Caii~K observed at optical wavelengths.
This implies that, while the overall rate of magnetic heating might be greater for ``active'' M dwarfs, the form of the magnetic heating is unchanged. 
This accords well with a model in which the inactive stars simply have a lower ``magnetic filling factor'' than active stars. 

The consistency of FFDs in relative units means their difference in absolute units could be predicted directly from their difference in quiescent \gfuv\ flux.
A further implication is that all M dwarfs, regardless of how ``inactive'' they are as gauged by chromospheric emission, will show vigorous flaring in lightcurves of FUV emission.
Future flare surveys should determine if this result is robust against larger sample sizes and whether it extends to other sources of flare emission, such as the blackbody flux predominantly emitted in the NUV.
In the meantime, this result has critical importance for exoplanets, as it implies that a single observation of an M dwarf's quiescent FUV  flux level will also constrain the energies  of that stars' FUV flares (e.g., a 100$\times$ greater FUV  flux  indicates $\sim$100$\times$ more energetic flares).
The consistency of equivalent-duration FFDs also means that conclusions drawn from the FFD presented in this work will likely apply to all M stars.
We pursue several such conclusions pertaining to stellar physics in the following subsections.

\subsection{Energy Emitted at FUV Wavelengths Could be Dominated by Flares}
\label{sec:fvq}

For the targets with the lowest detection limits in $\delta$ (GJ~176, GJ~832, GJ~876, and AD~Leo), the observed flares contributed 10~--~40\% of the total \gfuv\ energy emitted by the star.
This is a significant fraction; however, this value does not reflect the true contribution of flares to the overall energy budget of M dwarf \gfuv\ emission.
The observations were too limited in duration to capture infrequent, highly energetic flares, yet the $\alpha < 1$ slope of the FFD power laws implies these flares contribute more energy than the more frequent, lower energy flares.
The same is not true for G dwarfs, for which an analysis of Kepler data yielded  power-law slopes of -1~--~-1.2 \citep{shibayama13} to an energy FFD. 
For M dwarfs, rare, energetic, unobserved flares will significantly raise the relative contribution of flares to a star's FUV emission, potentially to a point where flares contribute as much or more energy in the \gfuv\ band than the star's quiescent emission (when considering timescales long enough to include such rare flares). 

As such, we pose the question ``how far must the power-law fit to an M-dwarf FFD be extrapolated before the energy emitted by flares will match that emitted by quiescence?"
This quantity can be derived from the power-law fit to the flare equivalent durations, and we term it the ``critical equivalent duration,'' $\dcrit$.
Starting from Eq. \ref{eqn:pow}, we obtain
\begin{equation}
\dcrit = \delta_\mathrm{ref} \left( \frac{1-\alpha}{\mu\alpha} \right)^{1/(1-\alpha)}.
\end{equation}

The critical equivalent duration is given for each object in Table \ref{tbl:pew_stats}.
It is very sensitive to uncertainty in $\alpha$, resulting in more than a 3-order-of-magnitude range in possible values for the power-law FFD fit of the aggregated flare sample.
If this power law extends unmodified to $\delta$ values of $10^{6.2}$, odds favor flares as contributing more energy than quiescence to M dwarf emission in the \gfuv\ band.  
Such energetic flares would occur every in the range of once per few weeks to once per year.
In conclusion, it seems possible that flares dominate the FUV emission of M dwarfs.
However, this begs the question, are flares with $\delta > \dcrit$ possible?
We address this in the next subsection.

\subsection{How Big Do Flares Get?}
\label{sec:elimit}
The upper limit on the energy of flares will determine both the relative fraction of flare energy that is missed by finite-duration observations and whether this energy dominates overall FUV emission. 
Yet such a limit is difficult to constrain, since the most energetic events are also the rarest, thereby difficult to observe. 
Here, we explore constraints on such a limit.
Because flare surveys generally employ energy rather than equivalent duration and because it is natural to expect a physical limit on flares to apply to energy rather than  equivalent duration, we frame much of the discussion of this section in terms of flare energies.

The most energetic flare spectrally and temporally resolved in the FUV is the Great Flare of 1985 on AD~Leo \citep{hawley91}.
This flare produced equivalent durations in \Cii\ and \Civ\ during its impulsive phase (start of the flare to the start of its gradual decay; estimated from the plots in \citealt{hawley91}) on the order of 40~ks (\Cii) and 70~ks (\Civ).
Considering only the impulsive phase of the AD~Leo and GJ~876 flares here presented, we find equivalent durations of 0.4~ks (AD~Leo, \Cii), 0.9~ks (AD~Leo, \Civ), and  3~ks (GJ~876, \Cii, no \Civ\ data).
These are 10~--~100$\times$ below that of the 1985 AD~Leo flare.
Observations of the 1985 AD~Leo flare saturated in the strongest emission lines, complicating the interpretation of its light curves in those lines.
However, \cite{hawley91} reconstruct the lines based on fits to their unsaturated wings.
The flux enhancements estimated from these reconstructions are similar to those observed for the GJ~876 and AD~Leo flares mentioned above, meaning the greater equivalent duration of the 1985 AD~Leo flare is predominantly due to the 1985 flare's 10~--~100$\times$ longer impulsive phase.

M dwarf flares of much greater energy have been observed in other bandpasses, such as a flare on AU~Mic radiating $3\sn{34}$ erg in the EUV \citep{cully93}, a flare on EV~Lac radiating $10^{34}$~erg in 0.3~--~10~keV X-rays \citep{osten10}, and two flares by DG CVn (a young M4 binary) radiating a few $10^{34}$~erg in the V band ($10^{36}$~erg in 0.3~--~10~keV X-rays; \citealt{osten16}).
Scaling to FUV emission based on the multiwavelength AD~Leo flare observations of \cite{hawley03}, these flares are 3-4 orders of magnitude more energetic than the most energetic inactive-star flare ($\sim10^{30}$ erg) and active-star flare ($\sim10^{31}$ erg), implying equivalent durations of $10^7$~--~$10^8$~s.

More complete flare samples are accessible through surveys using \textit{U} band and optical photometry.
The \cite{hawley14} anlysis of \textit{Kepler} data and \cite{hilton11} ground-based \textit{U}-band campaign yielded M dwarf flares that, again scaled based on \cite{hawley03}, are up to 2-3 orders of magnitude more energetic than the largest of this survey, i.e. equivalent durations of $10^6$~--~$10^7$~s.
If \textit{U} and \textit{Kepler} band scalings remain linear through this range, it would imply  the FFD we computed can be extrapolated to a limit beyond that which predicts equal contributions of flares and quiescence to FUV emission.

A much different approach to estimating an upper limit is to scale FUV equivalent duration with flare covering fraction and compute the energy of a flare covering the entire visible hemisphere.
The covering fraction of the largest AD~Leo flares we characterized was estimated by \cite{hawley03} to be roughly 0.01\% and we estimate equivalent durations of $\sim$1 ks for these flares.
This would imply, under the assumption  the FUV flare flux increases linearly with the flare covering fraction, the rate of flares would begin to fall below power-law predictions sometime before 3 orders of magnitude above the most energetic flares here characterized, about an order of magnitude above the most likely $\dcrit$. 
Similarly, another theoretical upper limit could be obtained through MHD modeling like that of \cite{aulanier13}, who estimate a theoretical upper limit of $6\sn{33}$~erg  for solar flares, but that is beyond the scope of this work. 
No matter which way the question is approached, it is reasonable to suspect that FUV flares can reach sufficient energy for the FUV energy budget of most M dwarfs to be dominated by flares.

\subsection{How Small do Flares Get? Microflares and Quiescent FUV Emission}
Whereas we have just discussed the implications of the high-energy end of the M-dwarf FFD, we now discuss the low-energy end. 
The cumulative effect of frequent, low-energy flares (``microflares'' or ``nanoflares'') has been suggested as a resolution to the coronal heating problem \citep{gold64,parker72}. 
For a power-law FFD,  if the index for the cumulative distribution ($\alpha$ in Eq. \ref{eqn:pow}) is $>1$, then the energy contributed by the smallest flares is unbounded.
Specifically, the integral yielding the total energy of all flares diverges as the low-energy bound on the integral approaches zero. 
This is not the case for the FUV flares we characterized.
Since $\alpha < 1$ for these flares, the contribution of weak flares hidden in the noise to the energy budget is bounded, even with the lower limit on integration set to zero.

Assuming the power-law FFD extends unmodified to infinitesimal flare energies, the contribution of undetected flares to the overall quiescent emission can be expressed as
\begin{equation}
\frac{F_{\mathrm{UF}}}{F_q} = \delta_\mathrm{lim} \frac{\mu\alpha}{1 - \alpha} \left( \frac{\delta_\mathrm{lim}}{\delta_\mathrm{ref}}\right)^{1-\alpha},
\end{equation}
where $F_{\mathrm{UF}}$ is the time-averaged flux from undetected flares, $F_q$ is the quiescent flux, and  $\delta_\mathrm{lim}$ is the equivalent duration detection limit. 
The FFD constrained in this work for flares aggregated from all M dwarfs predicts $F_\mathrm{UF}/F_q= 2-20$\% (with $\delta_\mathrm{lim}$ taken to be 200 s as a representative value), i.e unresolved flares do not account for quiescent FUV flux.
This conclusion is in line with those regarding solar EUV and X-ray flares, which cannot explain coronal heating (e.g., \citealt{hudson91}).

However, much of the \gfuv\ band includes a compendium of emission sources tracing different regions of the stellar atmosphere.
Isolating specific emission lines, essentially localizing the region of the stellar atmosphere being considered, yields differing results than considering the integrated \gfuv\ (Section \ref{sec:lineffds}).
At the extremes, analyzing flares in  \Nv\ emission yields $F_\mathrm{UF}/F_q = 1-6$\% whereas in \Siiv\ $F_\mathrm{UF}/F_q$ emission is essentially unity. 
Hence, unresolved flares could be directly responsible for quiescent \Siiv\ emission.

A value near unity for the \gfuv\ band would have tidily explained the consistency of the active- and inactive-star equivalent-duration FFDs. 
If the quiescent emission were merely unresolved flares, then it is natural that normalizing flare energies by such a quiescence would produce consistent results.
Because this is not the case, we conclude that the consistency of the equivalent-duration FFDs must be a result of some other underlying link between flares and quiescence.
That such a link would exist, given that both are likely powered by magnetic processes, is no great surprise.

To summarize Section \ref{sec:ffds}, we have shown that, while active-star flares might be generally an order of magnitude more energetic than inactive-star flares, in equivalent duration the two neatly match.
The consistent FFDs in equivalent duration imply that highly energetic flares are an important, perhaps dominant, contributor of FUV emission.
Meanwhile, unresolved, overlapping flares are insufficient to explain quiescent FUV emission.
Regardless, the energy of FUV flares cannot be disregarded when considering the transition-region emission of M dwarfs.

\begin{figure}
\includegraphics{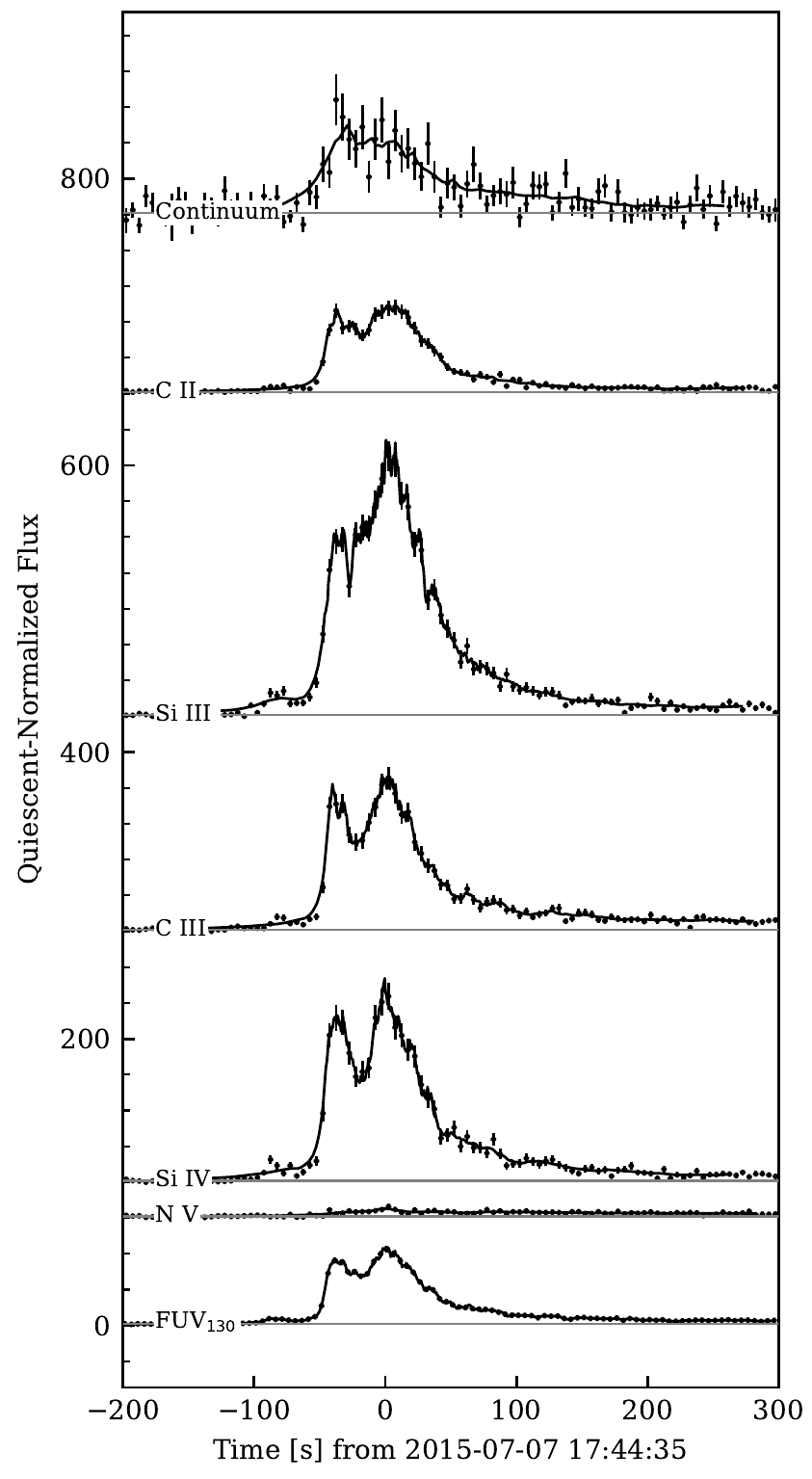}
\caption{Spectrophotometry of the most energetic flare observed on GJ~876.
Lightcurves have been normalized by the quiescent flux and offset vertically for display.
Underlying lines are ``count-binned'' (see Section \ref{sec:countbinning}) to provide adaptive time resolution.
The points are time-binned at a 5~s cadence.}
\label{fig:flare_gj876}
\end{figure}

\begin{figure*}
\includegraphics{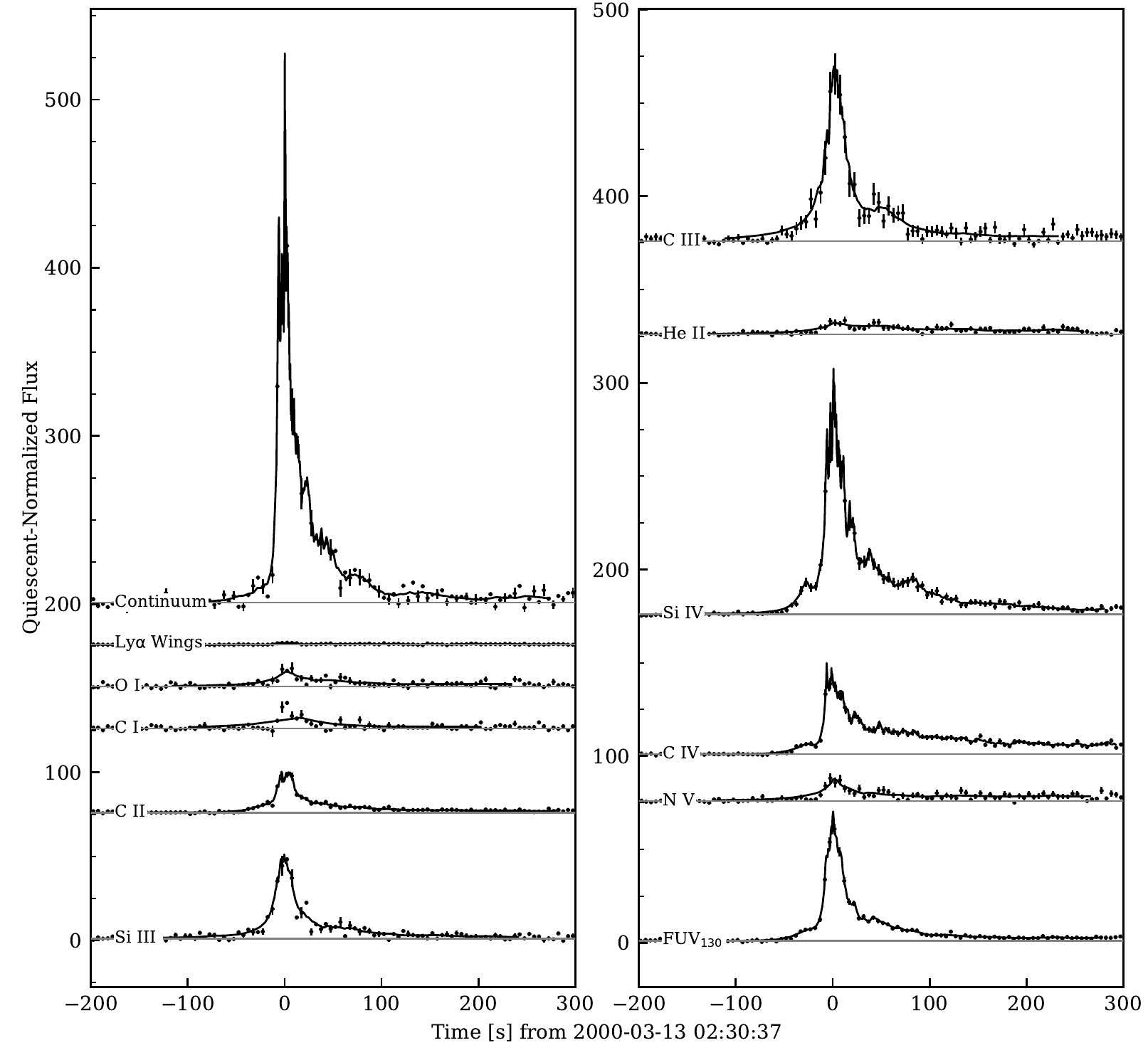}
\caption{Spectrophotometry of the most energetic flare observed on AD~Leo.
Lightcurves have been normalized by the quiescent flux and offset vertically for display.
Underlying lines are ``count-binned'' (see Section \ref{sec:countbinning}) to provide adaptive time resolution.
The points are time-binned at a 5~s cadence.
The figure has been split into two panels simply so lightcurves extracted from all the strong emission lines in the STIS E140M data could be included on a single page. 
A standout feature is the strong continuum response.
However, we find the energy emitted in the continuum is within the overall scatter in flare energy budgets described in Section \ref{sec:budget}.}
\label{fig:flare_adleo}
\end{figure*}

\section{Flares in Isolated Emission Lines}
\label{sec:lineflares}

Thus far, this paper has dealt only with the time dimension of the data.
The fact that flares are spectrally, as well as temporally, resolved in this dataset provides a wealth of additional information. 
It can be used to compare the effect of magnetic reconnection events on differing regions of the stellar atmosphere as manifested in the various sources of FUV emission, i.e. various lines and the continuum.
Similarly, by matching lines with counterparts formed in the same regions of the stellar atmosphere, comparisons can be made to flares observed in other wavelength regimes, such as solar flares observed in the EUV.
The spectral dimension also allows for the potential detection of mass flows related to flares.
These topics are explored in the following subsections.

As a basis for the discussion that follows, examples of how different sources of emission respond during a flare are shown in Figures \ref{fig:flare_gj876} and \ref{fig:flare_adleo} for two flares of particularly high S/N (and correspondingly high equivalent duration), showing the evolution of the flare in the broad \gfuv\ band, all major lines, and a compendium of narrow continuum bands hand-selected from a high S/N spectrum. 
Note that the GJ~876 flare plotted in Figure \ref{fig:flare_gj876} is the same flare analyzed in \cite{youngblood16}.

\subsection{Flare Frequency Distributions (FFDs) by Emission Line}
\label{sec:lineffds}

\begin{figure}
\includegraphics{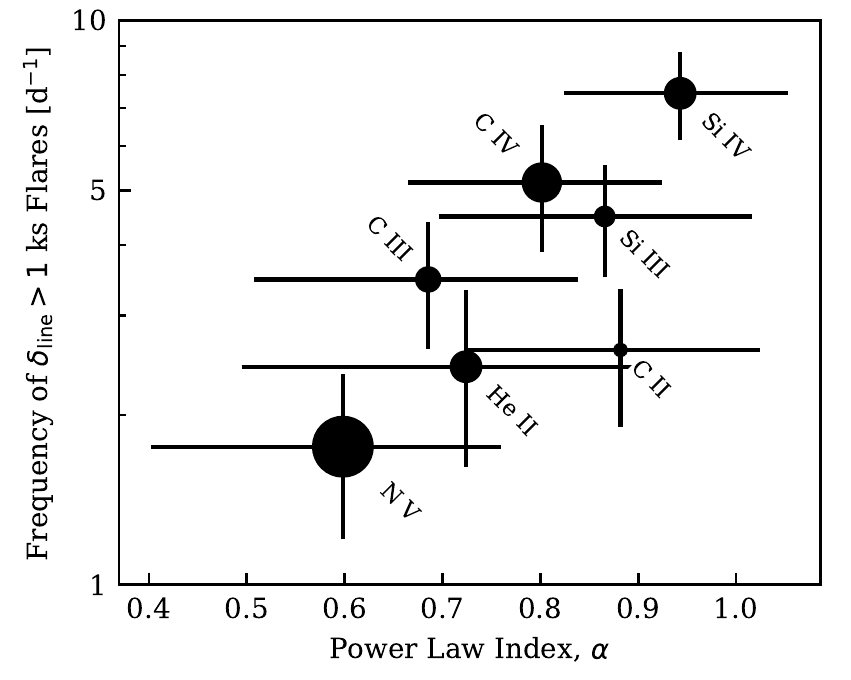}
\caption{Results of power-law fits to flares independently identified in differing emission lines.
Point radii are proportional to the line formation temperature.
The figure demonstrates the lack of any detectable relationship between flare rate, power-law index, and formation temperature, but does provide an ordering of flare activity by line (y-axis). 
 \Siiv\ shows the greatest flare activity, and therefore is an excellent probe of M dwarf flares.
Lines not shown provided too few flare detections to enable a power-law fit.
}
\label{fig:line_FFDs}
\end{figure}

We fitted the distributions of flares identified in each strong emission line with a power-law FFD in the same manner as with flares identified in broadband \gfuv\ emission. 
In Figure \ref{fig:line_FFDs}, we compare power-law fits to all major emission lines. 
For the lines not on the plot, namely \Oi, \Ci, and the wings of \lya, flares induce such a minimal response that too few flares are identified to enable a power law fit. 
The power-law indices are consistent with a single value, but flare rates are inconsistent at the 4.5$\sigma$ level.
Differences in flare rates span factors of a few between lines.
The relative ordering of these rates is reflected in the differing responses of emission in separate lines to the same events, as with the examples plotted in Figures \ref{fig:flare_gj876} and \ref{fig:flare_adleo}.
There is the hint of a relationship between the power-law index and the formation temperature of the emission, with cooler emission corresponding to lower power-law indices. 
The trend is not statistically robust, so further investigation is needed. 

The difference in flare rates between the \Siiv\ and \Heii\ lines is particularly noteworthy.
Individual flares illustrate the difference, sometimes quite dramatically, such as the AD~Leo flare shown in Figure \ref{fig:flare_adleo}.
Yet these lines have nearly identical peak formation temperatures in CHIANTI models \citep{dere09}.
A likely explanation is that the CHIANTI formation temperature is misleading and the regions of the stellar atmosphere in which \Heii\ and \Siiv\ actually form do not significantly overlap.
This is supported by nLTE modeling specifically of the formation of the \Heii\ 1640~\AA\ multiplet in the solar atmosphere by \cite{wahlstrom94}, who found  radiative ionization and pumping lower in the atmosphere at the 7,000~--~10,000~K level dominates over the peak of collisional ionization and excitation at the 70,000~K level in generating the line intensity.

In \cite{france16}, it was noted that \Siiii\ and \Siiv\ show the strongest response during the MUSCLES flares, suggesting that emission from these ions might be formed at a level in the stellar atmosphere where energy injection by reconnecting magnetic fields peaks.
Figure \ref{sec:lineffds} confirms this in a broader statistical sense with \Siiv\ exhibiting the greatest rate of  $\delta = 1$~ks flares. 
Energy injection would then drop off toward higher-temperature regions (\Nv ) and lower temperature regions (e.g., \Cii ). 
However, this conclusion is specific to emission from these optically-thin lines, as it is well-established that continuum flux accounts for the majority of the energy radiated by a flare \citep{hawley03,kowalski13,osten15}.
The disparity in flare rates as traced by differing emission lines could be fertile ground for future modeling of magnetic processes in M dwarf atmospheres, although precisely constraining the spatial distribution of injected energy would depend on the detailed properties of each emission process (i.e., care is required for cases like \Heii).

\subsection{\lya\ is a Gentle Giant}
Of particular importance in the behavior of isolated emission sources during M-dwarf flares is the muted response of \lya, the dominant source of flux in the FUV range for M dwarfs \citep{france12,france13}.
In flares producing peak fluxes in \Siiv\ 100$\times$ quiescence, flux in the \lya\ wings increases by only a factor of a few. 
The core of the line cannot be observed because it is absorbed by the ISM.
However, it could behave differently than the wings, an important consideration for planets orbiting the star that are exposed to the flux of the line core.
On average, photons in the core of the line originate higher in the stellar atmosphere than photons in the wings, which must undergo many inelastic scatterings to be shifted to the wing wavelengths.
Therefore, flare heating is likely to affect emission in the core of the line more than the wings.

In time-averaged spectra, the core of the \lya\ line can be reconstructed by fitting the wings with a model that parameterizes the ISM absorption (e.g., \citealt{youngblood16}).
However, these fits are not practical for time-series data. 
Therefore, the response of the \lya\ core must be inferred from the activity of related sources of emission.
We attempt to do so by using the \Oi\ lines at 1305~\AA\ and \Ci\ lines at 1657~\AA\ as proxies for emission by the \lya\ core.
Note that we use the term core to denote the region most impacted by ISM absorption, -100~--~100~km~\pers.
The central 10~km~\pers\ of the line is actually formed primarily in the transition region at temperatures above the formation temperatures of the \Oi\ and \Ci\ proxies in a solar model \citep{avrett08}.
However, in this model most of the central 100~km~\pers\ of the line is formed in the upper chromosphere, providing a reasonable match to the \Oi~1305~\AA\ and \Ci~1657~\AA\ lines in the same model.

\begin{figure*}
\includegraphics{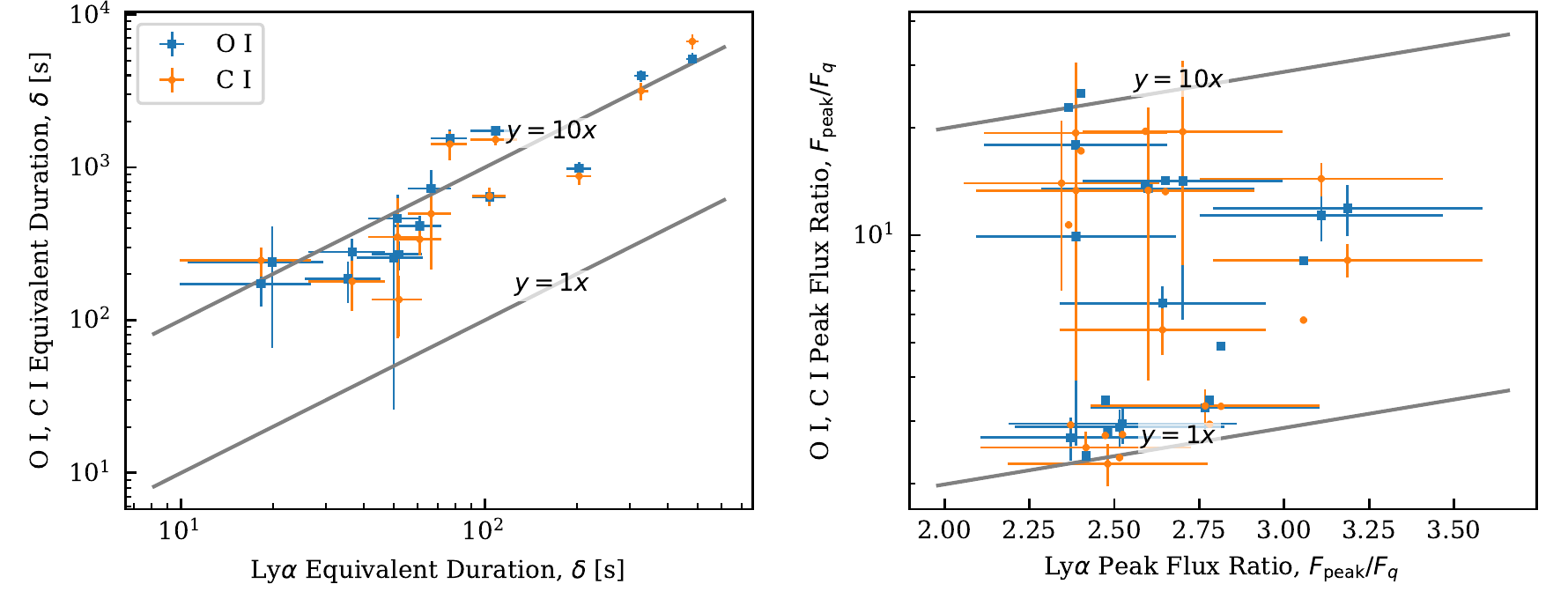}
\caption{Comparison of equivalent duration and peak flux measurements of flares using emission from the \lya\ wings and resonance lines expected to be formed cospatially with the \lya\ core.
The comparison lines are the \Oi~1305~\AA\ and \Ci~1657~\AA\ multiplets.
If \Oi\ and \Ci\ are taken as a proxy for the response in the core of the \lya\ line to a flare, then its relative increase during flares must exceed that of the wings by a factor of a few to ten.
Some error bars are omitted in the right panel to reduce clutter. 
}
\label{fig:lya_behavior}
\end{figure*}

We compare the equivalent duration and peak flux ratio of the proxy lines to that of the \lya\ wings during flares identified in the \gfuv\ bandpass in Figure \ref{fig:lya_behavior}.
Relative increases in the \Oi\ and \Ci\ lines during a flare exceed that of the \lya\ wings by a factor of one to ten, suggesting  the core of the \lya\ line responds substantially more strongly during a flare than the wings. 
This response would still be at least an order of magnitude below that of \Siiv.
The strength of the \lya\ line means that flaring emission will be an important source of photolysis in planetary atmospheres, even though flare increases are not as dramatic as in other emission lines.

\subsection{Comparison to Solar Flares}
\label{sec:solar}
Given the interest in the habitability of terrestrial M-dwarf planets versus Earth, it is worthwhile to compare, as best as is possible, the FFDs of M~dwarfs and the Sun. 
There is no solar dataset that is directly comparable to the spectrophotometric FUV data we analyzed for M dwarfs.
Specifically, we could find no FUV spectra of disk-integrated solar emission that spectrally and temporally resolves entire flares.
In the absence of directly comparable data, we used the flare catalog from the \textit{Solar Dynamics Observatory} Extreme-ultraviolet Variability Experiment (\textit{SDO} EVE) mission for comparison to M dwarf flare data \citep{hock12}.
This catalog contains measurements of flares in several EUV emission lines with formation temperatures similar to the FUV lines COS and STIS observe, all originating in the stellar transition region.

Since solar and M dwarf lines will have differing luminosities, we compared equivalent durations of flares rather than absolute energies.
This would not be an ideal comparison at FUV wavelengths because the solar photosphere contributes significant flux, but for the EUV lines all flux is from the upper atmosphere, just as with the FUV lines for the \mds. 
The EVE flare catalog provides flare energies and pre-flare fluxes (among other metrics), but no estimates of equivalent durations.
Therefore, we estimated these as
\begin{equation}
\delta = \frac{E}{F_\mathrm{pre} 4\pi (1\ \mathrm{AU})^2},
\end{equation}
where $F_\mathrm{pre}$ is the pre-flare flux at 1~AU.
From these estimates, we constructed FFDs in the same manner as with the \md\ data. 

\begin{figure}
\includegraphics{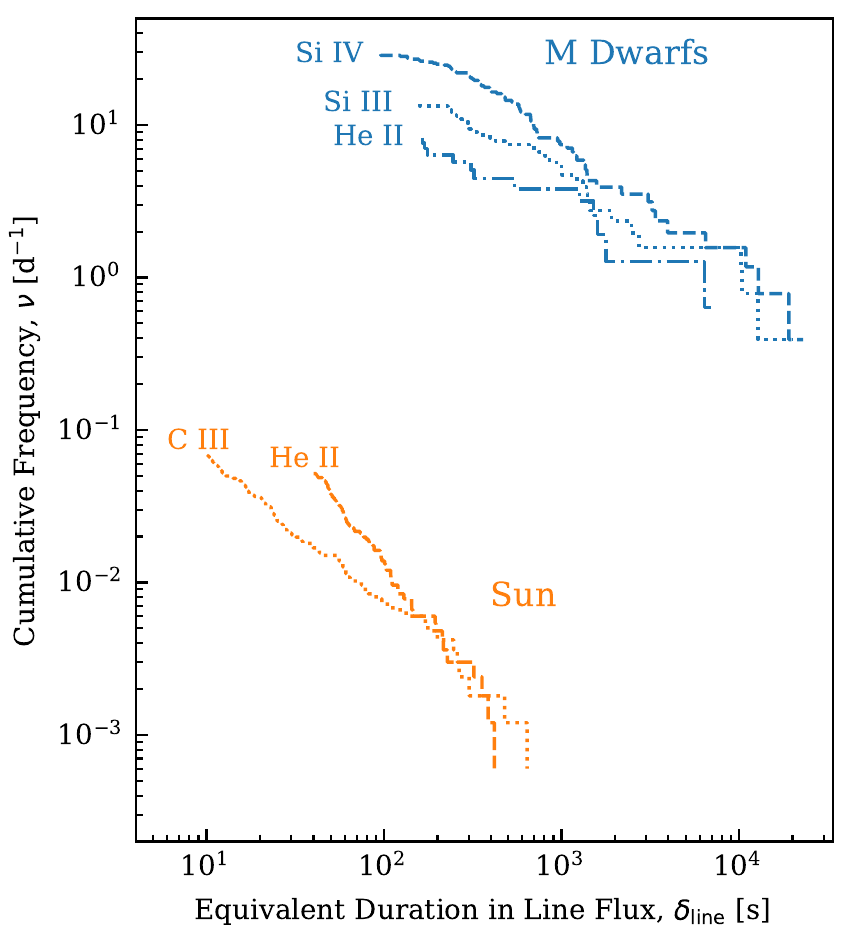}
\caption{Comparison of Solar (orange) and M dwarf (blue) flare rates (aggregating flares from all 10 stars) in transition region lines. 
Line wavelengths and formation temperatures are defined in the text.
Transition-region flares of a given equivalent duration occur on M dwarfs $\sim$3 orders of magnitude more frequently than on the Sun.}
\label{fig:ffd_sun}
\end{figure}

Figure \ref{fig:ffd_sun} shows the resulting cumulative FFDs for the solar \ion{C}{3}~977~\AA\ ($\tform = 10^{4.8}$~K) and \Heii~304~\AA\ ($\tform = 10^{4.9}$~K) lines compared to the cumulative FFDs for the \md\ \Siiii~1206~\AA\ ($\tform = 10^{4.7}$~K), \Siiv~1393,~1402~\AA\ ($\tform = 10^{4.9}$~K), and \Heii~1640~\AA\ ($\tform = 10^{4.9}$~K) lines.
The \Heii~1640~\AA\ data come from the flare stars only.
Flares in these lines occur $\sim$3 orders of magnitude more frequently on M dwarfs than on the Sun for a given flare equivalent duration.

An estimate of the absolute energies of solar flares in FUV lines can be made  under the assumption that the equivalent duration of solar flares (modulo the photospheric contribution) is of the same order of magnitude in FUV and EUV lines of equivalent formation temperatures.
Examining solar FUV data from the \textit{SORCE} spectrograph indicates that the Sun and inactive M dwarfs have comparable quiescent fluxes in transition region FUV lines relative to their bolometric luminosity, while the active M dwarfs have quiescent fluxes roughly an order of magnitude higher.
This implies that for planets receiving similar bolometric fluxes (e.g., habitable-zone planets), those orbiting inactive M dwarfs will experience $\sim$3 orders of magnitude more flare emission from these lines than those orbiting Sun-like stars.
For active M dwarfs, this ratio increases to $\sim$4 orders of magnitude. 

This result is also relevant to the discussion of Section \ref{sec:ffds} regarding the significance of flares to the overall transition-region emission of M dwarfs.
This is decidedly not the case for the Sun.
(We are careful to specify transition-region rather than FUV emission here because of the significant photospheric emission by the Sun at FUV wavelengths.)
Whatever is the underlying link that causes the consistency in equivalent-duration FFDs across M dwarf acitivity levels, it does not operate in the same way or with the same efficiency on the Sun.

It is worth noting, before concluding this subsection, that these types of solar-stellar comparisons would be greatly facilitated by disk-integrated, spectrophotometric FUV data of the Sun. 
Though the EUV is an excellent tool for the study of solar magnetic activity, observations of stellar EUV emission between $\sim$400~--~912~\AA\ are challenged by interstellar absorption and there are no operating observatories that can access the 120~--~400~\AA\ range where direct comparisons to solar observations could be made.
Thus, to strengthen the Sun-star connection, particularly as regards differences in the magnetic processes heating their upper atmospheres, spectrophotometric FUV observations of the Sun should be pursued.
However, it is already clear from the present observations that the flares of the Sun are much less important to both its atmosphere and the atmospheres of its planets than the flares of M dwarfs.

\subsection{An Inconsistent Relationship Between Flare FUV and X-ray Emission}
\label{sec:xrays}
\begin{figure}
\includegraphics{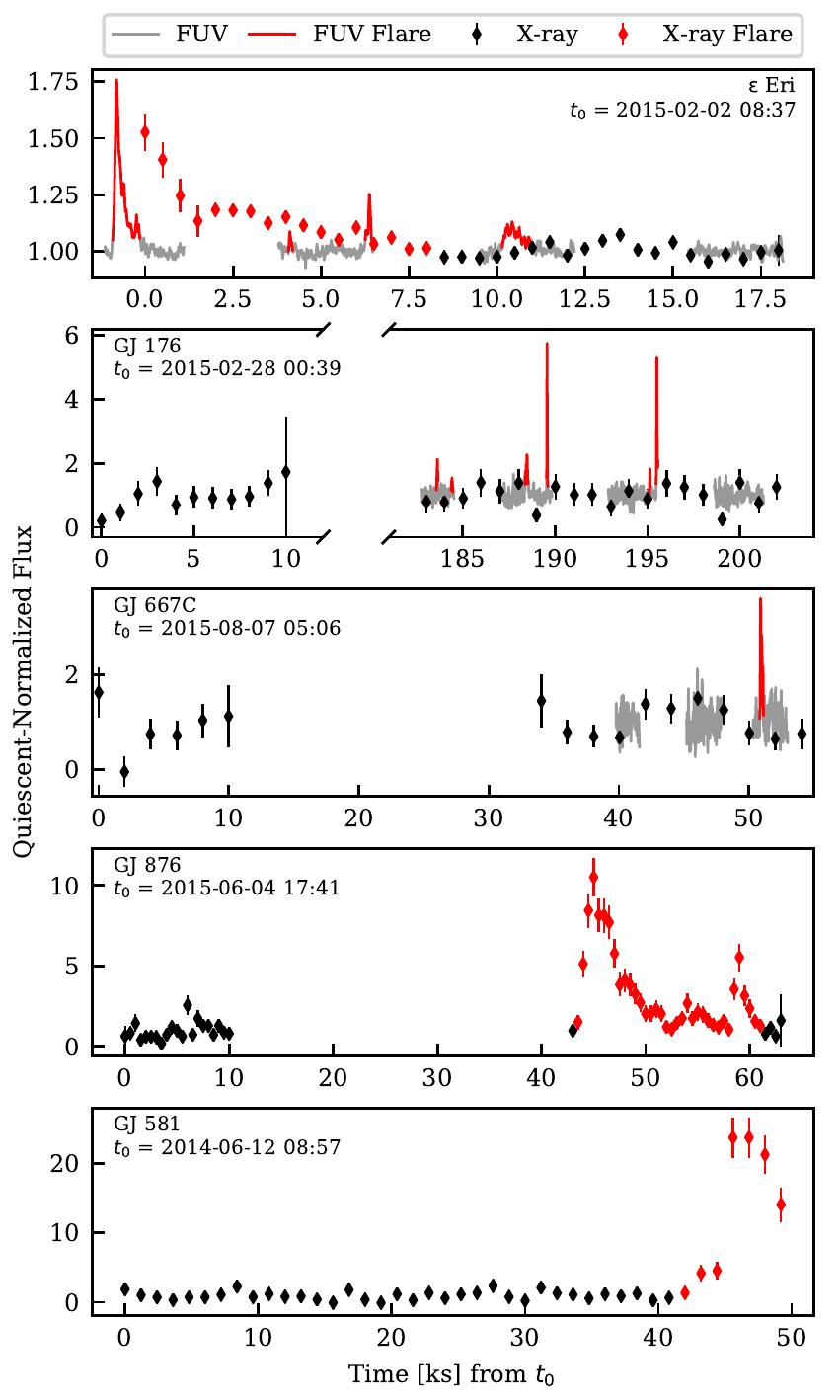}
\caption{X-ray data of those stars with overlapping FUV data or where X-ray flares were observed. Lightcurves are of \gfuv\ emission, except for \epseri\ where a narrower band spanning $\sim$1330~--~1430~\AA\ is used due to necessary differences in observing configurations.
X-ray data were also obtained for GJ~832, GJ~436, and the K dwarfs HD~40307 and HD~85512, but exhibited no flares and did not overlap with any FUV observations.}
\label{fig:xcurves}
\end{figure}

X-ray observations are sometimes the only data available to characterize the variability in an exoplanet host star's high energy radiation (e.g. \citealt{ribas16}).
Therefore, it is important to seek relationships between FUV and X-ray emission from flares.
The MUSCLES program obtained X-ray data for all M dwarf targets (Section \ref{sec:data}), and in Figure \ref{fig:xcurves} we plot lightcurves from these data for the cases where flares occurred or there was overlap with the MUSCLES FUV observations.
Although our analysis focused on \mds, we include the K dwarf \epseri\ in Figure \ref{fig:xcurves} because it is the only MUSCLES target for which the same flare appeared in both the FUV and X-ray data.
The X-ray observations captured only the declining phase of the flare, but they imply an equivalent duration at least several times as large as observed in FUV emission. 

\begin{deluxetable*}{rrrrl}
\tablewidth{0pt}
\tabletypesize{\scriptsize}
\tablecaption{X-ray flares \label{tbl:xflares}}
\tablehead{ \colhead{$\delta$} & \colhead{$t_\mathrm{peak}$} & \colhead{$\frac{F_{\mathrm{peak}}}{F_{q}}$} & \colhead{FWHM} & \colhead{Star} \\ \colhead{s} & \colhead{MJD} & \colhead{} & \colhead{s} & \colhead{} }
\startdata
$ 102904 \pm 7021 $ & 56820.9147 & $ 22.7 \pm 3.6 $ & 4910.87 & GJ 581\\
$ 41892 \pm 2027 $\tablenotemark{a} & 57178.2578 & $ 9.5 \pm 1.3 $ & 2914.99 & GJ 876\\
$ 1302 \pm 87 $ & 57055.3596 & $ 0.526 \pm 0.082 $ & 1194.94 & $\mathrm{\epsilon}$ Eri\\
\enddata

\tablenotetext{a}{Could be interpreted as three separate events.}

\tablecomments{The \epseri\ and GJ~581 flares are truncated by an exposure beginning and end near their peak flux, strongly affecting measurements of the flare properties.}
\end{deluxetable*}

We identified 2 additional flares in the \md\  X-ray data, cataloged in Table \ref{tbl:xflares}.
The flares on GJ~876 and GJ~581 have equivalent durations at least an order of magnitude larger than the most energetic FUV flares likely to have occurred during the duration of the X-ray observations.
Such a relationship between the X-ray and FUV equivalent durations, however, is inconsistent with X-ray data overlapping with the FUV flares of GJ~176 and GJ~667C.
If the scaling held, then the FUV flares of these two stars would have produced enhancements several times in excess of the scatter in the quiescent X-ray lightcurve.
However, X-ray flares are not observed in these data, and re-reducing the X-ray data to produce lightcurves with finer time sampling during these flares did not reveal any hidden flux enhancements. 
This implies  the flare energy budget between FUV and X-ray bands is not consistent.
If so, different energy dissipation mechanisms could be operating in different events, particularly in the brief FUV flares observed on GJ~176 and GJ~667C versus the more extended flare on \epseri\ and the extended, highly energetic flares on GJ~876 and GJ~581.

We posit that the differences could be related to the size of the magnetic structures in which reconnection occurs. 
This model would then predict a relationship between the relative emission of a flare at X-ray versus FUV wavelengths and the energy of a flare.
The premise is that more energetic flares are likely to result from larger magnetic structures that in turn could inject a greater fraction of their energy into the corona.
Low energy flares from small magnetic structures might deposit greater fractions of their energy into the stellar transition region and below.

This is not the first time  FUV-only events have been observed.
\cite{ayres15b} noted such events in observations of EK Draconis, a G dwarf, as well as a single precedent in X-ray, FUV, and NUV \textit{XMM-Netwon} observations of M stars by \cite{mitra05}.
Regardless of the physical explanation for how such isolated events come about, an important implication is that observations in X-ray bands cannot be relied upon to constrain the UV flaring of M dwarfs.

\subsection{Emission Line Profiles}
\label{sec:profiles}
\begin{figure*}
\includegraphics{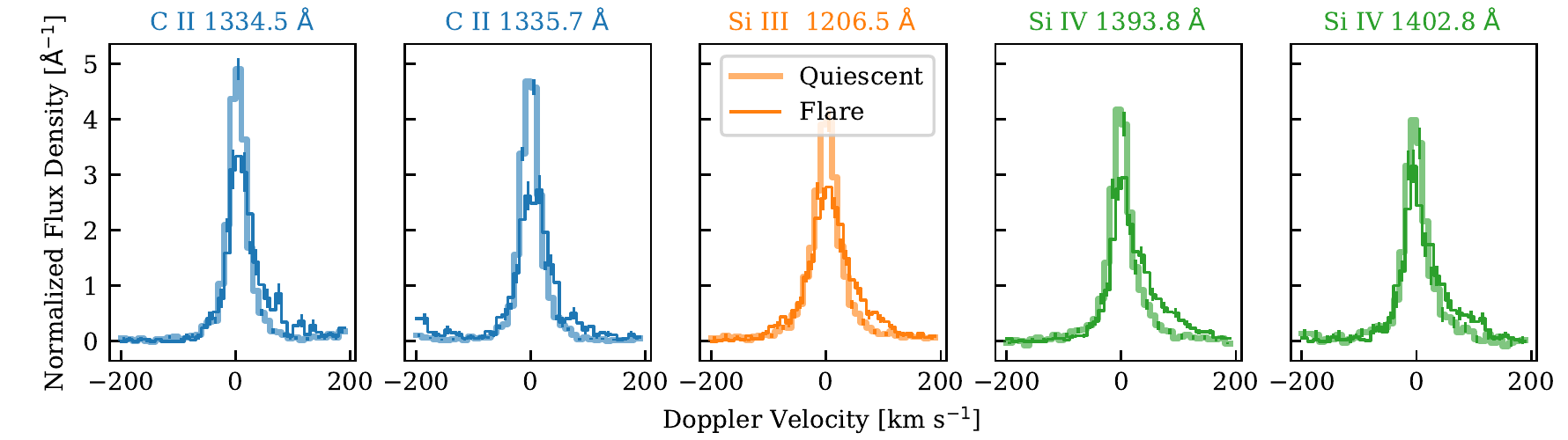}
\caption{Profiles of the five strongest transition-region emission lines of GJ~832 during its most energetic flare, compared to the quiescent profiles from the same epoch of data.
The profiles are normalized such that their wavelength integral over the plotted range is unity.
Note the excess of redshifted emission during the flare, particularly in the \Siiv\ lines.
The associated flare lightcurve is plotted in Figure \ref{fig:line_kin_curves}.}
\label{fig:line_kin_profs}
\end{figure*}

\begin{figure}
\includegraphics{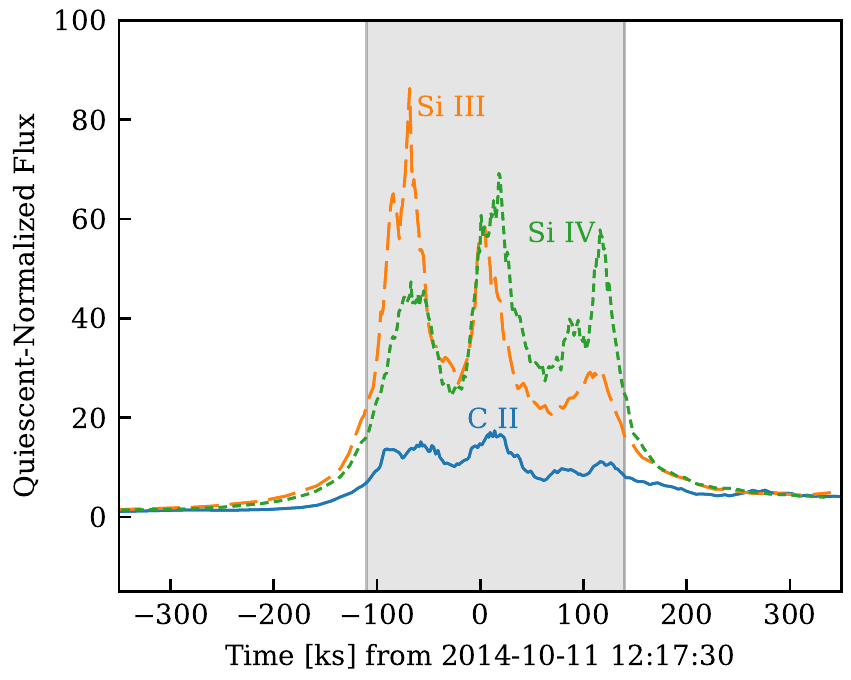}
\caption{Count-binned lightcurves of the most energetic flare on GJ~832.
The shaded region was used to produce the line profiles in Figure \ref{fig:line_kin_profs}.}
\label{fig:line_kin_curves}
\end{figure}

During flares, the changing profiles of emission lines can reveal heating, mass motions, and strong electric or magnetic fields associated with the events.
Figure \ref{fig:line_kin_profs} shows an example of emission line profiles during the peak of one of the most energetic flares  we observed compared with quiescence, characteristic of all the observations of the most energetic flares in the sample. 
The corresponding line lightcurves are shown in Figure \ref{fig:line_kin_curves}. 
The line profiles exhibit significant redshifted emission during the flare extending out to roughly 100~km~\pers.
Redshifted emission is common in \md\ flares.
\cite{hawley03} have previously cataloged the appearance of redshifted emission in AD~Leo data, finding flux-weighted line centers shifting by 30~--~40 km \pers.
\cite{redfield02} found redshifts out to 200 km \pers\ in \Ciii\ and \Ovi\ emission during AU~Mic flares in FUSE data.
Redshifted emission indicates a downflow of material toward the stellar surface that could be a result of ``chromospheric condensation'' like that seen in solar flares \citep{hawley03}.

To summarize Section \ref{sec:lineflares}, the spectral dimension of FUV flare observations from \textit{HST} reveals significant differences in the emission from various regions of the stellar atmosphere during flare events.
The strongest lines in the strongest flares consistently show a slight excess of redshifted emission extending to as much as 100~km~\pers.
These details contain information on the mechanisms and locations of energy deposition in the stellar atmosphere and are a promising avenue for constraining the stellar flare models.
Detailed modeling is beyond the scope of this work; however, the observed trends in FUV line emission and inconsistency in X-ray emission lead us to suggest the relative distribution of emission between FUV lines as well as X-ray emission during flares is correlated with the size of magnetic structures in which reconnection has occurred, a prediction that could be tested with future models and data. 

\section{Stellar Properties and Stellar Flares}
\label{sec:startrends}

\begin{figure*}
\includegraphics{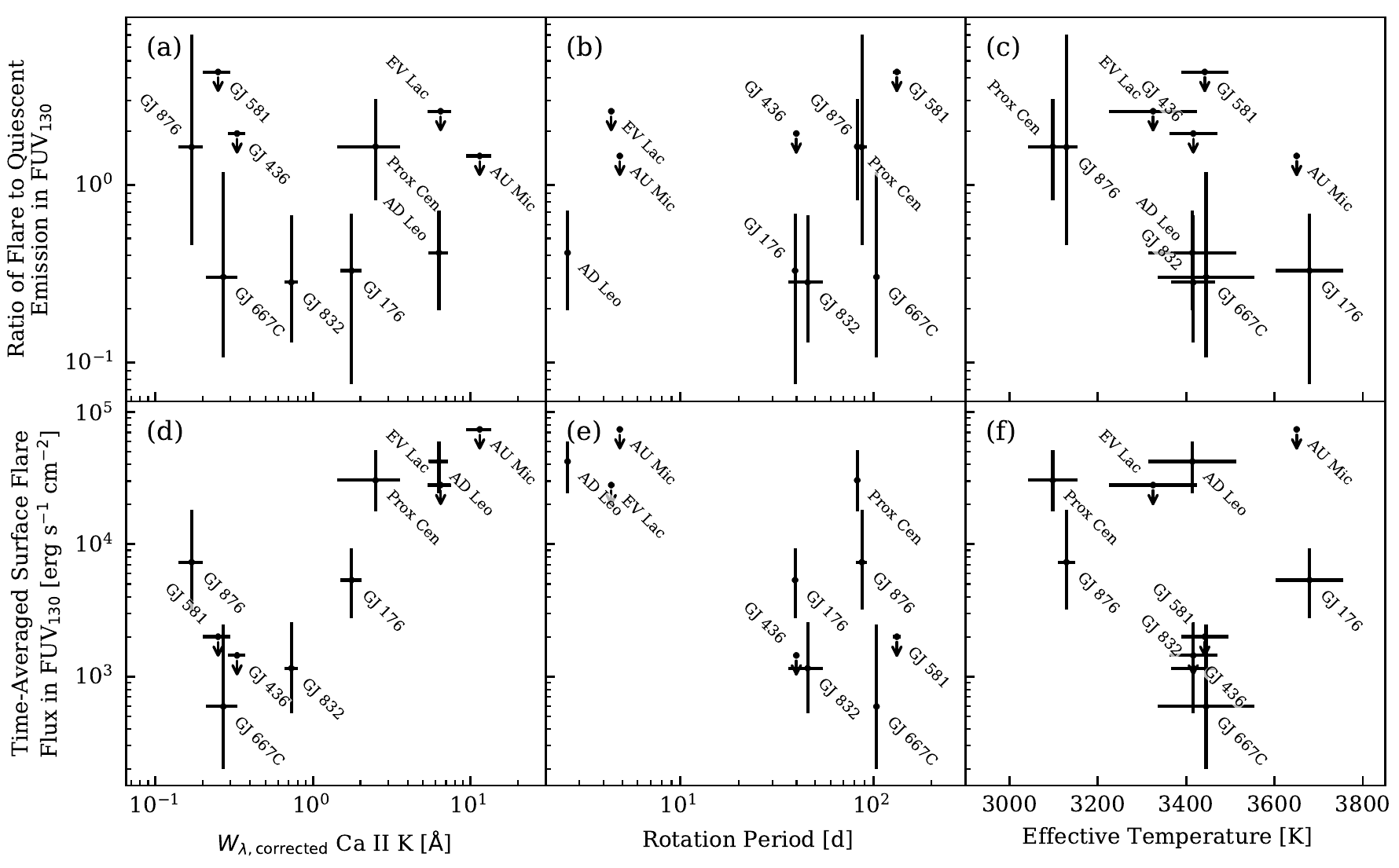}
\caption{The flare activity of each star, using quiescence-normalized (top) and absolute (bottom) metrics, as a function of chromospheric \Caii~K line emission, rotation period, and effective temperature.
The apparent correlation between the absolute metric and \Caii~K emission is suggestive and agrees with past results but is not statistically significant in this case.}
\label{fig:x_vs_flariness}
\end{figure*}

The constraints on the flare rates of individual stars can be used to explore relationships between stellar properties and flare activity in finer detail than simply comparing the active and inactive groups.
As with the FFDs, we quantify flare activity with both absolute and relative metrics for this purpose: the ``flare surface flux,'' \Fsfc, and the ratio of time-averaged emission by flares to emission by quiescence, \EfEq.
These quantities are defined in Section \ref{sec:metrics} and provided for each target in Tables \ref{tbl:pew_stats} and \ref{tbl:energy_stats}.
We compared these metrics with \Caii~K corrected equivalent widths, rotation period, and effective temperature (Table \ref{tbl:starprops}) and plot the comparison in Figure \ref{fig:x_vs_flariness}.
The \Caii~K equivalent widths are corrected for differences in the baseline continuum for stars of differing effective temperature.

Uncertainties in flare rates are roughly an order of magnitude in most cases because of the small number of flares detected.
This translates to similar uncertainties for \Fsfc\ and \EfEq.
Where no flares were detected, we can only place upper limits on the flare activity metrics. 
For a trend to be detectable through this noise, it would need to yield greater than an order of magnitude variation in these metrics.
Such a variation is not observed in \EfEq, in accordance with the similarity of the equivalent-duration FFDs for the active and inactive stars (Section \ref{sec:absrel}).

In \Fsfc\ there is the suggestion of a trend with \Caii~K equivalent widths (panel (d) of Figure \ref{fig:x_vs_flariness}).
However, when the data are allowed to vary within their uncertainties, the $p$-values on a Spearman rank-order correlation test are insignificant.
Nonetheless, a trend would be consistent with the separation of the inactive and active sample FFDs in absolute energy (Section \ref{sec:absrel}).
It is also consistent with the results of \cite{hilton11} and \cite{hawley14}.
Both groups found higher rates of flaring for stars classified as active versus inactive based on H$_\alpha$ emission. 
Therefore, we consider it likely that \Caii~K flux does in fact correlate with \Fsfc, but these data are too noisy for the trend to be clear. 
No other trends are apparent.

Previous surveys using \textit{SDSS} and \textit{2MASS} have found trends between \md\ flare activity and stellar subtype, with flare activity increasing toward later types \citep{kowalski09,hilton10}. 
\cite{audard00} studied flares on a sample of F~--~M stars with \textit{EUVE} data and found a strong relationship of the rate of $>10^{32}$ erg flares with rotation period, projected rotational velocity, Rossby number, and X-ray luminosity. 
Along these lines, it is notable in this study that the relative flare activity of the two stars in the sample most likely to be fully convective based on their effective temperature, Prox~Cen and GJ~876, exhibit the highest relative rate of flaring.
To attempt to uncover trends in FUV flare activity with stellar properties, staring observations at FUV wavelengths employing minute-timescale cadences for roughly a dozen or more targets are needed.

\section{Flare Lightcurves and Energy Budgets}
\label{sec:shapes}

\subsection{Flare Lightcurves}
The observed flares exhibited a diversity of lightcurve profiles, with some appearing symmetric and impulsive, others exhibiting a classic impulse-decay, and other exhibiting sustained emission followed by decay.
Examples of similar diversity can be found in Figure 13 of  \cite{loyd14} and is what motivated our shape-agnostic identification algorithm.
For the flares of this work, rise times  and FWHMs are all typically around tens of seconds and correlate with total event energy (Table \ref{tbl:flares}).
Decay times, as we have defined them, are generally hundreds of seconds for the more energetic flares. 
The most energetic events sometimes exhibit elevated flux before an impulsive increase that yielded longer rise times.
Often the flux increase from these events lasted until the exposure was ended due to Earth occulting the target (10s of minutes).

\begin{figure}
\includegraphics{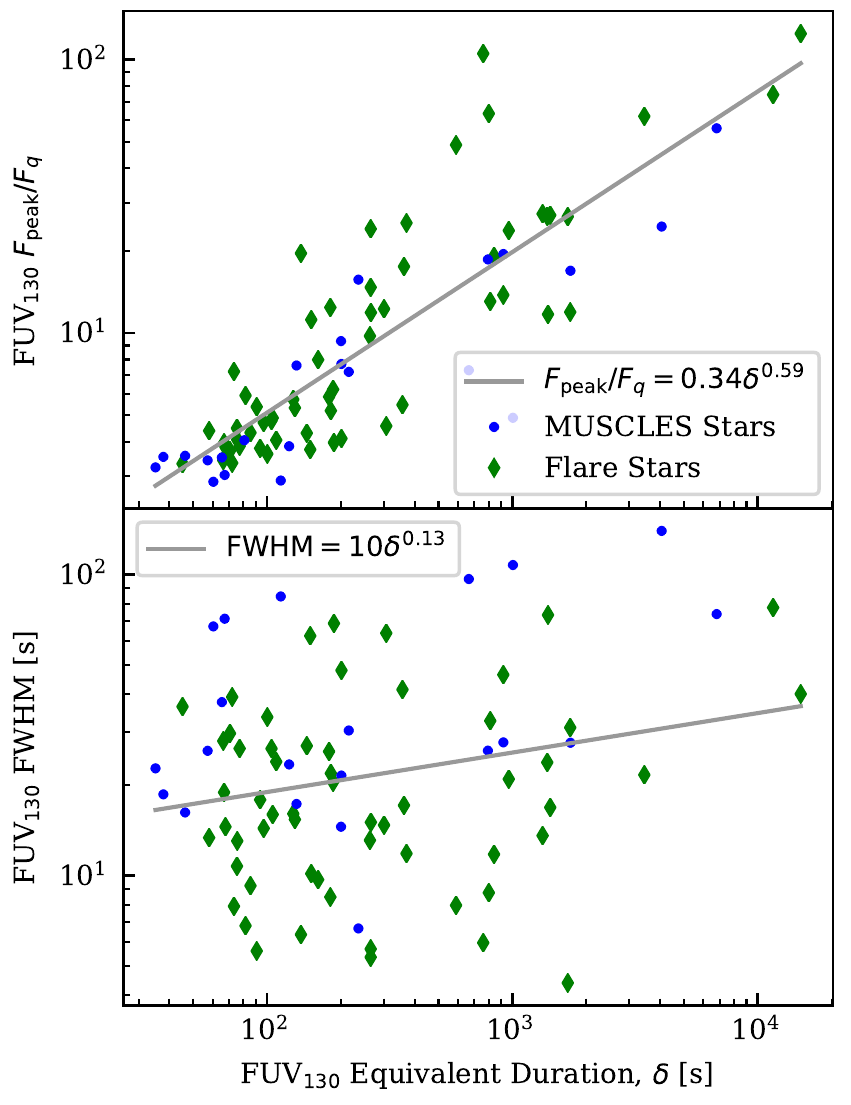}
\caption{Trends in the time profile of flares, as characterized by their ratio of peak to quiescent flux and FWHM (taken as the cumulative time that flux was above half peak to accommodate multipeaked flares). At lower $\delta$, the flare peak is less well-resolved, introducing a bias. 
The bias implies  the peak ratio trend could be more gradual and the FWHM trend steeper than those shown in the figure.}
\label{fig:shape_v_pew}
\end{figure}

We explore trends in flare lightcurves in Figure \ref{fig:shape_v_pew}.
The figure demonstrates that, for a given equivalent duration, the FWHM of the flares varies considerably, by roughly an order of magnitude.
We did not estimate uncertainty in the FWHM measurements, but the lower variability in peak fluxes and the consistency of this variability across three orders of magnitude of equivalent duration support the validity of this diversity.
The plot also suggests  the  active-star flares are very slightly more impulsive (lower FWHM, larger peak/quiescent ratio) than the MUSCLES stars.

Figure \ref{fig:shape_v_pew} depicts a clear trend in the peak flux ratio with equivalent duration and only a comparatively weak trend in FWHM.
Both measurements are biased at small $\delta$ by the difficulty in resolving the flare peak due to lower count rates in less energetic flares.
This would tend to artificially lower peak/quiescent ratio at low $\delta$.
Peak flux measurements biased to lower values would bias FWHM measurements to higher values (since the half-max level is underestimated).
Hence, the true flux ratio vs. $\delta$ trend is unlikely to be any steeper than that observed and the true FWHM vs $\delta$ trend is unlikely to be any shallower than that observed.
These trends provide the basis of an idealized lightcurve we present in a later subsection.

\subsection{Spectral Energy Budget}
\label{sec:budget}
In Table \ref{tbl:linecorrs}, we present empirical relationships of flare energy, equivalent duration, and the ratio of peak to quiescent flux as measured in a variety of bands versus the values measured in \Siiv.
We select \Siiv\ because equivalent durations are typically the largest in this line (Figure \ref{fig:line_FFDs}).
Further, although for most of this work we focus on the \gfuv\ band, this band is specific to the COS G130M spectrograph and so cannot be easily related to other datasets, whereas \Siiv\ can.
We include continuum and ``interline'' bands in these fits.
The continuum bands are carefully selected narrow bands free of lines in the highest S/N spectrum available, whereas ``interline'' regions merely avoid the major emission lines and include a mixture of continuum and weak or unresolved emission lines.

\begin{figure*}
\includegraphics{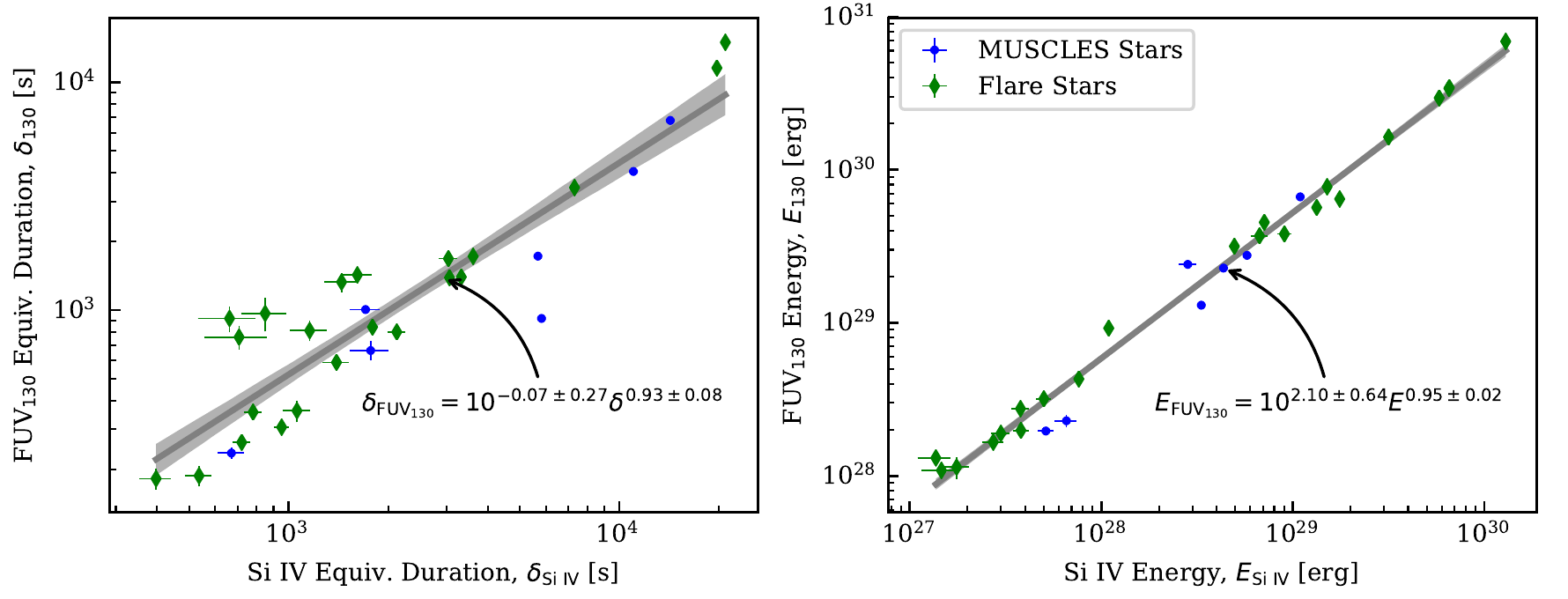}
\caption{Relationship between flare  equivalent duration and energies in \gfuv\ versus \Siiv\ emission, provided as an example of the empirical relationships computed between all major emission sources and \Siiv\ (Table \ref{tbl:linecorrs}).}
\label{fig:energycorr}
\end{figure*}

The energies and equivalent durations of the flares as measured in various bands are generally consistent (or nearly so) with being linearly related.
An example is shown in Figure \ref{fig:energycorr}, the relationship of absolute energy emitted in \gfuv\ to \Siiv. 
This suggests that the emission processes do not change appreciably over the range of flare energies observed.
However, it is possible that much more energetic flares might initiate a much different pattern of emission, deviating from these relationships.
For example, above a certain energy level flares might regularly eject emitting plasma from the higher temperature lines or produce continuum emission in \gfuv\ that would dominate over lines.

\begin{deluxetable*}{lrrrrrrrrrrrr}
\tablewidth{0pt}
\tabletypesize{\scriptsize}
\rotate
\tablecaption{Empirical fits relating flare energy, equivalent duration, and ratio of peak to quiescent flux in various bands to that of \Siiv.  \label{tbl:linecorrs}}

\tablehead{ \colhead{Band} & \multicolumn{4}{c}{Equivalent Duration, $\delta$} & \multicolumn{4}{c}{Energy, $E$} & \multicolumn{4}{c}{Peak/Quiescent Flux, $F_\mathrm{peak}/F_q$} \\ \colhead{} & \colhead{$m$} & \colhead{$b$} & \colhead{$\rho$} & \colhead{$\sigma_{O-C}$} & \colhead{$m$} & \colhead{$b$} & \colhead{$\rho$} & \colhead{$\sigma_{O-C}$} & \colhead{$m$} & \colhead{$b$} & \colhead{$\rho$} & \colhead{$\sigma_{O-C}$} }
\startdata
FUV$_{130}$ & $ 0.929 \pm 0.082 $ & $ -0.07 \pm 0.27 $ & -0.99002 & 0.19 & $ 0.952 \pm 0.023 $ & $ 2.10 \pm 0.64 $ & -0.99953 & 0.10 & $ 0.673 \pm 0.076 $ & $ 0.32 \pm 0.12 $ & -0.93911 & 0.20\\
G130M Interline\tablenotemark{a} & $ 0.99 \pm 0.12 $ & $ -0.2 \pm 0.4 $ & -0.99017 & 0.22 & $ 0.990 \pm 0.035 $ & $ 1 \pm 1 $ & -0.99959 & 0.14 & $ 0.67 \pm 0.15 $ & $ 0.36 \pm 0.23 $ & -0.94080 & 0.29\\
E140M Interline\tablenotemark{a} & $ 1.1 \pm 0.1 $ & $ -0.52 \pm 0.34 $ & -0.99038 & 0.19 & $ 0.999 \pm 0.033 $ & $ 0.50 \pm 0.95 $ & -0.99948 & 0.12 & $ 0.866 \pm 0.063 $ & $ 0.125 \pm 0.098 $ & -0.94021 & 0.13\\
G130M Continuum\tablenotemark{b} & $ 0.89 \pm 0.12 $ & $ 0.39 \pm 0.43 $ & -0.99145 & 0.19 & $ 1.016 \pm 0.077 $ & $ -1.0 \pm 2.2 $ & -0.99976 & 0.18 & $ 0.81 \pm 0.15 $ & $ 0.22 \pm 0.26 $ & -0.94490 & 0.28\\
Ly$\mathrm{\alpha}$ Wings & $ 0.66 \pm 0.11 $ & $ -0.3 \pm 0.4 $ & -0.99115 & 0.16 & $ 0.662 \pm 0.054 $ & $ 9.9 \pm 1.6 $ & -0.99944 & 0.16 & $ 0.030 \pm 0.016 $ & $ 0.374 \pm 0.024 $ & -0.93447 & 0.04\\
C I 1329 \AA & \nodata & \nodata & \nodata & \nodata & \nodata & \nodata & \nodata & \nodata & $ 0.33 \pm 0.15 $ & $ 0.29 \pm 0.27 $ & -0.97058 & 0.11\\
C I 1657 \AA & $ 0.91 \pm 0.15 $ & $ -0.31 \pm 0.53 $ & -0.98966 & 0.17 & $ 0.830 \pm 0.095 $ & $ 4.5 \pm 2.8 $ & -0.99976 & 0.14 & $ 0.376 \pm 0.067 $ & $ 0.1 \pm 0.1 $ & -0.92932 & 0.11\\
O I 1305 \AA & $ 0.87 \pm 0.12 $ & $ -0.1 \pm 0.4 $ & -0.98901 & 0.18 & $ 0.81 \pm 0.06 $ & $ 4.8 \pm 1.8 $ & -0.99969 & 0.13 & $ 0.473 \pm 0.065 $ & $ 0.0 \pm 0.1 $ & -0.92775 & 0.12\\
Si II 1265 \AA & $ 0.87 \pm 0.19 $ & $ 0.00 \pm 0.71 $ & -0.99614 & 0.18 & $ 0.923 \pm 0.054 $ & $ 1.0 \pm 1.6 $ & -0.99977 & 0.09 & $ 0.65 \pm 0.11 $ & $ -0.17 \pm 0.18 $ & -0.95841 & 0.16\\
C II 1335 \AA & $ 1.004 \pm 0.063 $ & $ -0.51 \pm 0.21 $ & -0.98970 & 0.18 & $ 0.921 \pm 0.028 $ & $ 2.1 \pm 0.8 $ & -0.99951 & 0.18 & $ 0.665 \pm 0.061 $ & $ -0.039 \pm 0.094 $ & -0.93911 & 0.16\\
He II 1640 \AA & $ 1.0 \pm 0.1 $ & $ -0.46 \pm 0.35 $ & -0.98849 & 0.18 & $ 1.009 \pm 0.071 $ & $ -0.3 \pm 2.1 $ & -0.99967 & 0.17 & $ 0.533 \pm 0.088 $ & $ 0.03 \pm 0.14 $ & -0.92775 & 0.16\\
C III 1175 \AA & $ 1.046 \pm 0.057 $ & $ -0.3 \pm 0.2 $ & -0.98962 & 0.12 & $ 0.974 \pm 0.038 $ & $ 0.8 \pm 1.1 $ & -0.99971 & 0.11 & $ 0.92 \pm 0.08 $ & $ -0.15 \pm 0.13 $ & -0.93827 & 0.19\\
Si III 1206 \AA & $ 1.031 \pm 0.072 $ & $ -0.21 \pm 0.24 $ & -0.99019 & 0.16 & $ 0.95 \pm 0.03 $ & $ 1.31 \pm 0.87 $ & -0.99958 & 0.12 & $ 0.9 \pm 0.1 $ & $ -0.12 \pm 0.16 $ & -0.94390 & 0.26\\
C IV 1549 \AA & $ 1.051 \pm 0.072 $ & $ -0.45 \pm 0.24 $ & -0.99041 & 0.14 & $ 0.995 \pm 0.027 $ & $ 0.62 \pm 0.77 $ & -0.99942 & 0.11 & $ 0.667 \pm 0.066 $ & $ 0.111 \pm 0.098 $ & -0.93447 & 0.15\\
N V 1240 \AA & $ 0.88 \pm 0.11 $ & $ -0.16 \pm 0.39 $ & -0.98935 & 0.25 & $ 0.875 \pm 0.061 $ & $ 3.2 \pm 1.8 $ & -0.99961 & 0.22 & $ 0.307 \pm 0.072 $ & $ 0.21 \pm 0.11 $ & -0.94241 & 0.17\\
Fe XII 1242,1349 \AA & \nodata & \nodata & \nodata & \nodata & \nodata & \nodata & \nodata & \nodata & $ 0.178 \pm 0.099 $ & $ 0.25 \pm 0.18 $ & -0.95207 & 0.11\\
Fe XXI 1354 \AA & $ 1.1 \pm 1.1 $ & $ -0.8 \pm 4.3 $ & -0.99574 & 0.39 & $ 0.63 \pm 0.16 $ & $ 9.4 \pm 4.6 $ & -0.99966 & 0.12 & $ 0.57 \pm 0.51 $ & $ -0.12 \pm 0.94 $ & -0.96674 & 0.34\\
\enddata

\tablenotetext{a}{Includes all flux between major emission lines. This flux likely includes unresolved lines and thus might not be considered a true continuum.}
\tablenotetext{b}{Includes only regions where no emission lines were found in the highest S/N spectrum available in the dataset.}

\tablecomments{All fits are of the form $\log y_\mathrm{band} = m \log y_\mathrm{Si IV} + b$, where $y$ stands for the measured quantity ($\delta, E$, or $F_\mathrm{peak}/F_q$). The correlation between the fit parameters is given in the $\rho$ columns and the scatter of points about the best-fit line in log space is given in the $\sigma_{O-C}$ (standard deviation of the observed -- computed values) columns. We recommend that $\sigma_{O-C}$ be used to estimate uncertainty in predictions made using the fits because the scatter in the relationships is likely physical. Fits were restricted to points with S/N $>$ 3 to avoid highly non-normal uncertainties when transforming to log space. When this yielded fewer than 5 points, no fit was attempted.}

\end{deluxetable*}

\begin{figure*}
\includegraphics{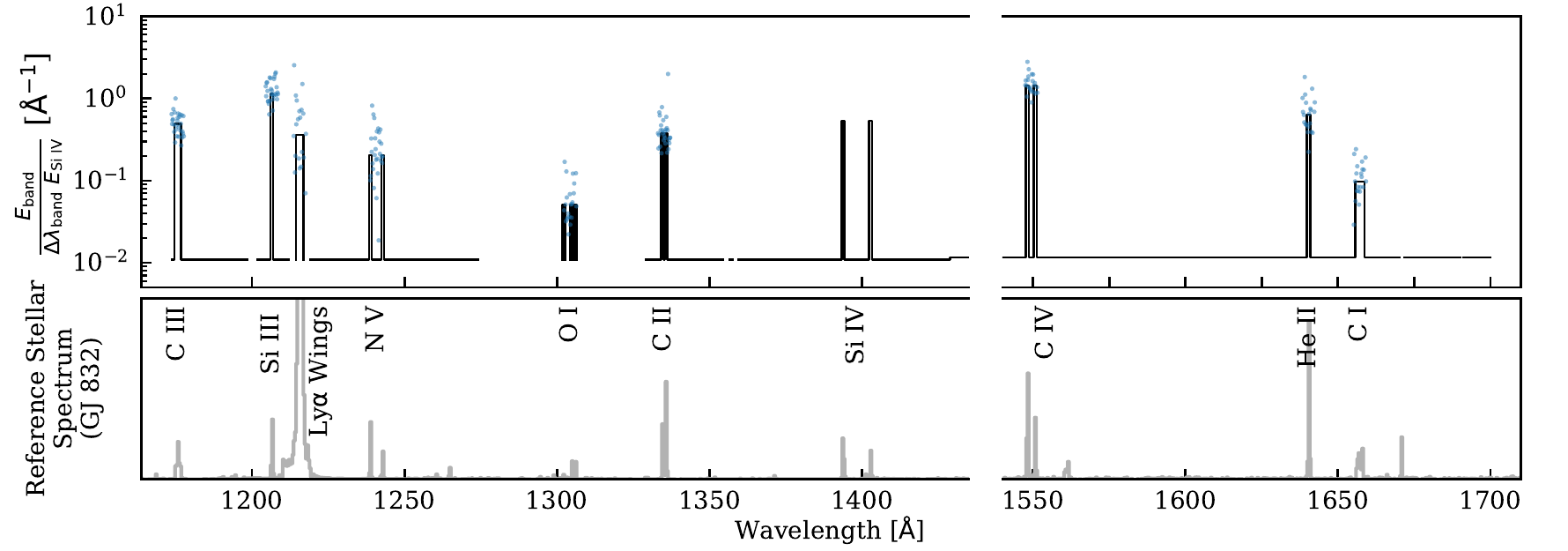}
\caption{Flare spectral energy budget. 
The lower panel shows the location of major emission lines in an M dwarf spectrum for reference (linear scale). 
In the upper panel, the scattered, translucent blue points represent the energies of single flares in the associated lines relative to \Siiv\ emission, divided by the line width.
The solid line traces the median value of these energy ratios.
Wide lines result from overlapping multiplets.
Gaps are due to airglow-contaminated regions and wavelengths not consistently covered by the COS or STIS spectrographs.}
\label{fig:budget}
\end{figure*}

The near-linear relationships between different FUV emission sources are convenient because they allow for  a ``typical'' flare energy spectrum to be defined that is valid across the range of observed flare energies. 
Such a spectrum, created by taking the median ratios of the energy in the major emission lines and the interline regions over the energy in \Siiv, is depicted graphically in Figure \ref{fig:budget}. 
Because flux is low in the interline regions, we combine flux from all the interline areas in two regions, one covering the COS~G130M band, \fuvlo~--~\interlineDivide, and another covering the remainder of the STIS~E140M band, \interlineDivide~--~\fuvhi.
Figure \ref{fig:budget} plots the energy budget as a spectral density (dividing by the integration bandpasses) for a more intuitive comparison to flux density spectra.
Thus, we use units of \perAA.
We culled data with S/N $<$ \statsSNcut. 
Above \interlineDivide\ all data is from only the active stars.
The solid line traces the median values.

The previous two subsections provide the basis for developing a standardized UV flare model for use in modeling flare impacts on planetary atmospheres.
The remainder of this paper presents such a model and the results from using it to asses the potential implications of \md\ UV flares on planetary atmospheres.

\section{A Fiducial UV Flare for Modeling Planetary Atmospheres}
\label{sec:fiducialflare}
Because stellar UV flares likely have a significant impact on atmospheric chemistry and mass loss on orbiting planets, we expect a great deal of modeling of these impacts in ensuing years.
Indeed, some modeling has already been done based on a single benchmark flare from AD~Leo that finds the compound effect of repeated flares can produce secular changes in atmospheric composition (\citealt{tilley18}; we direct the reader back to the introduction for more discussion).
As such, we created a fiducial UV flare for use in modeling to share with the community.
Although no such fiducial flare will perfectly represent a true stellar flare, establishing consistency across models in this early stage in the study of stellar flare impacts on planetary atmospheres could make future model comparisons more straightforward.

To simplify the implementation of the fiducial UV flare, we developed a short Python module that can be used to generate synthetic UV flares based on this template, available online.\footnote{\url{\fidurl}}
A text table of the spectral energy budget the code uses is included separately in the online repository so that this information can be accessed independent of Python. 
In addition to providing time-evolving spectra of the fiducial flare, this code can generate simulated series of flares based on power-law FFDs. 
This should allow modelers to realistically simulate the cumulative effect of FUV flares on atmospheric photochemistry.

\subsection{Observed and Unobserved FUV}
\begin{figure*}
\includegraphics{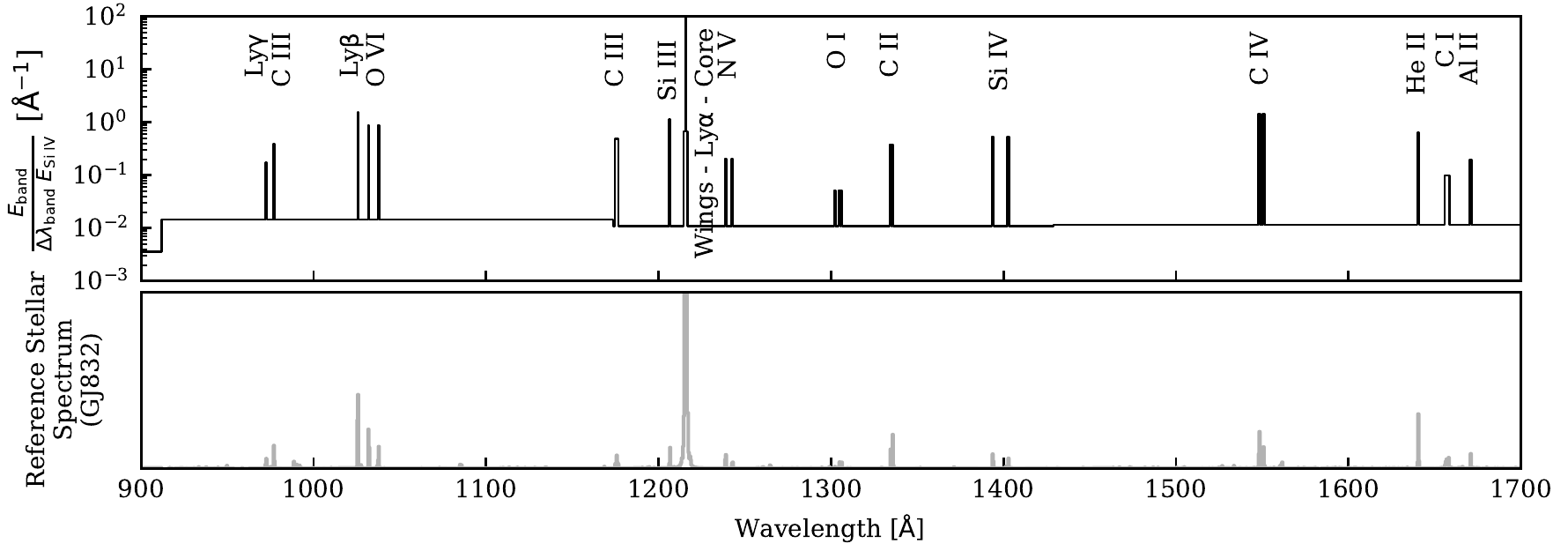}
\caption{Spectral energy budget of the fiducial flare (for photochemical modeling) over FUV wavelengths. 
The lower panel shows the location of major emission lines in an M dwarf spectrum for reference (linear scale).}
\label{fig:fiducialFUV}
\end{figure*}

We base the fiducial flare spectrum on the FUV energy budget in Figure \ref{fig:budget}.
Because this energy budget does not extend over all photochemically-relevant wavelengths, we extended it by  using lines of similar formation temperature as proxies for those not observed.
These unobserved lines are the \lya\ core, \lyb, \lyg, and \Mgii\ 2796, 2803 \AA\ (proxy \Oi~1305~\AA\ multiplet for all four preceding lines), \Alii\ 1671 \AA\ (proxy \Cii~1334, 1335~\AA), \Ciii~977~\AA\ (proxy \Ciii~1175~\AA\ multiplet), and \Ovi~1031, 1037~\AA\ (proxy \Nv~1238, 1242~\AA).
For wavelengths shortward of $\sim$\interlineLoLim, we use the interline (continuum + weak lines) COS~G130M flux as a proxy.
To compute the energy contribution of the unobserved lines, we assumed they had the same equivalent duration during flares as the proxy line, then adjusted according to the ratio of fluxes of the unobserved and proxy lines.
To compute these ratios, we used archival FUSE and STIS E140M data for AD~Leo (\Ciii, \lyg, \lyb, \Ovi, \lya\ core, and short wavelength interline regions), quiescent COS G130M and G160M data for GJ~832 (\Alii), and the panchromatic SED from the MUSCLES spectral atlas for Prox~Cen (\Mgii). 
Figure \ref{fig:fiducialFUV}  shows the spectral energy budget of the FUV portion of the fiducial flare. 

\subsection{Blackbody}
Stellar flares are known to be accompanied by a continuum source that closely resembles the spectrum of an A star \citep{kowalski13}, though it is generally treated as a blackbody.
The multiwavelength observations of AD~Leo flares by \cite{hawley03} provide estimates of the blackbody temperature and relative energy of this emission source in comparison to FUV lines. 
Based on these results, we add a \bbT\ blackbody with a bolometric energy relative to \Siiv\ of \bbratio\ to the fiducial flare.
(Note  the \textit{HST} STIS data used in that work are included in the datasets analyzed here.)
The added blackbody would contribute a few percent of the flux shortward of 1700~\AA; however, it is only included at wavelengths longward of 1700~\AA\ where there was no \textit{HST} data.

Observations at UV and optical wavelengths have shown variation in the blackbody temperature both between flares and over time during the same flare \citep{kowalski13,kowalski16}.
Notably, \cite{kowalski13} estimated blackbody color temperatures from 9000~--~14000~K at the peak of flares and 5000~--~9000 K during decay, though after correcting for absorption features in the spectrum the range of peak temperatures drops to 7700~--~9400~K.

\begin{figure}
\includegraphics{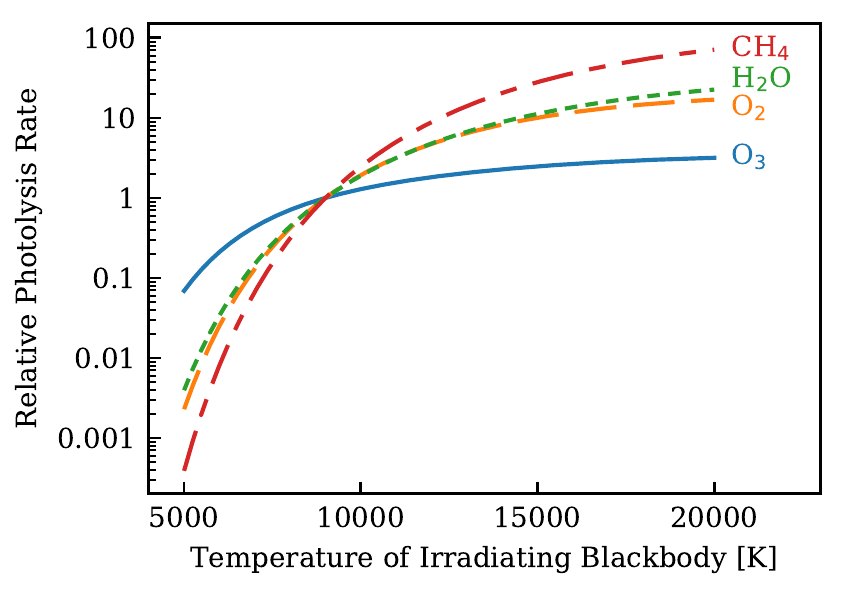}
\caption{The photolysis rate of several molecules exposed to blackbody radiation of constant bolometric power but varying temperature, relative to the value at 9000 K.
For molecules with photolysis cross sections that peak at FUV wavelengths, the variation in photolysis rate is dramatic in comparison with \OIII, which has a broad peak in photolysis cross section in the NUV.}
\label{fig:bb_v_diss}
\end{figure}

The blackbody included in the fiducial flare accounts for the bulk of the flux, and therefore is important to accurate modeling of photodissociation.
The relationship between the photolysis rate of various molecules directly exposed to blackbody emission of the same bolometric power but varying temperature is plotted in Figure \ref{fig:bb_v_diss}.
For \OIII, photolysis of directly exposed molecules varies by about a factor of two from 7000 to 9000 K.
Molecules with photolysis cross sections sharply peaking at FUV wavelengths are much more sensitive to changes in blackbody temperature, varying by an order of magnitude from 7000 K to 9000 K.
Nonetheless, for this work we chose to use a fixed \bbT\ blackbody for the fiducial flare for simplicity and ease of interpretation.
We leave it to future work to explore the influence of varying flare blackbody temperatures on more sophisticated atmospheric simulations. \\ \\ 

\subsection{EUV}
\label{sec:euv}
Constraints on EUV emission are critical to estimating EUV-driven atmospheric escape.
At present, quiescent EUV fluxes are typically estimated using empirical scalings to a star's reconstructed \lya\ flux \citep{linsky14,youngblood16}, X-ray flux \citep{chadney15}, or both \citep{louden17,king18}.
Some insight into EUV flares is accessible through \textit{EUVE} data.
\cite{gudel03} analyzed \textit{EUVE} data of AD~Leo flares and found a power-law index of -$1.1\pm0.1$.
However, the FFDs of lines formed in the chromosphere and transition region (i.e., Figure \ref{fig:line_FFDs}), have lower power-law indices generally inconsistent with a value of -$1.1\pm0.1$. 
Within the blue end of the EUV observed by \textit{EUVE}, stellar emission is mostly coronal, whereas at longer, ISM-absorbed wavelengths that cannot be observed from Earth, EUV emission is predominantly from the chromosphere and transition-region, similar to FUV emission \citep{linsky14}.
It is the ISM-absorbed portion of the EUV where ionization cross sections of H, neutral He, and \HII\ peak, and absorption by these same species in the upper atmosphere of a planet is what powers thermal escape \citep{lammer03,murray09,koskinen10}.  
Given that the intent of the fiducial flare is to provide input useful to modeling of planetary atmospheres, we consider it reasonable, in lieu of output of detailed models of stellar atmospheres, to approximate the contribution of the EUV to flares using observations of FUV transition-region emission. 

\begin{figure*}
\includegraphics{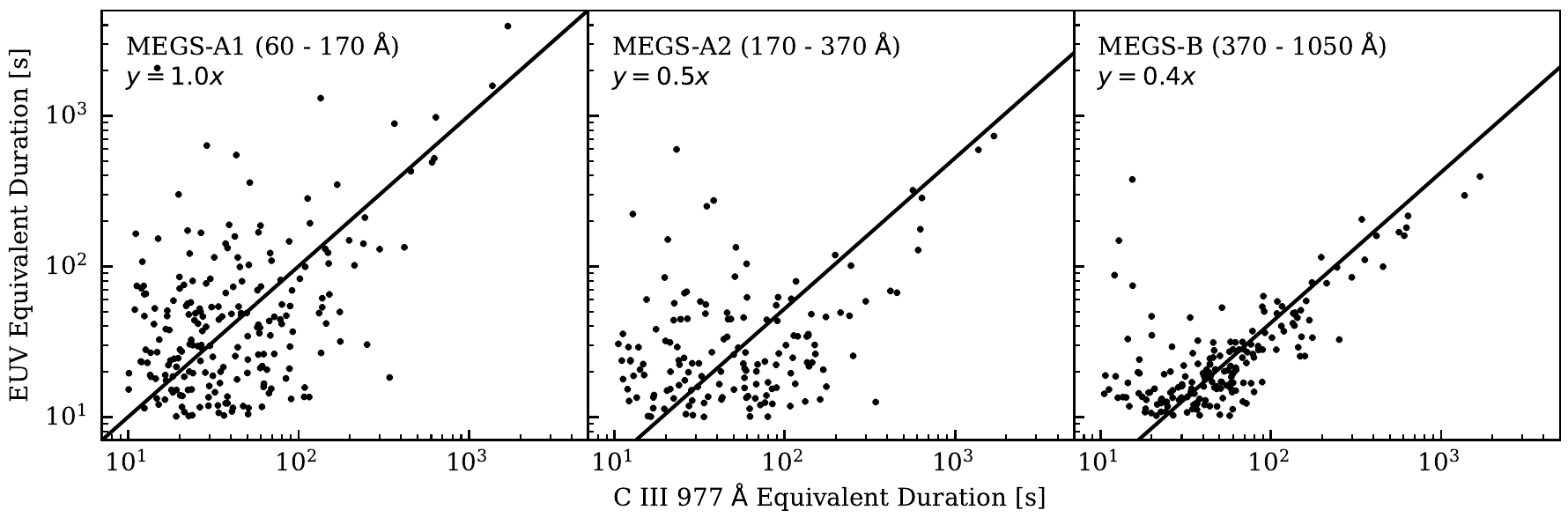}
\caption{Estimates of the equivalent duration of solar flares in three broad EUV bands versus the transition-region emission line \Ciii~977~\AA.
Lines show a linear scaling at the median ratio of the broadband to \Ciii\ equivalent durations.
Flares with equivalent duration estimates below 10 s are not shown and were omitted in computing the median.}
\label{fig:solar_euv_c3}
\end{figure*}

To this end, we estimated the energy of each observed flare in the EUV by scaling from \Ciii~1175~\AA.
This line has a proxy in the solar \textit{SDO} EVE bandpass, \Ciii~977~\AA.
Hence, we used the EVE flare catalog \citep{hock12} to determine a solar scaling in $\delta$ (equivalent duration) between three broad EUV bands and \Ciii~977~\AA\ (Figure \ref{fig:solar_euv_c3}), and took this to be representative of a scaling with \Ciii~1175~\AA\ for the \textit{HST}-observed M dwarfs.
From this scaling, we estimated the equivalent duration of flares in the three EVE bands (MEGS-A1, 60~--~100~\AA; MEGS-A2, 170~--~370~\AA; and MEGS-B, 370~--~1050~\AA)  for the M-dwarf flares identified in this study. 
The product of the equivalent duration and the quiescent EUV luminosities estimated by \cite{youngblood16} yielded the absolute EUV flare energy.
We excluded flares on AD~Leo, AU~Mic, and EV~Lac due to a lack of quiescent EUV estimates.
We then used the median ratio of these energies to \Siiv\ to include the EUV in the fiducial flare model, dividing the EUV emission into the same bands used for the quiescent estimates of \cite{youngblood16}.

For those readers desiring a more straightforward EUV flaring relationship, we distilled the above results into power laws describing an EUV FFD akin to the FUV FFDs presented in Section \ref{sec:absrel} for M dwarf flares.
These are 
\begin{equation}
\EUVpew
\end{equation}
for the equivalent duration of EUV flares from any M dwarf and
\begin{equation}
\EUVinactive
\end{equation}
for the energy of flares from inactive M dwarfs.
For active M dwarfs, we suggest the rate constant for the absolute energy FFD be increased by an order of magnitude (see Section \ref{sec:absrel}). 

The index of these power laws is appreciably shallower than those for FUV flares because the scatter in the quiescent EUV luminosity of the stars introduces additional scatter in the estimates of EUV equivalent durations and energies, stretching the FFD over a wider range.
Whether this reflects a physical reality or is merely a systematic should be tested whenever a next-generation EUV observatory becomes available.
Until such a time, the above relations provide a stopgap solution for predicting EUV flares on \mds.

\begin{figure}
\includegraphics{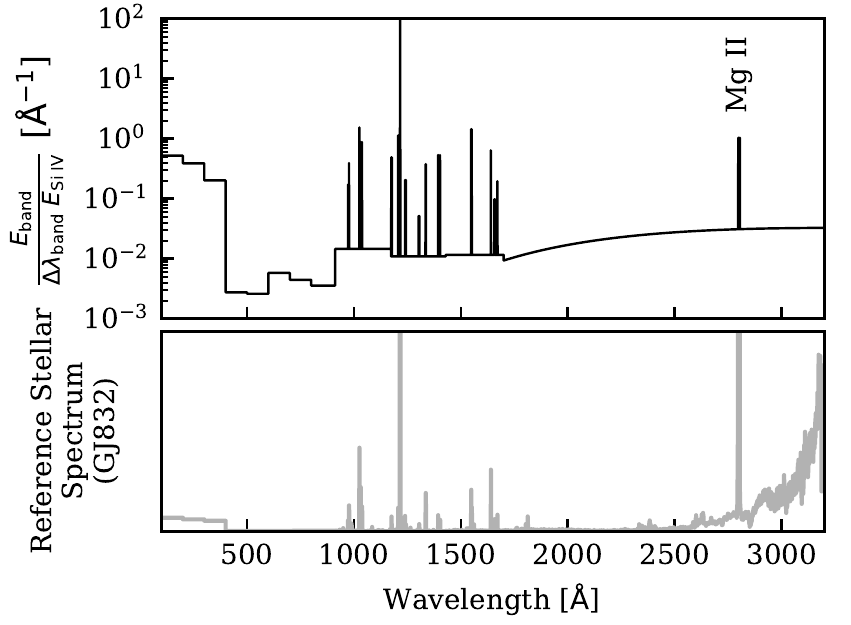}
\caption{Spectral energy budget of the fiducial flare (for photochemical and evolutionary modeling) over the full UV. 
The lower panel shows the location of major emission lines in an M dwarf spectrum for reference (linear scale).} 
\label{fig:fiducial}
\end{figure}

This completes the definition of the fiducial flare energy budget in wavelength.
The full UV energy budget is plotted in Figure \ref{fig:fiducial}.
We now move on to defining a simple but realistic distribution of energy in time.

\subsection{Lightcurve}
\label{sec:fflightcurve}
\begin{figure}
\includegraphics{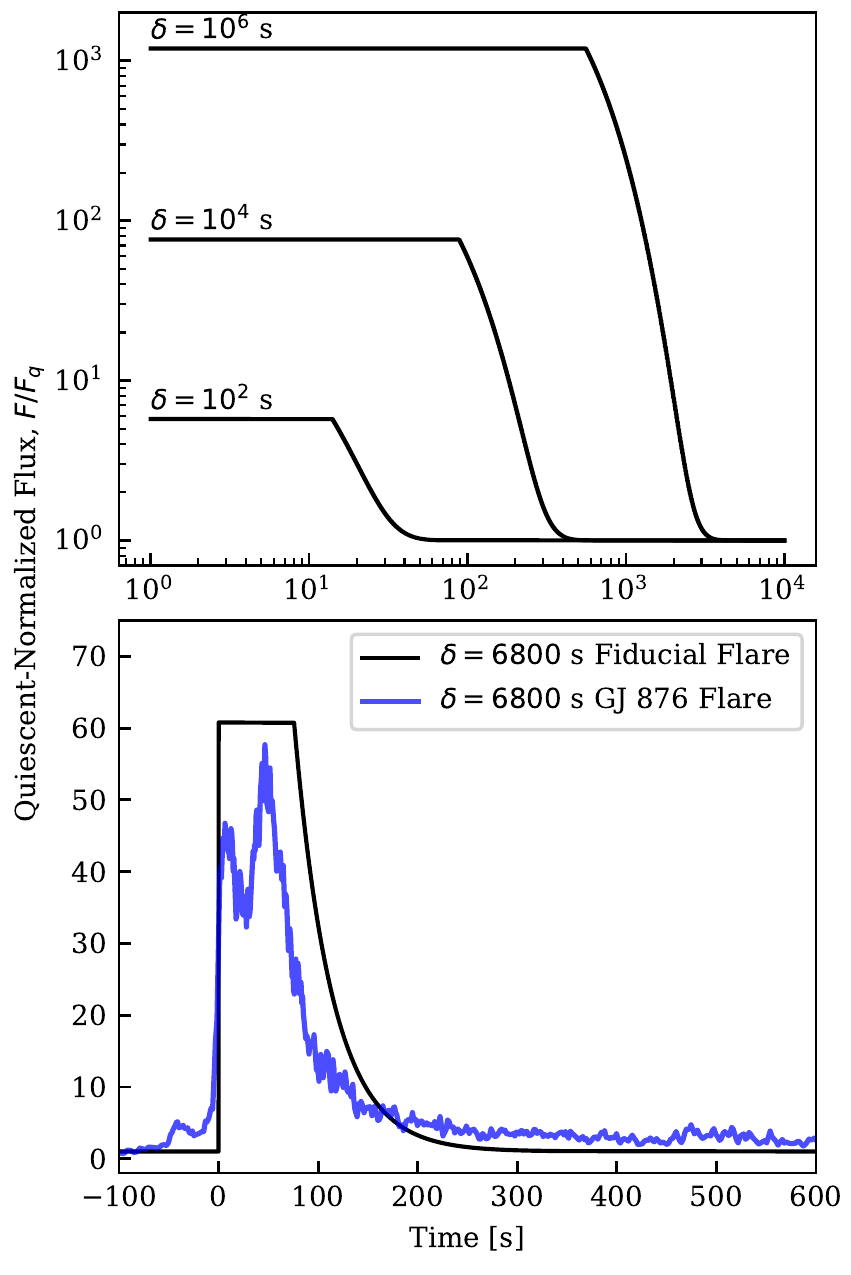}
\caption{Adopted temporal profile of the fiducial flare, shown for several values of equivalent duration, $\delta$, in logarithmic coordinates in the top plot.
The bottom plot compares the fiducial flare profile to that of a true flare on a linear scale.
The noticeable difference in areas is accounted for in the elevated flux of the GJ~876 lightcurve beyond the range of the plot.} 
\label{fig:fiducial_lightcurve}
\end{figure}

The temporal profile of the fiducial flare is a boxcar followed by an exponential decay, chosen as a simple analytic form to approximate the diverse and complex lightcurves of the observed flares.
In particular, the boxcar encapsulates the multiple, sustained peaks that accompany higher-energy FUV flares, deviating from the more canonical model of an impulsive (or Gaussian) rise followed by an exponential decay. 
The decay phase produces half the energy of the boxcar, which has a height equal to the peak flux predicted by the power-law fit of Figure \ref{fig:shape_v_pew}.
The adopted shape is plotted in Figure \ref{fig:fiducial_lightcurve}.
To maintain some semblance of simplicity in an already complex model, we assume  all bands follow the same temporal evolution. 
Both the spectral and temporal flare profiles are normalized to the \Siiv\ fluence (flux time-integrated over the entire flare), so a value for the baseline \Siiv\ flux must be specified to produce a simulated flare in absolute flux units.
We suggest adopting a quiescent \Siiv\ flux of \fidSiivqM\ for a generic inactive M dwarf and \fidSiivqF\ for a generic active M dwarf at the distance where the bolometric stellar flux is equivalent to Earth's insolation.

With this template for a UV M dwarf flare, modelers can generate consistent input to experiment with the effects of flares on planetary atmospheres.
We present results from a foray into such experimentation intended to gauge the ``photochemical power'' of such flares with respect to an Earth-like atmosphere in the next section of the paper. 

\section{M Dwarf UV Flares and Planets}
\label{sec:planets}

\subsection{An Impulsive Approximation to Flare Photolysis}
Models of planetary atmospheres incorporating the effects of flares have found significant, possibly measurable changes in atmospheric composition, but have relied on sparse UV datasets (\citealt{segura10,venot16,tilley18}; see Section \ref{sec:intro}).
To assess the potential for the flares characterized in this work to drive atmospheric photochemistry through photolysis, we created a custom ``impulsive photolysis'' model.
This model computes the  photolysis of \NII, \OII, \OIII, \HIIO, \COII, CO, \CHIV, and \NIIO\ in an Earth-like atmosphere in response to a single flare idealized as an instantaneous event. 
The model incorporates scattering by \lya, \lyb, and \lyg\ and the resonant \Oi~1305~\AA\ multiplet through the plane-parallel, quadrature two-stream radiative transfer formulation of \cite{toon89}.

In typical photochemistry models that step through time, care must be taken in non steady-state scenarios to ensure  time steps are short enough to resolve the flare and associated changes in atmospheric composition.
The ``impulsive'' model differs in that the event is treated as a single impulse, i.e., photons are treated to all arrive at essentially the same instant.
(Note  this model is therefore agnostic to the lightcurve, but we included such a lightcurve in the fiducial flare for use in time-dependent models.)

The impulsive model tracks the absorption and scattering of the flare photons and terminates when all photons have been either absorbed within the atmosphere, absorbed by the ground (assumed to be perfectly absorbing), or have exited the top of the atmosphere (scattered out). 
The result represents an upper limit on the dissociation of various species: in reality recombination reactions will be happening during any true flare, reducing the maximum quantities of dissociated species.
However, an exception to this could occur if secondary reactions amplify dissociation, such as catalytic destruction of \OIII\ by \NOx\ species.

We use the inactive star fiducial flare model of Section \ref{sec:fiducialflare} to set the spectral content of the flare input.
The modeled atmosphere is that of Earth (1 bar surface pressure) and receives a bolometric flux equivalent to Earth's insolation.
We consider only the substellar point in the atmosphere. 
Its compositional profile is a combination of several empirical models of Earth's atmosphere covering different height regimes and including different constituents:
\begin{itemize}
\item NRLMSISE-00 \citep{picone02} is a global atmospheric model commonly used for engineering purposes. We use it to specify \NII, \OII, O, H, N, Ar,  and He densities from 0~--~1000~km.
\item NASA Earth GRAM 2016\footnote{\url{https://software.nasa.gov/software/MFS-32780-2}} is a goal atmospheric model intended for environmental science. We use it to specify \HIIO, \OIII, \COII, CO, \CHIV, and \NIIO\  densities from 0~--~1000~km.
\item \cite{hu12} compiled atmospheric data from multiple sources on a variety of trace species to validate the general purpose atmospheric photochemistry model they developed. We use their profiles to specify OH, NO, and \NOII\ densities from 0~--~80~km.
\end{itemize}
\HII\ and \OID\ were included purely as dissociation products.
We use a grid of 400 layers, with thicknesses of 0.5~km up to 100~km and 4.5~km up to the top of the model at 1000~km.
The high upper boundary to the model allows important resonant scattering by atomic H and O to be included. 
This yields scattering columns of  $2.9\sn{13}$~H and $8.5\sn{17}$~O atoms~cm$^{-2}$ above 100 km (300 nbar) and $1.1\sn{14}$~H and $1.1\sn{18}$~O atoms cm$^{-2}$ at ground level. 

\begin{figure*}
\includegraphics{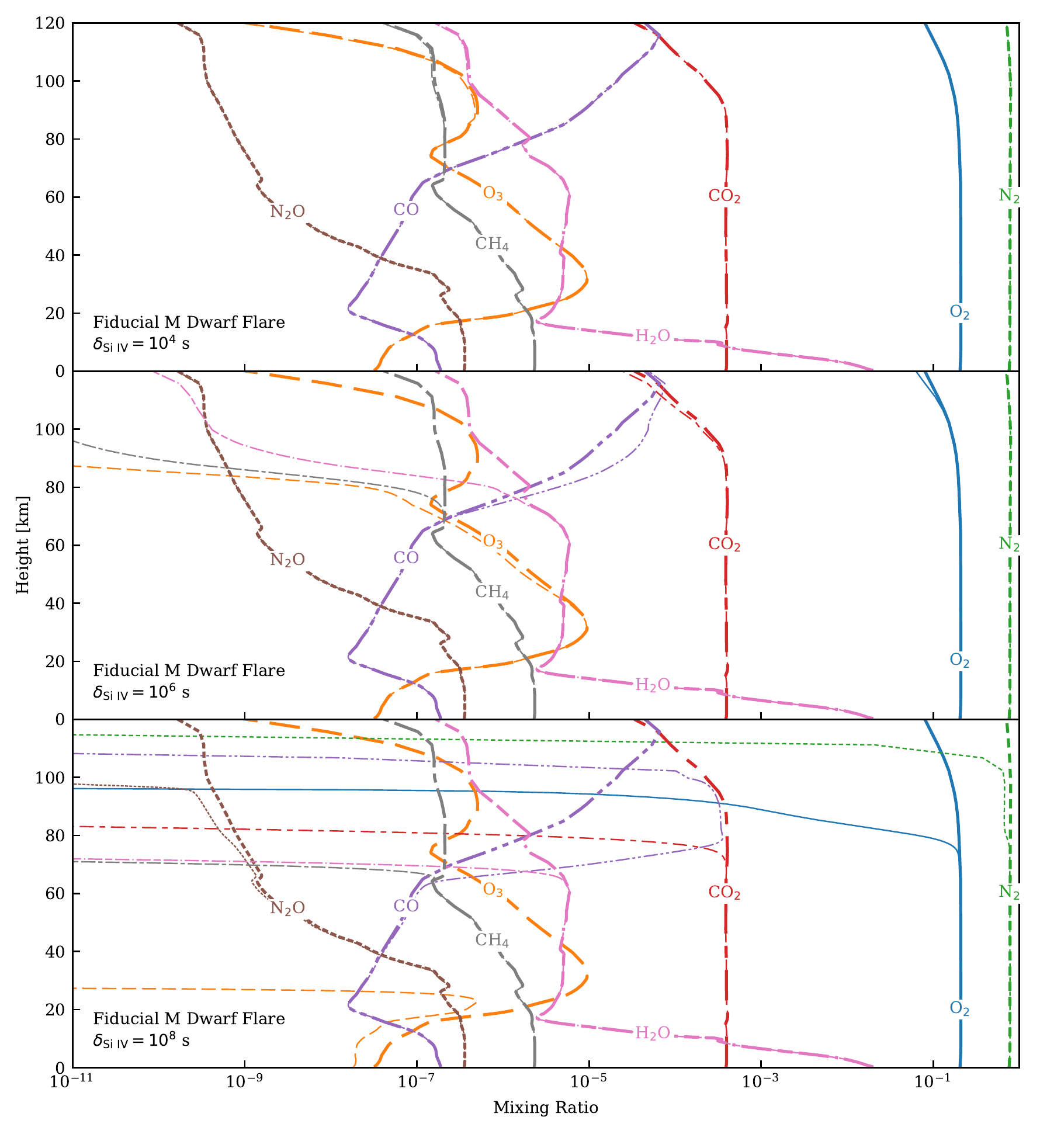}
\caption{Composition of an Earth-like atmosphere at the substellar point following a fiducial flare from a generic, inactive M dwarf for three representative flare equivalent durations.
The model incorporates scattering by resonant FUV lines and treats all photons as arriving at the same instant.
Thick lines show species concentrations before the flare and thin lines after the flare.}
\label{fig:ex_diss}
\end{figure*}

Figure \ref{fig:ex_diss} shows the state of the atmosphere before and after exposure to flares of three representative equivalent durations.
The $\delta=10^4$~s flare is akin to the largest observed in this survey, whereas the  $\delta=10^8$~s flare is likely about an order of magnitude less energetic than the largest M dwarf flare observed to date (the DG CVn flare described by \citealt{osten16}; see Section \ref{sec:elimit}).
The effects progress from negligible to dramatic over this range.
At the highest energy, the atmosphere exhibits clear dissociation fronts among the different species.
A potentially surprising feature is the increase in CO over pre-flare levels immediately below its dissociation front, but this is merely a result of the production of CO from the photolysis of \COII.

\begin{figure}
\includegraphics{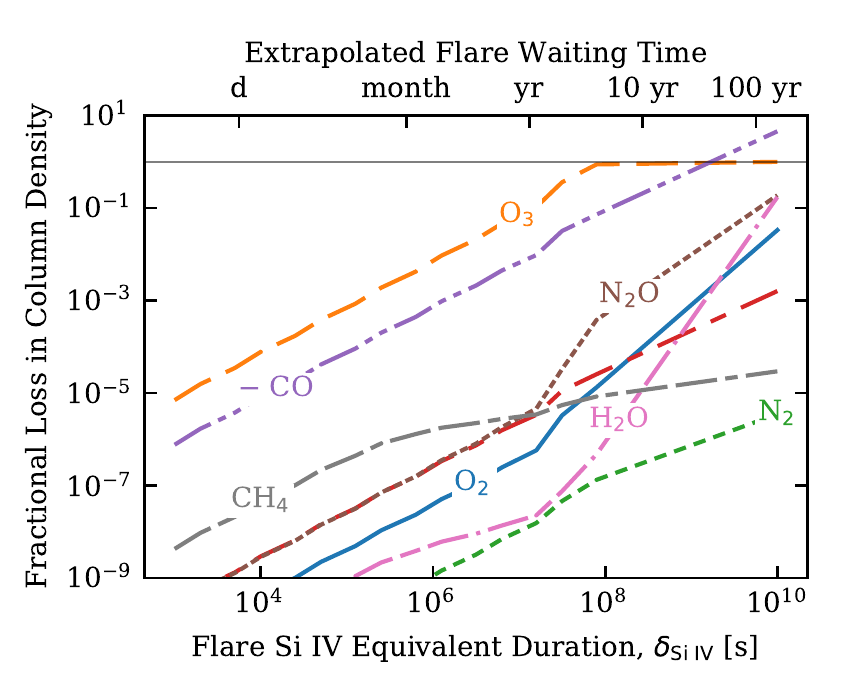}
\caption{Fractional loss in column density as a function of flare energy.
Changes in atmospheric opacity and scattering are primarily responsible for the nonlinear behavior of some species.
CO is given as the negative of the true value because more CO is created from the dissociation of \COII\ than is lost.
The top axis gives the expected time between successive flares of energy greater than that shown on the bottom axis, using on an extrapolation of a power-law fit to the FFD of all flares (Table \ref{tbl:pew_stats}).
The thin gray line corresponds to full dissociation.}
\label{fig:diss_col}
\end{figure}

\begin{figure}
\includegraphics{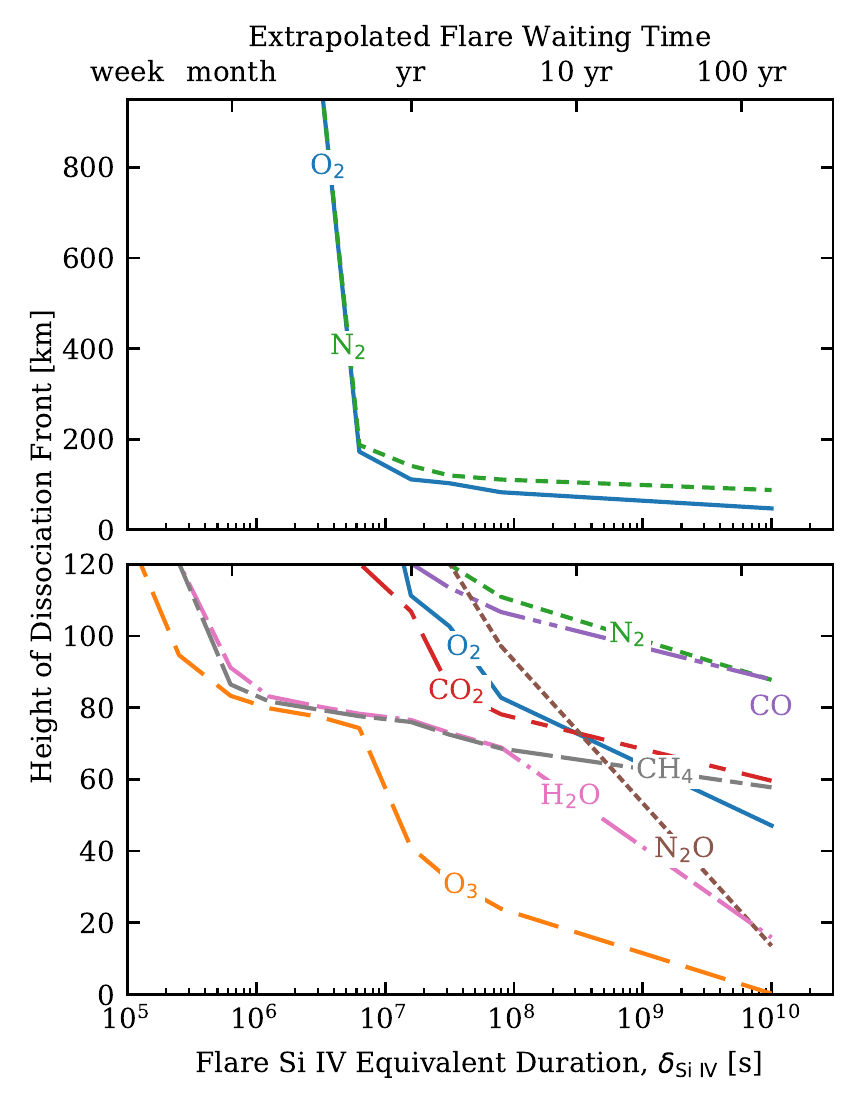}
\caption{Depth of the dissociation as a function of flare energy, defined as the altitude above which at least 90\% of the species has been dissociated. } 
\label{fig:diss_front}
\end{figure}

To examine the relationship between flare energy and the total effect on various species, we plot the fractional change in column density in Figure \ref{fig:diss_col} and the altitude of the dissociation front in Figure \ref{fig:diss_front} as a function of flare energy.
At present, the dissociation code is computationally costly, so we used only a coarse grid of 25 flare energies.
The total fraction of the column that is dissociated increases roughly linearly with flare equivalent durations, though with some additional structure due to changes in atmospheric opacity as the flare photons essentially drive deeper into the atmosphere.
No species is ever completely dissociated save \OIII\ beyond $\delta=10^8$~s.
The expected time between events of similar magnitude is included in the plots.
The flare waiting times could be compared with estimates of the time required for molecules to recombine to estimate the effects of repeated flaring, however recombination is complex and secondary reactions can be important.
Hence, a meaningful treatment requires modeling chemical networks and is beyond the scope of the present work. 

The regular flares observed in the \textit{HST} UV data to date that exhibit $\delta_\mathrm{Si\ IV} \lesssim 10^4$ s have a negligible effect on an Earth-like atmosphere.
However, within the range of flare energies observed by surveys in the \textit{U} and \textit{Kepler} bands ($\delta = 10^6-10^9$ s; Section \ref{sec:elimit}), flares begin to have a substantial effect.
For a $\delta = 10^9$~s flare, the full \OIII\ column is dissociated.

The ``impulsive dissociation'' approximation provides essentially an upper limit on the degree to which an atmosphere could be pushed out of its steady-state composition by a flare. 
The true evolution of atmospheric composition of an atmosphere over the course of one or more flares will be much more complex.
Throughout the flare(s), thermochemical reactions are ongoing.
For thermochemical reactions with short timescales relative to the flare, a quasi-equilibrium will be achieved with concentrations of reactants and products varying in lock-step with the flare.
For reactions with timescales much longer than the time between significant flares, the time-varying radiation field could be treated simply as an average of the flare and quiescent emission and yield the same results.
Reactions with timescales between these extremes and secondary reactions connecting short and long-timescale chemical pathways complicate the situation.

More sophisticated models of flare effects on planetary atmospheres that rely on UV observations of a single event, the 1985 Great Flare of AD~Leo \citep{hawley91}, have previously been conducted by \cite{segura10,venot16} and \cite{tilley18}.
The analysis by \cite{segura10} predicts a rapid recovery of the \OIII\ column density in an Earth-like atmosphere after photolysis by a flare, with the column density eventually overshooting its initial value.
We find  the initial drop in \OIII\ column  they compute roughly agrees with our results using the impulsive approximation. 
The overshoot is due to free O atoms from other species (e.g., \OII\ and \HIIO) combining with \OII\ to form an excess of \OIII.
This result was reproduced by \cite{tilley18} and extended to realistic time-series of simulated stochastic flaring, showing that secular declines in \OIII\ can be produced by the combined effect of repeated flares.
\cite{venot16} found  repeated flares incident upon hypothetical hot super-Earth and mini-Neptune atmospheres would produce short-timescale oscillations and secular deviations that could be detectable in transmission spectroscopy with the \textit{James Webb Space Telescope}.
Further exploration, guided by the M dwarf FUV flare properties characterized in this work, is merited.

\subsection{A Discussion of Mass Loss}
In addition to inducing photochemistry in an atmosphere, flares will enhance atmospheric escape.
The escape of exoplanet atmospheres, namely of hydrogen, can inhabit many different regimes according to the bottleneck of the process. 
Earth's hydrogen escape, for example, is diffusion limited: as fast as hydrogen can diffuse through the gas just above the homopause such that it is free to reach the exobase, it escapes through a balance of thermal and nonthermal processes \citep{hunten73}. 
Escape can also be limited by the rate at which atomic hydrogen is photolyzed from a heavy molecule \citep{hunten82}.

For planets orbiting very near their star, photoionization heating of the upper atmosphere by the stellar EUV can be intense enough to power hydrodynamic escape, i.e., an outflow wind \citep{lammer03}.
For \HII-dominated atmospheres, hydrodynamic escape is predicted to be bottlenecked by the rate of energy absorption (energy-limited), the rate of radiative cooling (recombination-limited), or, less-commonly, the rate of ionizations (photon-limited; \citealt{murray09,owen16b}). 
The rate of escape is energy-limited (or photon-limited) at low levels of irradiation and scales linearly with the received ionizing radiation flux $F$.
At high levels of irradiation, ionizing photons penetrate deep enough in the atmosphere to reach densities where the recombination timescale exceeds the flow timescale. 
When this happens, radiative cooling from recombinations dominates over adiabatic expansion and the mass loss rate scales roughly as $F^{1/2}$.
However, new simulations suggest that, even in hot Jupiter atmospheres, cooling from H$^+$ recombination might be secondary to cooling from other species \citep{liu17}.

If these relationships hold for the impulsive energy input of flares, then the impact of additional ionizing flux from flares is clear: it will increase the overall escape according to the $F^{1/2}$~--~$F$ scaling of the mass loss rate.
Assuming the distribution of EUV flare energies follows roughly that of FUV flare energies, then, as discussed in \ref{sec:elimit}, heating from the most energetic flares could account for most of the energy needed to liberate atmospheric mass.

Indeed, flare driven escape has been conjectured to be the cause of variability observed in the \lya\ transit of the hot Jupiter HD~189733b \citep{lecavelier12}.
Similarly, the \Ha\ transit of HD~189733b also varies significantly, with variations in the planetary atmospheric escape rate suggested as the most likely cause \citep{cauley17b}.
These variations do not correlate with the star's \Caii\ H \&\ K activity, but this would be consistent with a situation in which stochastic flaring versus steady activity is leading to variations in escape. 

Detailed modeling of HD~189733b's atmospheric response to a flare suggests  particles, rather than photons, are needed to power a sufficient increase in escape rate to explain the \lya\ transit observations \citep{chadney17}.
Further modeling is needed to determine how this result extends to other stellar types and planetary masses and whether the cumulative effect of flares will indeed play a significant role in long-term atmospheric loss.

\subsection{A Glaring Unknown in the Planetary Impacts of Stellar Flares: Particle Events}
\label{sec:particles}
While flare radiation might well dominate over quiescent radiation in powering escape, associated particle events could dominate over both.
On the Sun, coronal mass ejections are a daily event \citep{yashiro04}, and nearly all solar flares above a particular energy level (roughly GOES X class) are accompanied by CMEs \citep{aarnio11}.
Both the CMEs and solar energetic particles accelerated by CMEs and other processes (e.g., \citealt{ryan00}) can impact planets and increase escape (e.g., \citealt{jakosky15}).
This could potentially remove entire planetary atmospheres \citep{khodachenko07a,lammer07,airapetian17}.

Further, energetic particles could initiate nonthermal chemistry (beyond that initiated by the FUV photons), synthesizing precursors to organic molecules, generating greenhouse gases, and diminishing ozone \citep{segura10,airapetian16}, with effects potentially lasting for thousands of years \citep{segura10,youngblood17}.
Recent modeling by \cite{tilley18} shows  the cumulative effects of proton events based on solar scalings to M dwarf flares would destroy all but trace amounts of \OIII\ relative to the starting condition of an Earth-like atmosphere, exposing the planetary surface to extreme levels of UV flux. 

At present, there is no widely-accepted observational evidence of stellar CMEs or energetic particle events, despite some tentative evidence (e.g., \citealt{haisch83}) and some recent dedicated searches for radio bursts associated with energetic particles \citep{crosley16,villadsen17a}.
Estimates of particle fluxes associated with individual flares must therefore rely on solar scalings, such as that of \cite{youngblood17}. 

However, there are physical reasons to believe that M dwarf flares could behave quite differently than solar flares in ejecting mass.
Mass and energy budgets inferred from solar scalings produce unphysical results for M dwarfs \citep{drake13}, and strong overlying magnetic fields have been proposed as a mechanism by which M dwarfs might contain flare plasma \citep{osten15,harra16,drake16,alvarado18}.
This phenomenon has been observed for some high-energy solar flares that were not accompanied by CMEs \citep{thalmann15,sun15}.

While individual CMEs cannot yet be observed, a constraint on their collective effect is provided by the measurements of stellar astrospheres, such as those of \cite{wood05b}.
The size of the stellar astrosphere depends on the stellar outflow, itself a blend of the steady stellar wind and CMEs.
Therefore, mass loss rates inferred from stellar astrosphere measurements serve as upper limits on the CME mass loss rate, though the fractional contribution of CMEs versus the steady wind is unknown \citep{drake13}. 
\cite{wood05b} find  such mass loss weakens at an activity threshold that could indicate the emergence of strong dipolar magnetic fields suppressing outflows.
This supports the hypothesis that dipolar fields could also suppress CMEs. 

The particle events associated with flares could preclude planetary habitability, so an essential step in this field is the development of an observational means of detecting or firmly constraining the individual particle events associated with stellar flares.

\section{Summary}
\label{sec:conclusions}
We analyzed spectrophotometric FUV data from 6 M dwarfs from the MUSCLES survey and 4 M dwarf flare stars with archival data to identify and characterize flares. 
The MUSCLES stars comprised an ``inactive'' sample ($\mathrm{EW}_\mathrm{Ca\ II\ K} < 2$~\AA), whereas the flare stars comprised an active sample ($\mathrm{EW}_\mathrm{Ca\ II\ K} > 10$~\AA). 
All M dwarfs flared at least once.
The fractional contribution of these flares to the stars' FUV emission exhibited no detectable correlation with \Caii~K flux, rotation period, or effective temperature.
In other words, all M dwarfs, even those with comparatively low levels of optical chromospheric emission, flare vigorously in FUV emission.

The independence of relative flare strengths on stellar activity is reflected in the flare frequency distributions (FFDs) that relate flare energies and occurrence rates. 
When flare energies are normalized by the host star's quiescent emission using the ``equivalent duration'' metric, the FFDs of the active and inactive distributions are identical. 
This is in spite of an order-of-magnitude difference in the typical absolute energies of flares between the two groups.
Specifically, flares occurring roughly hourly on the active stars have a typical energy of $10^{28.6\pm0.2}$~erg and those occurring on the inactive stars have a typical energy of  $10^{27.7\pm0.2}$~erg, whereas in equivalent duration these values are $10^{2.0\pm0.2}$~s and $10^{1.8\pm0.2}$~s. 
However, this consistency does not span spectral types.
Comparing to the Sun, an analysis of flares using emission lines of similar formation temperatures reveals flares of similar equivalent duration are 3 orders of magnitude more frequent on M dwarfs.

A power-law fit to the cumulative FFD of flares aggregated from all stars has an index of -0.76, implying large flares are energetically more important than small ones. 
If the power-law continues to describe the FFD at flare equivalent durations 2$_{-0.5}^{+3}$ orders of magnitude beyond the largest observed, then flares will contribute more to the long-term \gfuv\ emission of M dwarfs than quiescent emission. 
This is an important result due to its implications for atmospheric photochemistry and atmospheric escape of orbiting planets.
Consequently, the exclusion of UV emission by flares will constitute a systematic error in models of atmospheric photochemistry and mass loss for orbiting planets.
However, whether flares indeed account for more \gfuv\ energy than quiescence will not be known until FUV staring observations can sample the FFDs out to flares occurring as rarely as once every 10 to 100 days. 

The spectral distribution of energy from the flares was generally consistent between events relative to the large spread  (nearly four orders of magnitude) in event energies. 
We quantified the typical flare energy budget over the observed wavelength range, then used various scaling relations to extend it from the EUV through the NUV so  it can be applied to studies of atmospheric photochemistry.
A Python module is available to generate consistent model input data from this energy budget.\footnote{\url{\fidurl}}
The energy budget consistency does not appear to extend to X-ray wavelengths, though the data are too scant for a firm conclusion. 

We applied the fiducial flare to a model of impulsive photolysis of species in an Earth-like atmosphere receiving a bolometric flux equivalent to Earth and found  a significant change in ozone begins for flares about 10$\times$ as energetic (in absolute energy, not equivalent duration) as the largest detected in this survey.
At 10$^3\times$ as energetic as the largest flare of this survey, ozone is completely dissociated.
However, most other species remain essentially unaffected above $\sim$20 km in the Earth-like atmosphere.
Nonthermal chemistry from associated particle events was not considered, but could have greater effects than photons alone. 
To comprehensively assess the climate implications of \md\ flares, a means of observationally constraining particle events and CMEs associated with stellar flares is an essential future step.

\acknowledgements
Parke Loyd wishes to express sincere gratitude to Adam Kowalski for illuminating discussions regarding stellar flares and Ignasi Ribas for similar discussions occasioned by an analysis of data on Proxima Centauri. 
Thanks is expressed to Ben Dichter for the use of the \texttt{brokenaxes} code (\url{github.com/bendichter/brokenaxes}).
This work was supported by HST grant HST-GO-13650.01 and Chandra grants GO4-15014X and GO5-16155X to the University of Colorado at Boulder.

\facilities{HST (COS, STIS), CXO, XMM}

\appendix

\section{Flare Identification Algorithm}
\label{app:flare_ident}
For the identification of flares, FUV lightcurves were created with uniform 5 s time binning, chosen to allow the detection of short, impulsive flares.
Flare identification begins by fitting quiescent variations in the stellar flux with a Gaussian process (GP), masking out points $>$2.5$\sigma$ deviant from the median.
The initial sigma clip is necessary to prevent strong flares from driving the fit to quiescence (discussed shortly).
Runs of points above and below quiescence are identified and  their area computed.
If this area is positive with value greater than 5$\sigma$ above quiescence, it is marked as a flare.
The 5$\sigma$ cutoff is a reasonable choice to minimize the possibility that the event is a more frequent, lower-energy  event that a  chance combination  of noise has pushed over the 5$\sigma$ values \citep{murdoch73}.
Any run with greater than 3$\sigma$ area was identified as anomalous.
After masking out the flares and anomalies, the quiescence was fit again and new anomalies and flares identified.
These steps were then  iterated to convergence.

Flare identification is sensitive to  the method and parameters used for fitting the quiescence, particularly for low-energy  flares. 
Underneath the white noise of the measurements are true variations in the stellar flux that are correlated in time.
This correlated astrophysical noise is prone to producing false alarms. 
Therefore, we employed a GP via the \texttt{celerite} code \citep{foreman17} so that this correlated noise could  be modeled and  false positives mitigated. 
The GP uses a covariance kernel of the form $\sigma_x^2 e^{-\Delta t / \tau}$, where $\Delta t$ is the difference in time between data points and $\sigma^2$ and $\tau$ are parameters specifying the variance and decorrelation timescale of the data to model autocorrelations in the quiescent lightcurve.
At each iteration, the algorithm  finds the best fit values of $\sigma_x$ and $\tau$ and uses these for a GP regression fit, masking  out the flare and anomalous runs.

We restricted $\tau$ to the interval [100 s, 10 d].
The low end helps avoid overfitting while the high end helps avoid a scenario where the model chooses to fit correlated noise as white noise. 
A penalty in the calculation of the data likelihood is applied based on the GP power at high frequency (0.1 Hz) to favor more gradual variations in the quiescence.
This also helps avoid overfitting, which is a known issue with GPs and was sometimes present when the penalty was not applied.
If a GP does not yield a data likelihood at least twice that of a constant value with added white noise, then the white noise model is used instead.
The regression provides a statistical estimate of the uncertainty in the quiescent flux that grows according to distance from  the nearest quiescent points, and this uncertainty is  included when  determining whether a run's area is anomalous given the noise and in the final measurements of flare parameters (e.g. energy).

We did not allow runs to extend over exposure gaps.
However, we found that the performance of the algorithm was significantly increased by expanding the span  of data masked for each anomalous run.
Specifically, we masked data starting 30~s prior to the start of the run and increased the total span  masked by a factor of two.
If the spans overlapped, we combined them.

Our algorithm occasionally reached a steady-state oscillation, associating points with a flare in one iteration, then de-associating them in the next iteration using the new version of the smoothed lightcurve, then associating them again, ad infinitum.
We programmed the pipeline to  identify  these oscillations and use the average mask  over one oscillating period as the final mask for identifying flares and fitting quiescence. 

Code for our flare identification algorithm, which we name FLAIIL (FLAre Identification in Intermittent Lightcurves) is available online.\footnote{\url{https://www.github.com/parkus/flaiil}}

\subsection{Varying Flare Identification}
\label{app:analysis_vars}
The energies and total number of flares identified by FLAIIL were somewhat sensitive to the parameters we chose constraining the algorithm.
Hence, to mitigate overprecision in the eventual fits to the flare distribution, we ran the FLAIIL nine times, each time varying one of the parameters to either the minimum or maximum value we consider reasonable while keeping all other parameters nominal.
Below, we list the parameters as (min, nominal, max):
\begin{itemize}
\item Initial sigma clip threshold: (2, 2.5, 3)
\item Sigma threshold to flag run as anomalous: (2, 3, 4)
\item Factor by which a flare or anomaly was extended when flagging data: (0.5, 1. 1.5)
\item Lower limit on $\tau$ in GP fit: (0, 100, 300).
\end{itemize}

\section{FFD Power Law Fits}
\label{app:powfit}
We here provide details on the power-law fitting.
Though the methods presented here are not new, many different techniques are employed in the literature and we wish to be explicit in how we treated the fits in this work.

FFD fits were carried out by sampling the joint likelihood of the flaring rate and the index of a power-law distribution describing the flares, i.e., $\mu$ and $\alpha$ in the  equation for the cumulative distribution,
\begin{equation}
\nu = \mu E^{-\alpha}, \label{eqn:ffd}
\end{equation}
where $\nu$ is the rate of flares with energy greater than $E$.
We assumed probability of $n$ events occurring to be independent of the probability of a given set of event energies, $\vec{E}$, based on the power-law distribution.
Thus, for a single observation, the likelihood of the data is
\begin{equation}
p(n,\vec{E};\mu,\alpha) = p(n)p(\vec{E}).
\end{equation}

For $p(n)$, we took probability of $n$ events occurring to be given by a Poisson distribution.
While this inaccurately assumes  flare events are always independent, such a distribution nevertheless describes event rates well \citep{wheatland00}.
The number of expected events, $k$, is determined by the expected rate given in Eq. \ref{eqn:ffd}, the duration of the observations ($\Delta T$), and the detection limit ($E_\mathrm{lim}$) of the observations as
\begin{equation}
k = \Delta T E_\mathrm{lim}^{-\alpha},
\end{equation}
assuming no upper limit on flare energies (or detectable energies).
This assumption is allowable given that highly energetic flares do not contribute much to the total event rates.
Then, from the Poisson distribution,
\begin{equation}
p(n) = e^{-k} \frac{k^n}{n!}.
\end{equation}

For $p(\vec{E})$, the likelihood of the observed event energies is, from the power-law distribution,
\begin{equation}
p(\vec{E}) = \prod_i \frac{\alpha-1}{E_\mathrm{lim}} \left(\frac{E_i}{E_\mathrm{lim}}\right)^{-\alpha},
\end{equation}
where $i$ indexes the event energies in $\vec{E}$ and the power-law has been appropriately normalized.
When no events were detected, we took 
\begin{equation}
p(n,\vec{E};\mu,\alpha) = p(n).
\end{equation}

The flares analyzed in this work were detected in a variety of observations with varying total duration and detection limit.
Therefore, we further assumed these observations to be independent and simply took the product of the data likelihood for each separate observation to compute a final likelihood for use in fitting and posterior sampling.
We enforced a detection limit for each dataset based on the results of injection/recovery tests (Appendix \ref{app:inject}).
These tests showed a rapid rise in completeness over a short range of flare equivalent durations to 70~--~90\% completeness levels, after which completeness slowly approached 100\% with increasing flare equivalent duration. 
We took the equivalent duration at which the rapid rise in completeness stopped as the detection limit for the dataset. 
We created a short Python code to keep track of separate observations of flares, plot detection-limit-corrected FFDs, and carry out power-law fits using the data likelihood computations described above and have made this code available online.\footnote{\url{\ffdurl}}

We used this process to generate MCMC chains fitting the flares produced by each variation of the identification procedure (Appendix \ref{app:analysis_vars}). 
We then simply stacked these chains in order to determine the median values, errors bars, and compute derived quantities.
This mitigated possible overprecision in the fits. 

\section{Injection-Recovery Tests and Systematics}
\label{app:inject}
To determine flare detection limits and assess biases, particularly  those  introduced by gaps in the data, we preformed injection--recovery tests. 
We also tested for false-positives by generating random lightcurve series from the GP quiescence model, but found that these become negligible well before survey completeness reaches moderate values, so we do not discuss false positives further.
The injection/recovery tests require  an assumption of  the  flare time profile for a given  energy or equivalent duration.
For this, we used the time-profile specified for the the Fiducial Flare (Section \ref{sec:fiducialflare}).
Before injecting simulated flares, we cleaned the lightcurves of all flares and any runs $>$3$\sigma$ from quiescence (to be conservative in the cleaning).
The gaps were then filled by pulling random data from the GP used to fit the quiescent variations using the remaining data as a prior on the random draws (with no such prior, significant discontinuities occur).
Functions for this process are contained in the FLAIIL code.
New draws to fill the gaps were made with each injection/recovery trial to avoid the propagation of any single artifact that might be introduced in this process.

We found that the presence of multiple flares in a light curve significantly affected their mutual detectability.
Hence, we injected multiple flares for each trial, using earlier results to set the input for randomly drawing flares to inject.
We drew from the FFD
\begin{equation}
\nu = 4\ \mathrm{d}^{-1}\ \left( \frac{\delta}{1000\ \mathrm{s}}\right)^{-0.75}
\end{equation}
based on the equivalent duration FFD for all stars (Table \ref{tbl:pew_stats}).
We did not include the code for this in FLAIIL as we suspect our application is too specific to be of general use. 

\begin{figure}
\includegraphics{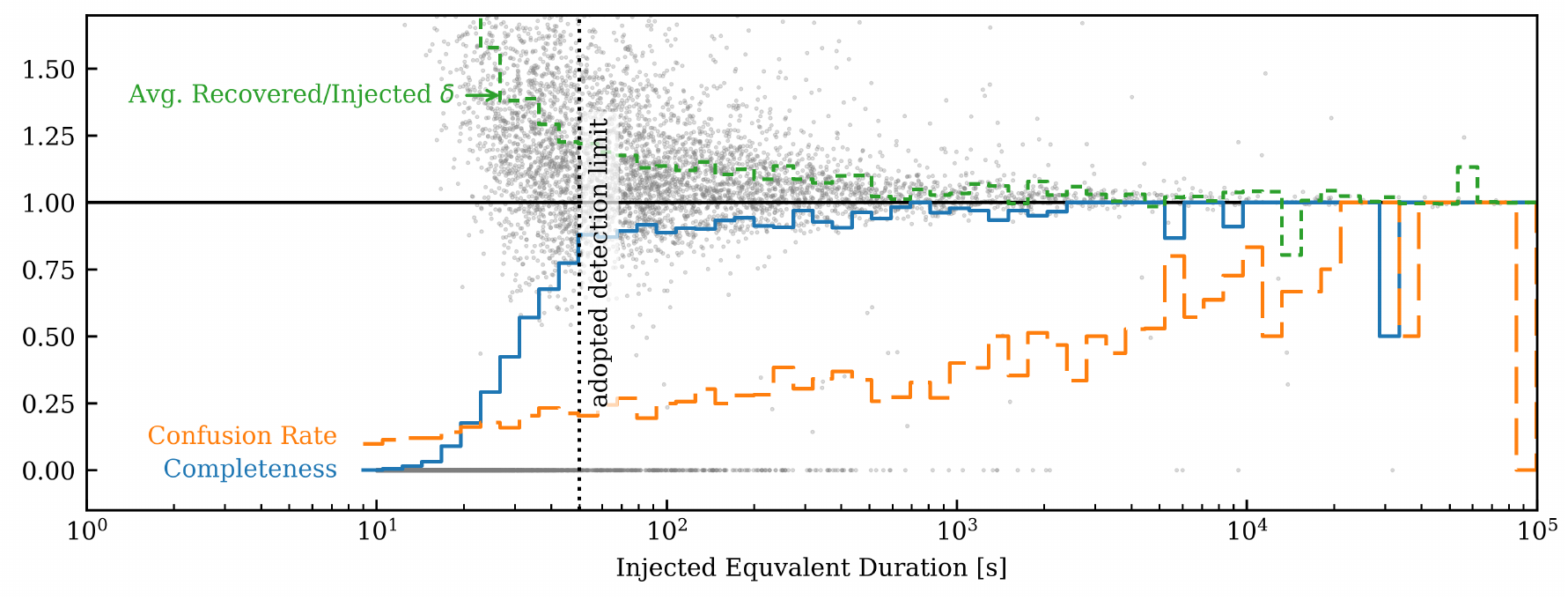}
\caption{Results of injection recovery tests for one of the datasets, the later epoch of the GJ~832 data. The background points are individual events, giving the ratio of the recovered $\delta$ to the injected $\delta$, where zeros indicate nondetections. The lower limit of injected flare $\delta$ was 10~s. See text for further discussion. The increase in completeness plateaus at around $\delta \approx 50$~s and 90\%, and this transition was adopted as the detection limit.}
\label{fig:ex_completeness}
\end{figure}

Figure \ref{fig:ex_completeness} depicts the results of an example injection/recovery analysis.
Because we drew flares from a realistic FFD for these tests, there are many more small events than large events (background points on the plot).
Confusion, here defined as the detection of multiple events as one single flare, is rampant in the tests (long-dashed orange line).
As one might expect, more energetic flares are more likely to experience confusion since their longer duration provides more time for other independent events to occur.
However, the effects of confusion are much greater for the lower energy flares.
Although less frequently confused, the energy added by a confusion is much greater for the smaller events.
This confusion results in a substantial positive bias in the recovered versus the injected $\delta$ (short-dashed green line).
However, much of this bias abates by the point where completeness (i.e. the fraction of injected flares retrieved by the algorithm, solid blue line) begins to approach 100\%.

Completeness rises rapidly over the range of a factor of a few in $\delta$, then plateaus, slowly approaching 100\% completeness over the next several orders of magnitude in $\delta$.
This behavior was present in the results from each test, though the onset of the plateau varied from 70~-~90\% depending, primarily, on the S/N of the data. 
The assumed flare light curves, rates, and distribution in $\delta$ affect the location of the sharp rise in completeness.
The steepness of the rise means that small changes in the $x$ values correspond to substantial changes in $y$ values.
Therefore, completeness at $\delta$ values within the sharp rise are in reality quite uncertain.
In consideration of this fact, we chose \textit{not} to use the completeness estimates to directly correct flares rates, even though this would have substantially increased the number of events that could be used to constrain a power law.
Instead, we used the onset of the plateau in the injection/recovery results for each dataset as a hard detection limit for flares to be considered in fitting a power law to the FFD (Section \ref{sec:ffd_fits}).
We consider this an appropriate balance between the risk of error resulting from including events where completeness errors might be large versus the reduced precision resulting from having fewer events to constrain the FFD. 

\subsection{Accuracy of Power Law Fits}
The substantial errors in the retrieved $\delta$ (or energy) of flares that can result from confusion of multiple overlapping events and the truncation by exposure gaps could lead to systematic errors in the power law fits that are not accounted for in in the statistical uncertainties listed in Tables \ref{tbl:pew_stats} and \ref{tbl:energy_stats}.
However, there is no straightforward means of correcting for these effects. 
Therefore, we used the injection/recovery tests to asses the error and possible bias in the parameters of power law fits and determine if they are problematically large.
These tests assumed a power law of the form $\nu = 4\ \mathrm{d}^{-1}\ \left(\delta/1000\ \mathrm{s}\right)^{-0.75}$ to generate random events. 

\begin{figure}
\includegraphics{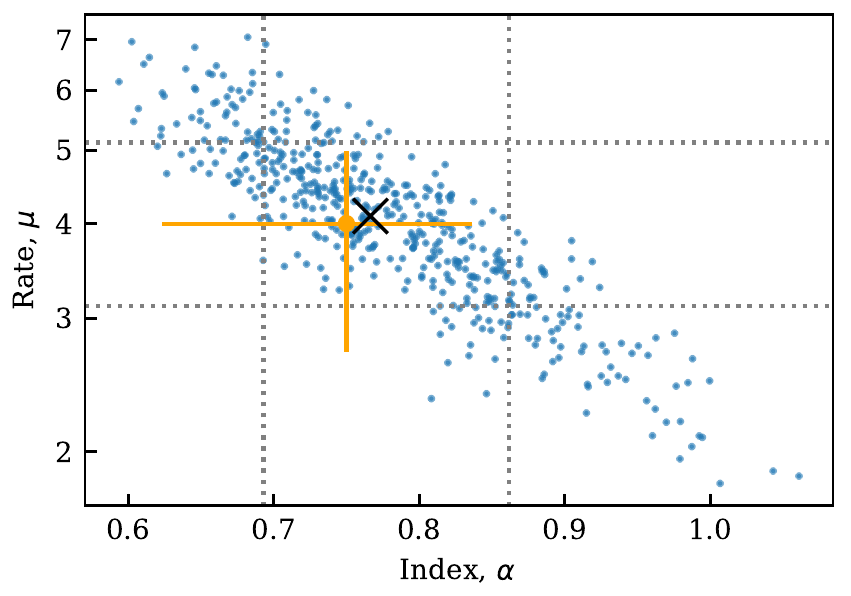}
\caption{Results of power law fits to the flares retrieved from injection/recovery tests. The orange dot marks the values used to generate flares for injection. The error bars are those of the fit to the real flare sample. Dotted lines show the \nth{16} and \nth{84} percentiles of the results and the black x shows the median values.}
\label{fig:ffd_bias}
\end{figure}

Figure \ref{fig:ffd_bias} shows the parameters resulting from fits to the retrieved flares of each injection/recovery trial.
The injection recovery trials were very computationally expensive, so the sample is limited to roughly 500 trials.
Results show the potential bias is well within the estimated uncertainty on the fit to the real-world flares presented in this work, and the error bars on that fit roughly match the \nth{16} and \nth{84} percentiles of the fit parameters estimated from the random trials.
Uncertainties were also estimated for each fit to the random trials, and in roughly 2/3 of trials the retrieved FFD parameters were within 1-$\sigma$ of truth.

\end{document}